\documentclass[twocolappendix]{emulateapj}
\usepackage{natbib}
\usepackage{graphicx}
\usepackage[space]{grffile}
\usepackage{latexsym}
\usepackage{multirow}
\usepackage{amsfonts,amsmath,amssymb}
\usepackage{url}
\usepackage[utf8]{inputenc}
\usepackage{fancyref}
\usepackage{hyperref}
\usepackage{dcolumn}
\usepackage{verbatim}

\hypersetup{colorlinks=false,pdfborder={0 0 0},}
\slugcomment{DRAFT \today}
\shorttitle{Wide DWDs and the IFMR}
\shortauthors{Andrews et al.}
\begin{document}

\newcommand{\vdag}{(v)^\dagger}
\newcommand{\truncateit}[1]{\truncate{0.8\textwidth}{#1}}
\newcommand{\scititle}[1]{\title[\truncateit{#1}]{#1}}
\newcommand{\Msun}{\ifmmode {{\rm M}_{\odot}}\else M$_{\odot}$\fi}
\newcommand{\Mdot}{\ifmmode {{\rm M}_{\odot}}\else M$_{\odot}$. \fi}
\newcommand{\Mwd}{\ifmmode {M_{\rm WD}}\else $M_{\rm WD}$\fi}
\newcommand{\tauc}{\ifmmode {\tau_{\rm cool}}\else $\tau_{\rm cool}$\fi}
\newcommand{\tauca}{\ifmmode {\tau_{\rm cool,1}}\else $\tau_{\rm cool,1}$\fi}
\newcommand{\taucb}{\ifmmode {\tau_{\rm cool,2}}\else $\tau_{\rm cool,2}$\fi}
\newcommand{\ta}{\ifmmode {\tau_1}\else $\tau_1$\fi}
\newcommand{\tb}{\ifmmode {\tau_2}\else $\tau_2$\fi}
\newcommand{\Rsun}{${R_{\odot}}$ }
\newcommand{\lapprox }{{\lower0.8ex\hbox{$\buildrel <\over\sim$}}}
\newcommand{\gapprox }{{\lower0.8ex\hbox{$\buildrel >\over\sim$}}}
\newcommand{\amin}{\ifmmode {^{\prime}\ }\else$^{\prime}$\fi}
\newcommand{\asec}{\ifmmode {^{\prime\prime}}\else$^{\prime\prime}$\fi}
\newcommand{\degree}{\ifmmode {^\circ}\else$^\circ$\ \fi}
\newcommand{\halpha}{$H\alpha$\ }
\newcommand{\Ro}{$R_o\ $}
\newcommand{\Teff}{\ifmmode {T_{\rm eff} }\else $T_{\rm eff}$\fi}
\newcommand{\logg}{\ifmmode {{\rm log}\ g }\else log~$g$\fi} 
\newcommand{\bs}[1]{\boldsymbol{#1}}
\newcommand{\given}{\,|\,}
\newcommand{\der}{\ifmmode {\rm d}\else d\fi}

\title{Constraints on the Initial-Final Mass Relation from Wide Double White Dwarfs}

\author{Jeff J.\ Andrews\altaffilmark{1}, 
Marcel A.\ Ag\"ueros\altaffilmark{1},
A.\ Gianninas\altaffilmark{2},
Mukremin Kilic\altaffilmark{2},
Saurav Dhital\altaffilmark{3},
Scott F.\ Anderson\altaffilmark{4}} 

\altaffiltext{1}{Department of Astronomy, Columbia University, 550 West 120th St., New York, NY 10027, USA}
\altaffiltext{2}{Department of Physics and Astronomy, University of Oklahoma, 440 West Brooks St., Norman, OK, 73019, USA}
\altaffiltext{3}{Embry-Riddle Aeronautical University, 600 South Clyde Morris Blvd., Daytona Beach, FL 32114, USA}
\altaffiltext{4}{Department of Astronomy, University of Washington, Box 351580, Seattle, WA 98195, USA}

\begin{abstract}
We present observational constraints on the initial-final mass relation (IFMR) using wide double white dwarfs (DWDs). We identify 65 new candidate wide DWDs within the Sloan Digital Sky Survey, bringing the number of candidate wide DWDs to 142. We then engage in a spectroscopic follow-up campaign and collect existing spectra for these objects; using these spectra, we derive masses and cooling ages for 54 hydrogen (DA) WDs in DWDs. We also identify one new DA/DB pair, four candidate DA/DC pairs, four candidate DA/DAH pairs, and one new candidate triple degenerate system. Because wide DWDs are co-eval and evolve independently, the difference in the pre-WD lifetimes should equal the difference in the WD cooling ages. We use this to develop a Bayesian hierarchical framework and construct a likelihood function to determine the probability that any particular IFMR fits a sample of wide DWDs. We then define a parametric model for the IFMR and find the best parameters indicated by our sample of DWDs. We place robust constraints on the IFMR for initial masses of 2--4 \Msun. The WD masses produced by our model for stars within this mass range differ from those predicted by semi-empirical fits to open cluster WDs. Within this mass range, where there are few constraining open cluster WDs and disagreements in the cluster ages, wide DWDs may provide more reliable constraints on the IFMR. Expanding this method to the many wide DWDs expected to be discovered by {\it Gaia} may transform our understanding of the IFMR.
\end{abstract}


\section{Introduction}
Accurate mass measurements for large numbers of hydrogen-atmosphere (DA) white dwarfs (WDs), which dominate the WD population, became commonplace with the advent of spectroscopic surveys such as the Palomar-Green Survey \citep{green86} and the Sloan Digital Sky Survey \citep[SDSS;][]{york00}. These confirmed that the DA mass distribution is strongly peaked at 0.6~\Msun\ \citep[e.g.,][]{kepler15}. However, matching these final WD masses to initial, zero-age-main-sequence masses is challenging, and large uncertainties about which main-sequence stars evolve into which WDs are still the norm. For example, the data cannot accurately tell us which stars produce those 0.6~\Msun\ DAs, with estimates ranging from 1.0 to 2.5~\Msun\ \citep{weidemann00}. This implies that predictions for mass loss as stars evolve into 0.6 \Msun\ WDs can differ by $>$1~\Mdot

\citet{sweeney76} pioneered the most commonly used method for constraining the initial-final mass relation (IFMR). The cooling age (\tauc) of a WD in an open cluster is derived from spectroscopy, and this age is subtracted from the cluster's age to determine the WD progenitor's main-sequence lifetime. Using stellar evolution codes, this lifetime is converted into an initial mass $M_{\rm i}$; paired with its spectroscopically determined mass, this $M_{\rm i}$ then provides a constraint on the IFMR \citep[e.g.,][]{weidemann83,weidemann00}.

In practice, this method is often difficult to implement: the open clusters must have accurate ages, member WDs must be identified and separated from contaminating objects, and these often faint WDs must be observed with high-resolution spectrographs. Furthermore, most accessible open clusters are $\lapprox$600 Myr old, so that only stars with $M_{\rm i}\ \gapprox\ 3.5$ \Msun\ have evolved into WDs \citep[there has been some recent work to identify WDs in older clusters; e.g.,][]{kalirai08,kalirai14}.

\citet{finley97} used a different approach. The co-eval WDs in the wide double WD (DWD) PG 0922$+$162 can be considered to have evolved independently. By comparing the more massive WD to massive WDs with accurate $M_{\rm i}$ determinations in open clusters, these authors assigned PG 0922$+$162B a $M_{\rm i}$ of 5.5--7.5~\Msun. Adding the corresponding main-sequence lifetime to the massive WD's \tauc, \citet{finley97} derived a system age of $\approx$320$\pm$32 Myr. Finally, these authors calculated the pre-WD lifetime of the less massive WD by subtracting its \tauc\ from the system age, and, using stellar evolution codes, found $M_{\rm i} = 3.8$$\pm$0.2 \Msun\ for this WD. The uncertainty on this point in the initial-final mass plane is comparable to that for the best open cluster data.  

While promising, this study has not been widely replicated. Until recently, there were only $\approx$35 known wide DWDs, and many were poorly characterized. Most lacked spectra, and even those with spectra were ill-suited to this analysis because of large uncertainties in their \tauc. Furthermore, it has not been clear how to convert observations into robust constraints on the IFMR.\footnote{We have found three other instances where this method was applied to a wide DWD. It was first used by \citet{greenstein83} on Gr 576/577, but one of the WDs in this pair is composed of an unresolved double degenerate \citep[][]{maxted00}. Because of the potential for mass transfer within the unresolved binary system, this is not a good system for constraining the IFMR. \citet{girven10} applied the method to PG 1258$+$593 to constrain the initial mass of its magnetic companion SDSS J130033.48$+$590407.0. While of use for studying the origin of magnetic WDs \citep[e.g.,][]{dobbie12}, this is another instance where the system does not place useful constraints on the IFMR. Recently, \citet{catalan15} developed her own form of this method. Although promising, it has generated meaningful constraints for only four systems so far.} 

In this work, we develop a statistical model that allows any well-characterized wide DWD to constrain the IFMR. We first construct a likelihood function to determine the probability that any particular IFMR fits a sample of wide DWDs, while taking into account observational uncertainties by marginalizing over the underlying parameters from the observables. We then develop a four-parameter piecewise-linear model for the IFMR, and iterate over the model parameters using a Markov Chain Monte Carlo technique to find the best parameters indicated by our wide DWDs. 

In \citet[][hereafter Paper I]{andrews12} we presented the results of a search for new DWDs in the SDSS Data Release 7 \citep[DR7;][]{DR7paper}. Here, we begin by describing in Section~\ref{search} the results of a search for DWDs in the larger SDSS Data Release 9 \citep[DR9;][]{DR9paper}. We assemble a catalog of 142 candidate and confirmed DWDs that includes new systems, those found in Paper I and in the literature, and those recently published by \citet[][]{baxter14}. In Section~\ref{sec:spec}, we discuss our spectroscopic follow-up observations of a subset of WDs in these binaries, and present the results of our model fits to these spectra. In Section~\ref{sec:meth}, we revisit the \citet{finley97} result and develop our hierarchical Bayesian model. We test our model on mock data, then apply it to a well-characterized subset of wide DWDs. In Section~\ref{sec:dis} we discuss our resulting constraints on the IFMR; we conclude in Section~\ref{sec:concl}.

\begin{figure}[t!]
\begin{center}
\includegraphics[width=0.73\columnwidth,angle=90]{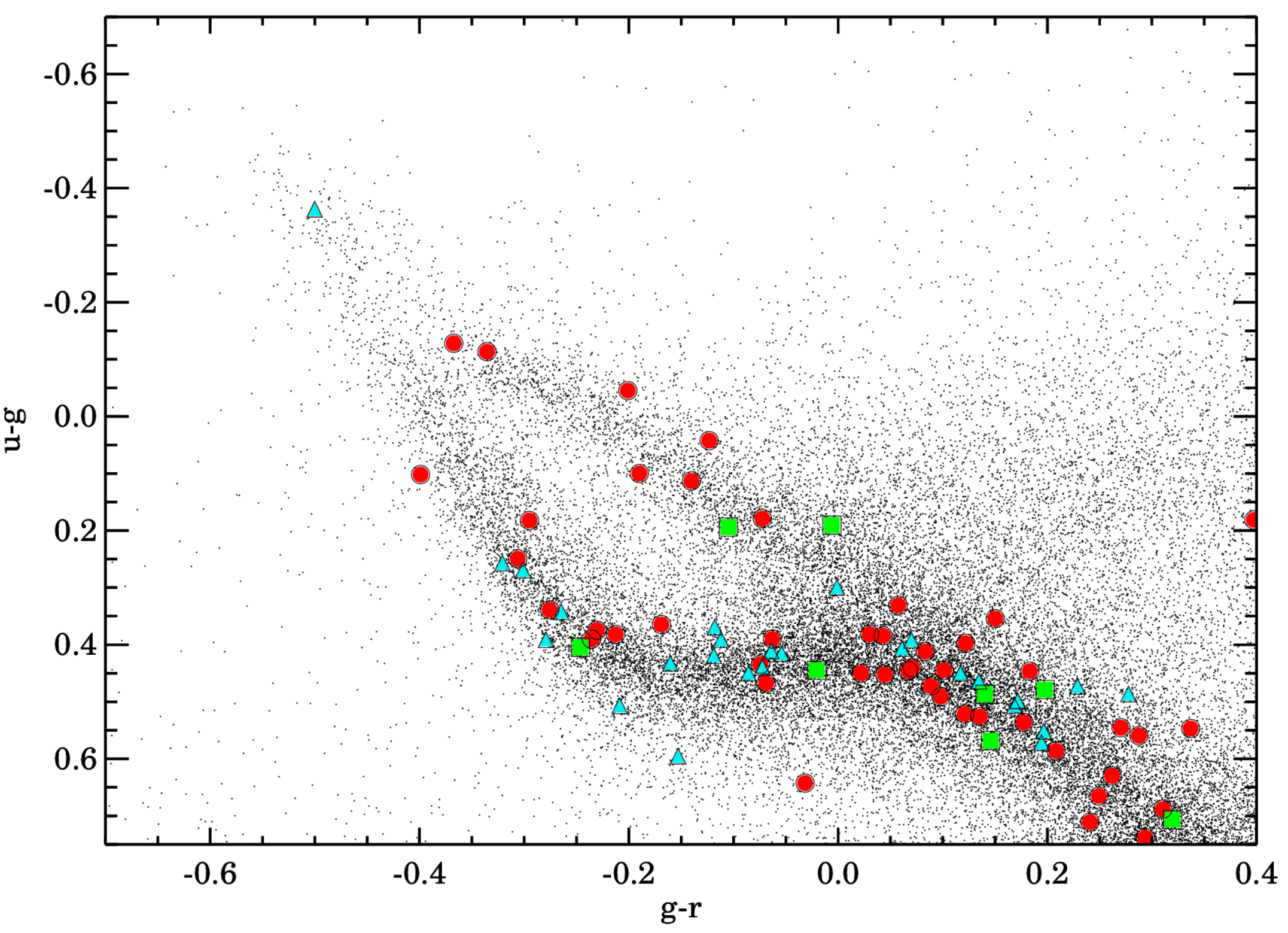}
\caption{$(u-g)$ versus $(g-r)$ for $\approx$4$\times10^4$ SDSS objects with $\sigma_{ugr} < 0.15$ mag, $\sigma_{iz} < 1.0$ mag, $\mu > 35$ mas~yr$^{-1}$, and $\sigma_{\mu} < 10$ mas yr$^{-1}$. The plot boundaries define the box in color space within which we sought to identify new WDs. The sickle-shaped band corresponds to the DA cooling sequence, with hotter, younger WDs toward the upper left, and cooler, older WDs toward the bottom right. The width of the band is primarily due to variations in the DA masses: for example, WDs slightly more massive than the canonical 0.6~\Msun\ have slightly higher surface gravities, and therefore slightly bluer $(u-g)$. The band of objects with bluer $(u-g)$ for a given $(g-r)$ is the cooling sequence for non-DA WDs. Hot DB WDs lie at the upper left of this band; as they cool to $\lapprox$12,000~K they appear as DC WDs with featureless blackbody spectra. The red circles are the candidate WDs in 23 new candidate wide DWDs identified here; the green squares are candidate WDs in four additional new candidate systems, while the cyan triangles are the WDs in the 13 known systems we re-detect.}
\label{fig:cpm_color}
\end{center}
\end{figure}

\section{Searching for Wide DWDs}\label{search}
\subsection{Common Proper Motion Pairs} \label{search:dr9}
We first search for DWDs by matching proper motions of WD candidates in SDSS, which requires accurate photometry and astrometry. Although the SDSS Data Release 8 is nearly triple the size of the DR7 photometric catalog, the astrometric solutions are not calibrated against the USNO CCD Astrograph Catalog data \citep{zacharias04}, causing a systematic shift of $\approx$50~mas~yr$^{-1}$ \citep{munn08}. DR9, however, includes an expanded photometric catalog and improved astrometric solutions (with a precision of a few mas~yr$^{-1}$), which allowed us to expand the search for DWDs described in Paper I. From the $>$9$\times10^8$ primary photometric objects in DR9, we selected those classified as stars (\texttt{ptype $=6$}) and matching our photometric and proper motion ($\mu$) quality constraints ($\sigma_{ugr} < 0.15$ mag, $\sigma_{iz} < 1.0$ mag, $\mu > 35$ mas~yr$^{-1}$, $\sigma_{\mu} < 10$ mas yr$^{-1}$). 

In Paper I, we used the color-color regions defined by \citet[][]{girven11} to identify WDs in $(u-g)$ versus $(g-r)$ space (the regions are shown in Figure~\ref{fig:nopm_color}). To include helium-atmosphere DB WDs in this search, we used a more liberal color constraint, selecting those stars with $-0.7 < (g-r) < 0.4$ and $-0.7 < (u-g) < 0.75$. Figure \ref{fig:cpm_color} shows the $(u-g)$ versus $(g-r)$ colors of the $\approx$4$\times10^4$ SDSS objects that met our quality constraints and fell within this region of color space. 

Quasars (QSOs) are by far the biggest contaminant in this region of color-color space. \citet{munn04} found that requiring $\mu > 10$~mas~yr$^{-1}$ eliminated 95\% of QSOs with $r<$ 20 mag. Our $\mu > 35$~mas~yr$^{-1}$ criterion should eliminate nearly all QSOs. Contaminating main-sequence stars and halo subdwarfs (metal-poor, Population II main-sequence stars) are more difficult to remove, as these objects may overlap with WDs in color and may have $\mu > 35$ mas~yr$^{-1}$. As shown in \citet{kilic06}, however, WDs can be effectively separated from blue stars in a reduced proper motion (H$_{\rm r}$) diagram.\footnote{${\rm H}_r = r + 5~{\rm log}~\mu + 5.$} 

\begin{figure}[!b]
\begin{center}
\includegraphics[width=0.76\columnwidth,angle=90]{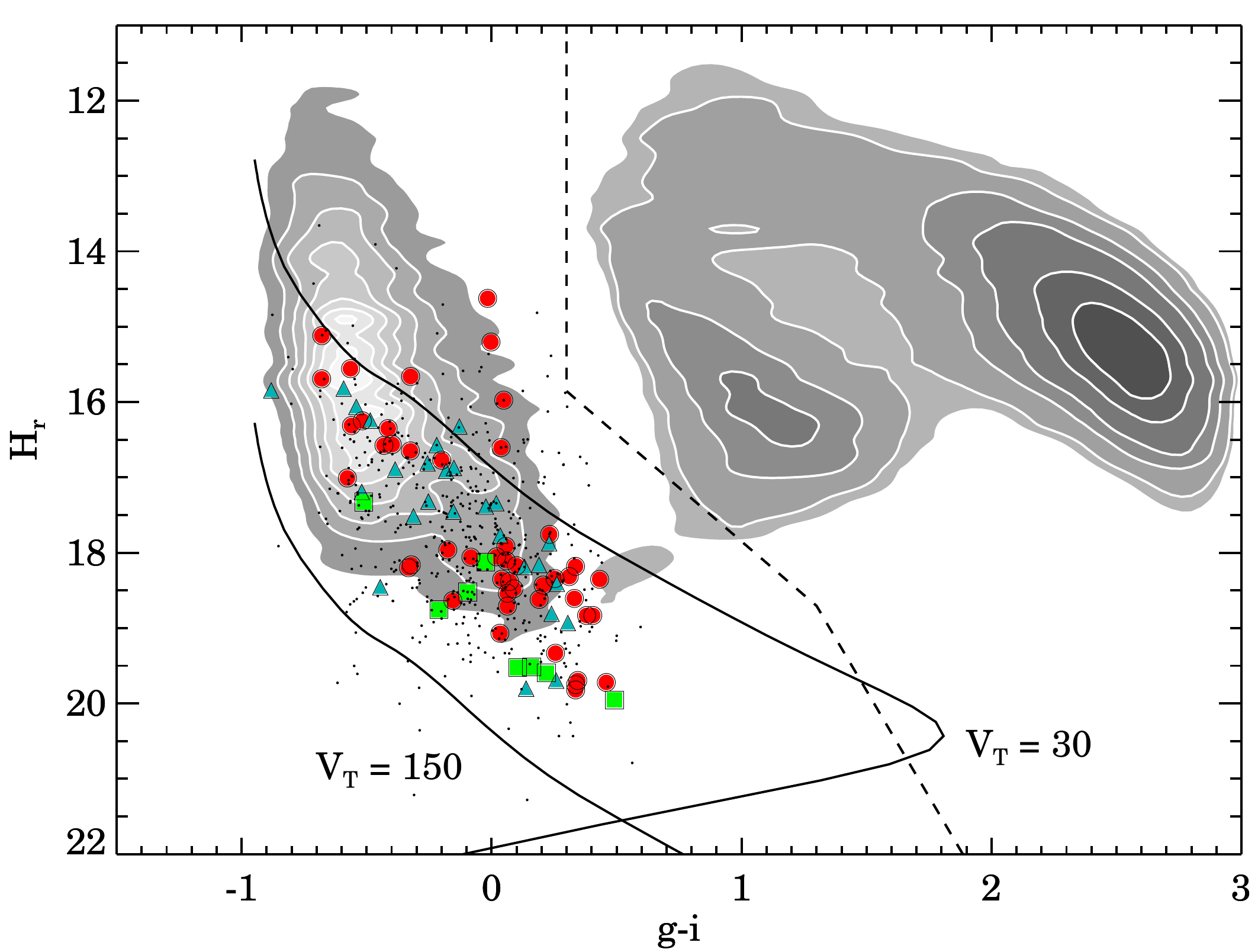}
\caption{Reduced proper motion H$_{\rm r}$ versus $(g-i)$ for the objects in Figure~\ref{fig:cpm_color}. The locus at $(g-i) \approx 2.5$ is of main-sequence stars, while that at $(g-i) \approx 1$ is of halo subdwarfs. The spectroscopically confirmed WDs in the \citet{kleinman13} SDSS catalog with a measured $\mu$ are shown by the contours peaking at $(g-i) \approx -0.5$. We used the dashed line, adapted from \citet{smith09}, to separate halo subdwarfs from candidate WDs. The solid lines represent the locations of WDs for transverse velocities V$_{\rm T}$ = 30~km~s$^{-1}$ (corresponding to the disk population) and 150~km~s$^{-1}$ (the halo population), and show that our candidates are likely in the Galactic disk. The symbols are the same as in Figure~\ref{fig:cpm_color}. Our candidates have systematically larger H$_{\rm r}$ because we require that they have relatively large $\mu$ and because the SDSS photometric catalog extends to fainter magnitudes than the spectroscopic catalog used by \citet{kleinman13} to identify WDs. }
\label{fig:rpm} 
\end{center}
\end{figure}

Figure~\ref{fig:rpm} is H$_{\rm r}$ versus $(g-i)$ for the objects in our sample. Because of their smaller radii, WDs are clearly separated from main-sequence stars and halo subdwarfs. We used the dashed line in Figure~\ref{fig:rpm}, adapted from \citet{smith09}, to separate halo subdwarfs from WDs, thereby reducing our sample to $\approx$34,000 objects.

Next, we searched for common proper motion matches within these data. We defined a match as occurring when two WDs had an angular separation $\theta < 5$\amin, and, following \citet{dhital10}, that their proper motions have a matching parameter $\Sigma^2 < 2$.\footnote{$\Sigma^2 =\left( \Delta \mu_{\alpha}/\sigma_{\Delta \mu_{\alpha}} \right)^2 + \left( \Delta \mu_{\delta}/\sigma_{\Delta \mu_{\delta}} \right)^2 $, where $\Delta \mu$ is the difference in $\mu$ measured in right ascension ($\alpha$) and declination ($\delta$), and $\sigma_{\mu}$ is the error in $\mu$.} Among our $\approx$34,000 objects, we thus identify 57 candidate DWDs.

Some of these candidate pairs may contain WDs at different distances and with different radial velocities, which, when projected on the sky, happen to result in $\Sigma^2 < 2$. Since the probability of finding a randomly aligned pair depends on the volume of phase space being searched, the random alignment likelihood should scale linearly with $\theta$, and a critical $\theta$ should separate pairs more likely to be real from those more likely to be contaminants. In Paper I, we found that pairs with $\theta < 1$\amin\ are most likely real. However, because our selection criteria have changed, we re-estimate this critical $\theta$.

\begin{figure}
\begin{center}
\includegraphics[width=0.7\columnwidth,angle=90]{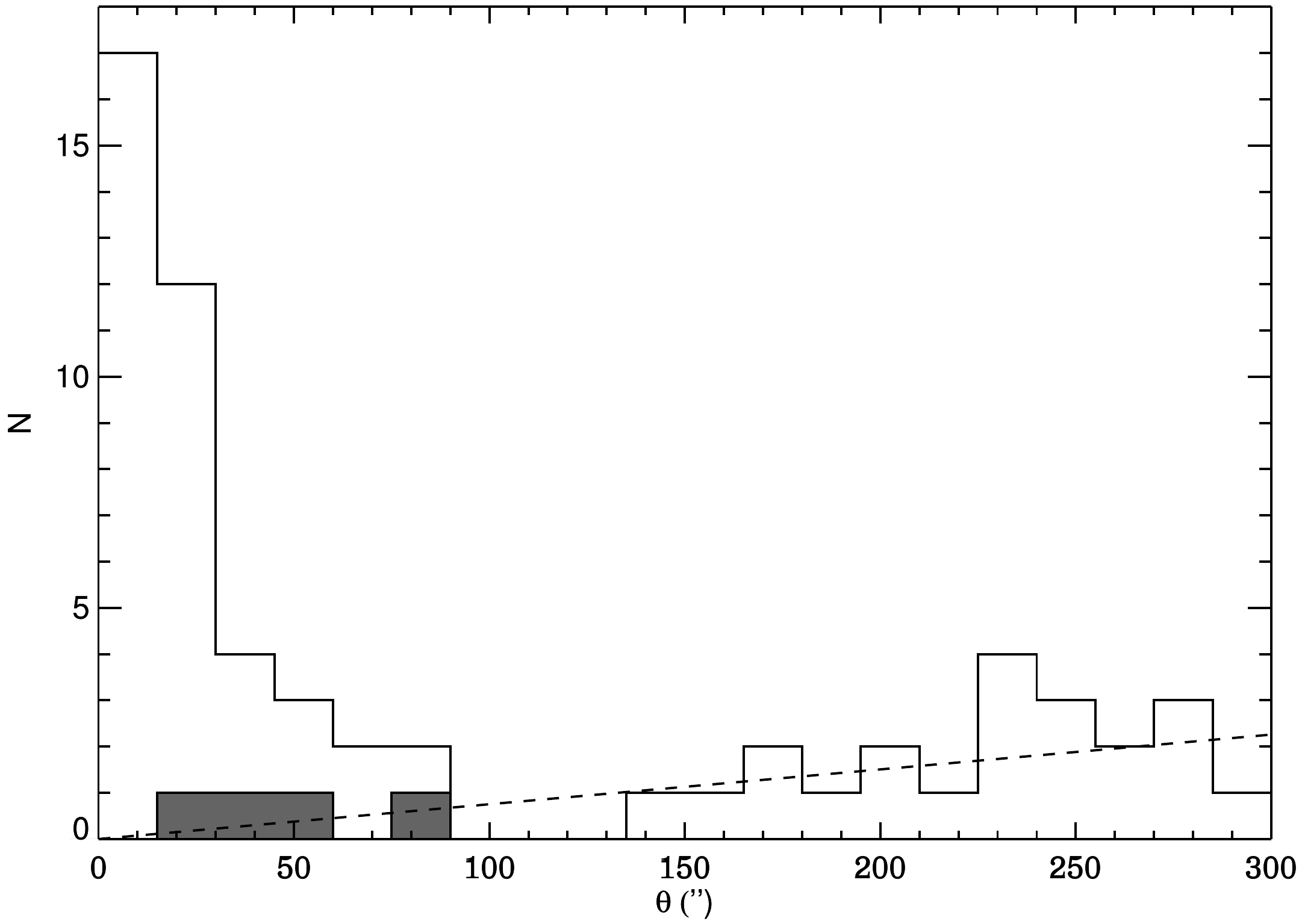}
\caption{Angular separation ($\theta$) distribution of our candidate common proper motion DWDs. The four pairs with $\Sigma^2 > 2$ but $\mu>80$ mas yr$^{-1}$ are shown in gray. The probability that a given pair is due to a random alignment should increase roughly linearly with $\theta$. The dashed line shows the best fit line to the distribution beyond 100\asec, where the noise dominates, assuming N($\theta$) $\propto \theta$ and that the line goes through the origin. Extrapolating this fit to smaller $\theta$ suggests that pairs with $\theta < 100$\asec\ are real pairs, while pairs at larger $\theta$ are physically unassociated. }\label{fig:cpm_theta}
\end{center}
\end{figure}

Figure \ref{fig:cpm_theta} shows a clear excess in the number of candidate pairs with $\theta < 100$\asec. We determine a linear fit to the distribution of systems with $\theta > 100\asec$, those that are most likely to be random alignments. Extrapolating this fit to those pairs with smaller $\theta$ suggests that $\lapprox$2 of the systems with $\theta < 100$\asec\ are likely to be random alignments. We conclude that the 36 pairs with $\theta < 100$\asec\ are high-probability DWD candidates (13 are re-detections of previously known systems).

The likelihood of random alignments drops dramatically as $\mu$ becomes larger. We therefore also searched for pairs with $\mu>80$ mas yr$^{-1}$, requiring only that $\Sigma^2<10$. We identified an additional four pairs in this manner, bringing the total number of newly identified candidate common proper motion DWDs to 27. 

\subsection{Astrometrically Close Pairs} \label{search:no_pm}
The population synthesis simulations described in Paper I predicted that the population of wide DWDs should have a minimum orbital separation of a few 10$^2$ AU. At typical distances to photometrically identified WDs, this corresponds to $\theta \approx 1$\asec, within the resolving limit of SDSS photometry. There should therefore be a substantial population of wide DWDs in SDSS with $\theta \leq 7$\asec, the minimum separation identifiable through proper motion matching to the USNO-B photometric plates. Can these pairs be identified through other means?

\citet{dhital15} search for photometrically resolved pairs of low-mass stars in SDSS with small $\theta$. These authors identify $>$40,000 binaries with $\theta$ of 0$\farcs$4$-$10\asec\ and argue that wide pairs can be efficiently identified without having to match proper motions. Similarly, \citet{baxter14} identified a set of wide DWDs with $\theta\ \lapprox\ 30$\asec\ in DR7 based exclusively on photometry. 

\begin{figure}[b!]
\begin{center}
\includegraphics[width=\columnwidth,angle=90]{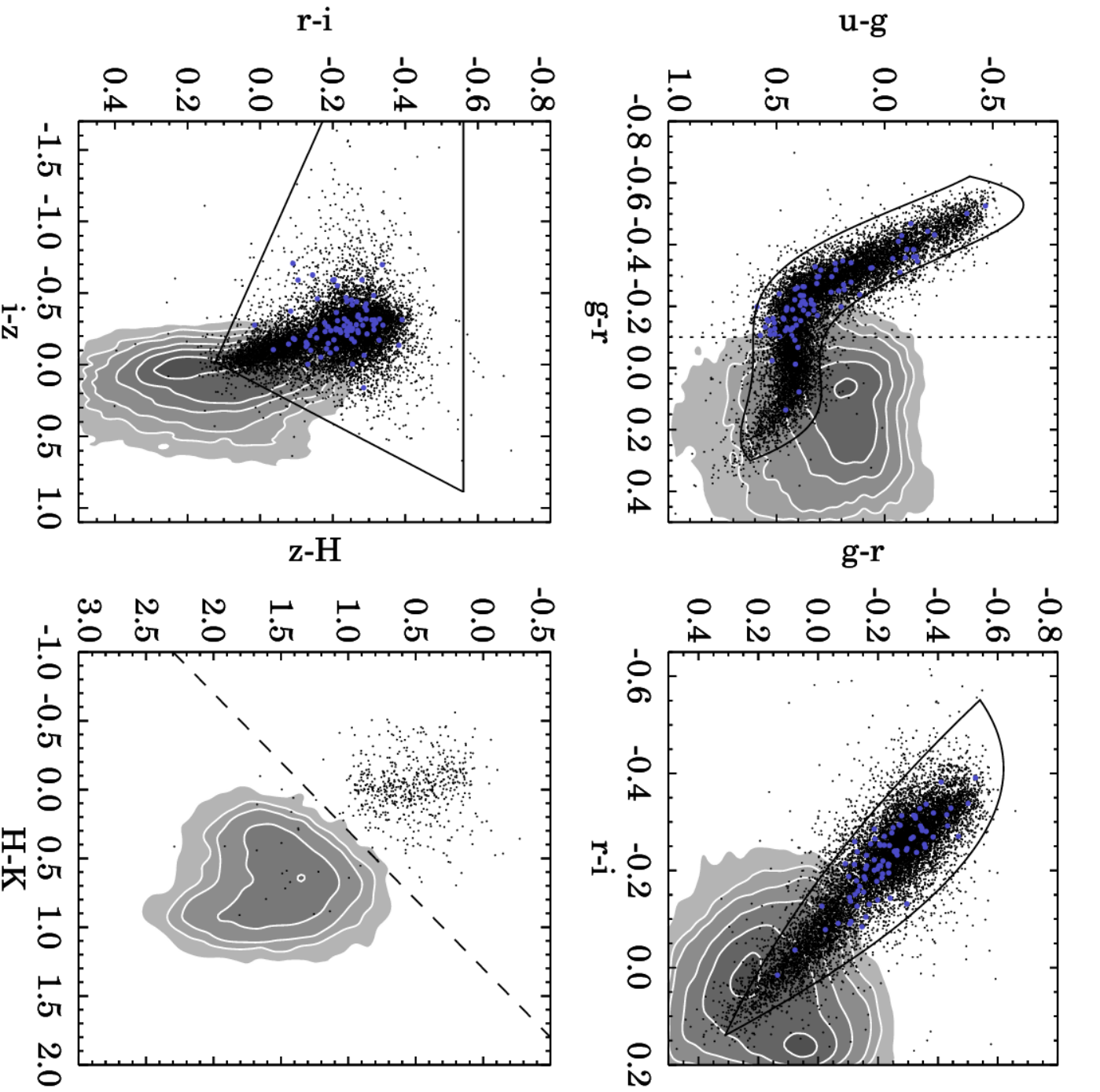}
\caption{Distribution in color space of spectroscopically confirmed QSOs (contours) and WDs (black dots) from SDSS \citep{schneider10,kleinman13}. The \citet{girven11} WD regions are overplotted. Blue dots indicate WDs from the 43 candidate wide DWDs that pass the selection criteria described in Section \ref{search:no_pm}. The objects in both SDSS catalogs with UKIDSS \citep{lawrence07} counterparts are also plotted in $(z-H)$ versus $(H-K)$ in the bottom right panel. The dashed line in that panel is $(z-K) = 1.3$ mag. While this line cleanly separates WDs from QSOs, the majority of our candidate WDs lack UKIDSS counterparts. Using SDSS photometry alone, $(u-g)$ versus $(g-r)$ colors provide the best constraints to separate QSOs from WDs.}
\label{fig:nopm_color} 
\end{center}
\end{figure}

To identify such pairs in DR9, we again started with the photometric catalog and extracted a sample of candidate WDs. Since $\mu$ measurements are generally unavailable for objects with nearby companions in SDSS, halo subdwarfs and QSOs are now significant sources of contamination. To reduce the contamination due to halo subdwarfs, we applied a more stringent color-color cut to our sample, selecting only those objects in the photometric catalog that fall in the \citet[][]{girven11} DA WD region (cf.~discussion in Paper I).

\citet{girven11} estimated that 17\% of the objects falling within this color-color region are QSOs. However, this was based on a $g<19$ mag sample of objects with SDSS spectra. Our sample extends to $g=21$, and we expected QSOs to be a more significant contaminant. 

To determine the extent of the overlap between QSOs and WDs, we examined the distribution in $ugrizHK$ color space of spectroscopically confirmed SDSS QSOs \citep{schneider10} and WDs \citep{kleinman13}. While QSOs and WDs can be cleanly separated using SDSS$+$infrared colors (see bottom right panel of Figure \ref{fig:nopm_color}), the majority of the SDSS photometric catalog lacks UKIDSS counterparts and infrared photometry.\footnote{The UKIRT Infrared Deep Sky Survey (UKIDSS) project is defined in \citet{lawrence07}. UKIDSS uses the UKIRT Wide Field Camera \citep[WFCAM;][]{casali07}. The photometric system is described in \citet{hewett06}, and the calibration is described in \citet{hodgkin09}. The pipeline processing and science archive are described in \citet{hambly08}. We used the Eighth Data Release.} 

\begin{figure}[b]
\begin{center}
\includegraphics[width=1.32\columnwidth,angle=90]{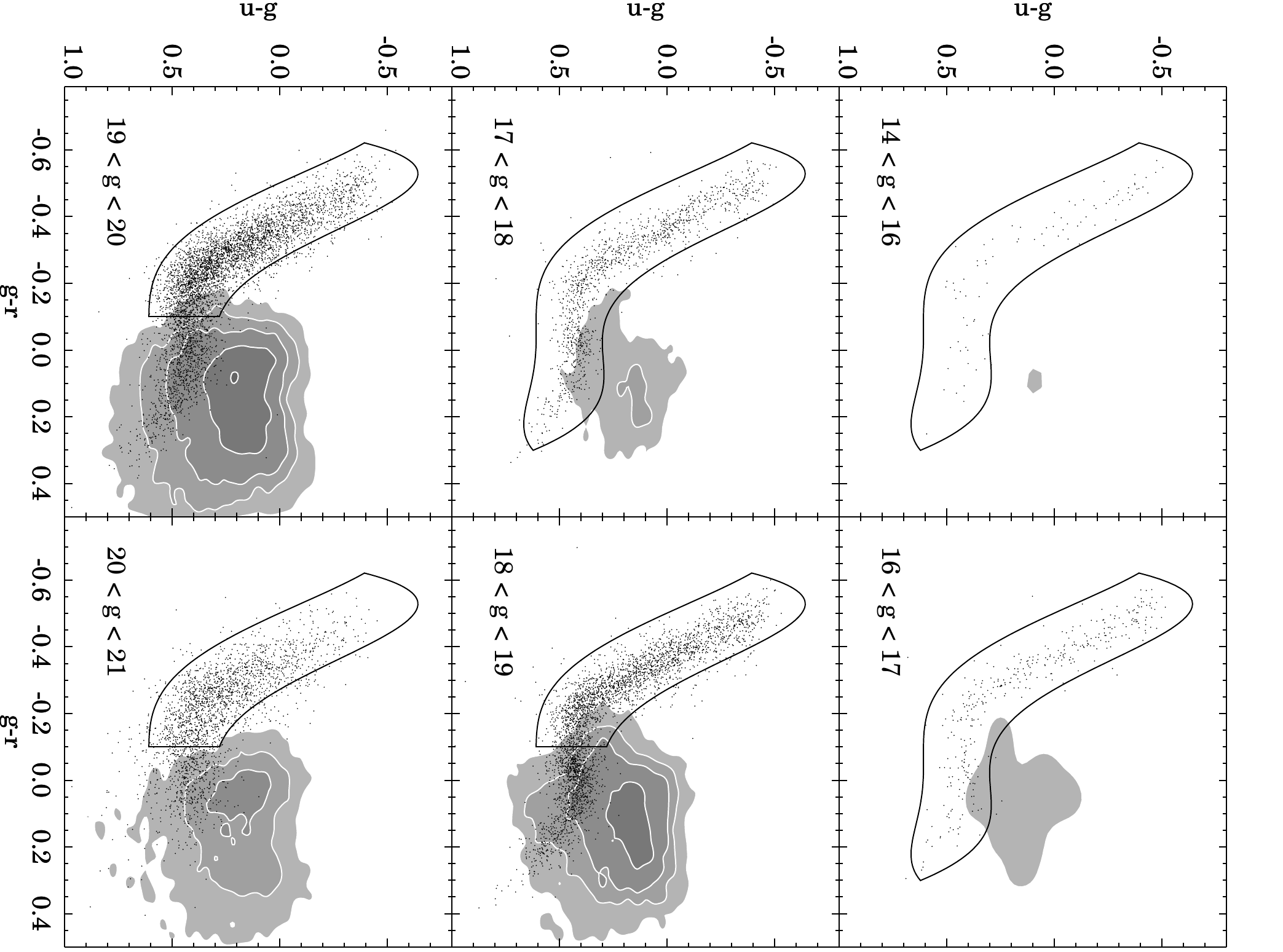}
\caption{$(u-g)$ versus $(g-r)$ for photometrically identified candidate WDs binned by magnitude. The contours are the spectroscopically confirmed QSOs from the \citet{schneider10} catalog, while the solid line in the top two and middle left panels is the \citet{girven11} WD region. QSOs become a major source of contamination at fainter magnitudes and redder $(g-r)$. To identify fainter candidate WDs, we therefore added the constraint that for $g>18$ mag, $(g-r)<-0.1$, resulting in the regions described by the solid line in the middle right and bottom two panels. }
\label{fig:nopm_uggr}
\end{center}
\end{figure}

Of the $ugriz$ color-color pairings, the least overlap between QSOs and WDs occurs in $(u-g)$ versus $(g-r)$. The bottom panels of Figure~\ref{fig:nopm_uggr} show that, in this color combination, the number of QSOs falling in the \citet{girven11} region increases at fainter magnitudes, but primarily at redder $(g-r)$. In addition to using the \citet{girven11} color-color regions to identify candidate WDs, for objects with $g>18$ we added the additional requirement that $(u-g)<-0.1$. There are 67,640 objects in DR9 that satisfy these photometric constraints.

\begin{figure}[t!]
\begin{center}
\includegraphics[width=0.72\columnwidth,angle=90]{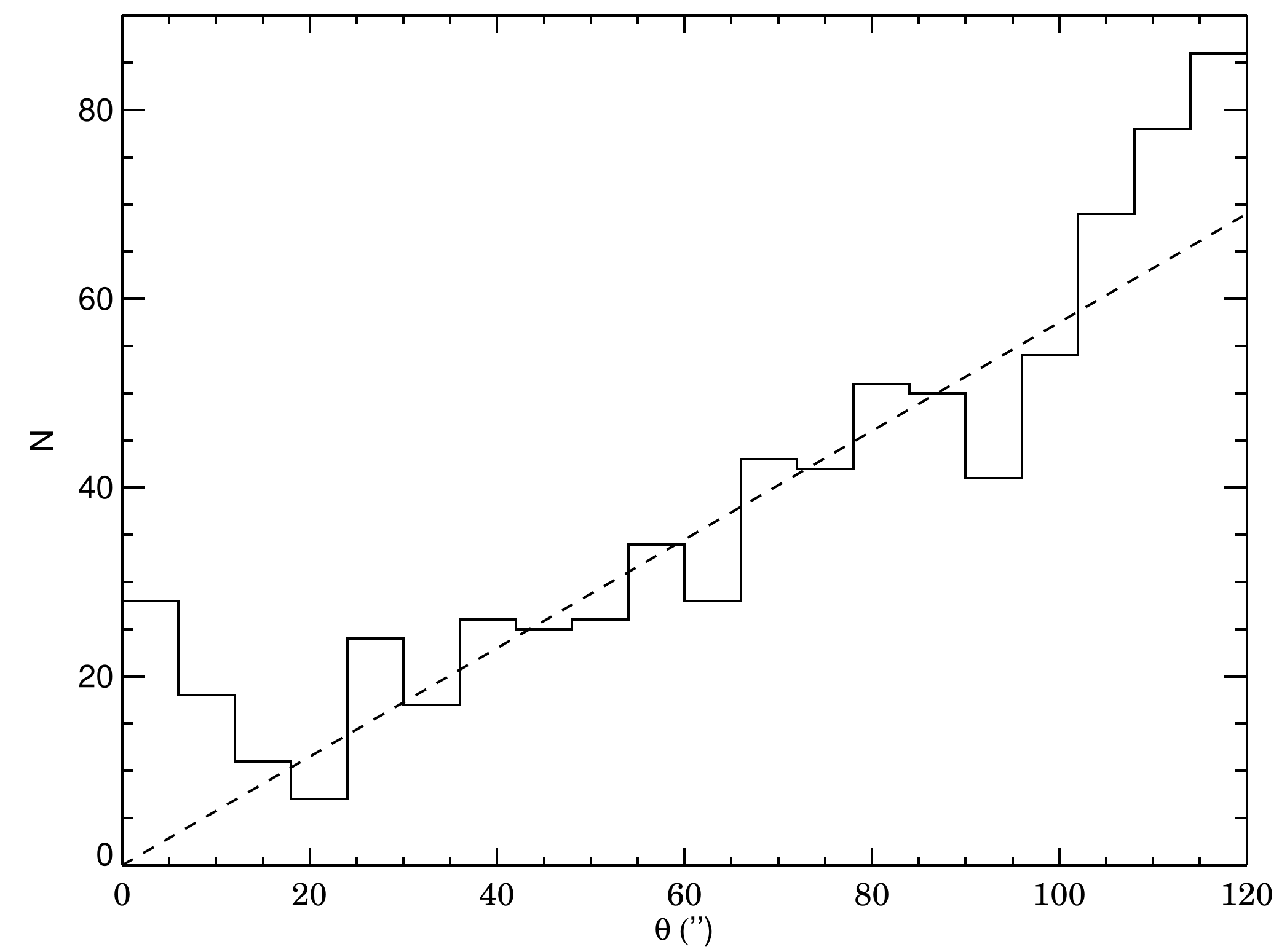}
\caption{Angular separation distribution for astrometrically selected candidate DWDs. For $\theta\ \gapprox\ 20$\asec, the number of matches increases linearly with $\theta$, as expected for random alignments. The dashed line shows the best fit line to the distribution of random alignments beyond 50\asec\ where random alignments dominate, again assuming N($\theta$) $\propto \theta$ and that the line goes through the origin. Extrapolating this fit to smaller $\theta$ shows that pairs with $\theta < 10$\asec\ are excellent candidates for follow-up. }
\label{fig:nopm_theta} 
\end{center}
\end{figure}

We searched for pairs of objects within this sample with $\theta < 2$\amin, removing pairs within crowded fields. Figure~\ref{fig:nopm_theta} is a histogram of the resulting $\theta$ distribution. For $\theta\ \gapprox\ 20$\asec, the rate of matches increases linearly with $\theta$, as expected for random alignments. We fit a line through the origin to the $\theta$ distribution for pairs with $\theta > 50$\asec. When extrapolating to smaller $\theta$ values, the line suggests that $\lapprox$5 of the systems with $\theta < 10$\asec\ are likely to be random alignments. Accordingly, we selected the 43 pairs with $\theta < 10$\asec\ as high-confidence candidate DWDs. Five of these were previously known, so that we have 38 new candidate DWDs.

Among the previously known wide DWDs are the 11 systems identified in Paper I and 36 systems identified elsewhere in the literature. To that sample we add 27 systems identified by common proper motions and 38 identified by their small astrometric separations. \citet{baxter14} found 53 wide DWDs in SDSS and spectroscopically confirmed 26 of these (one additional pair was found to be a contaminant). Thirty of the 53 DWDs are new, and 11 of these are spectroscopically confirmed. In Table~\ref{tab:DWD_obs} we provide positions, $g$ magnitudes, and $\mu_{\alpha}$ and $\mu_{\delta}$, when available, for the combined catalog of 142 candidate and confirmed wide DWDs, including the 19 candidate and 11 spectroscopically confirmed pairs from \citet{baxter14}.

\subsection{Comparison with Previous Samples}
\begin{figure}
\begin{center}
\includegraphics[width=0.72\columnwidth,angle=90]{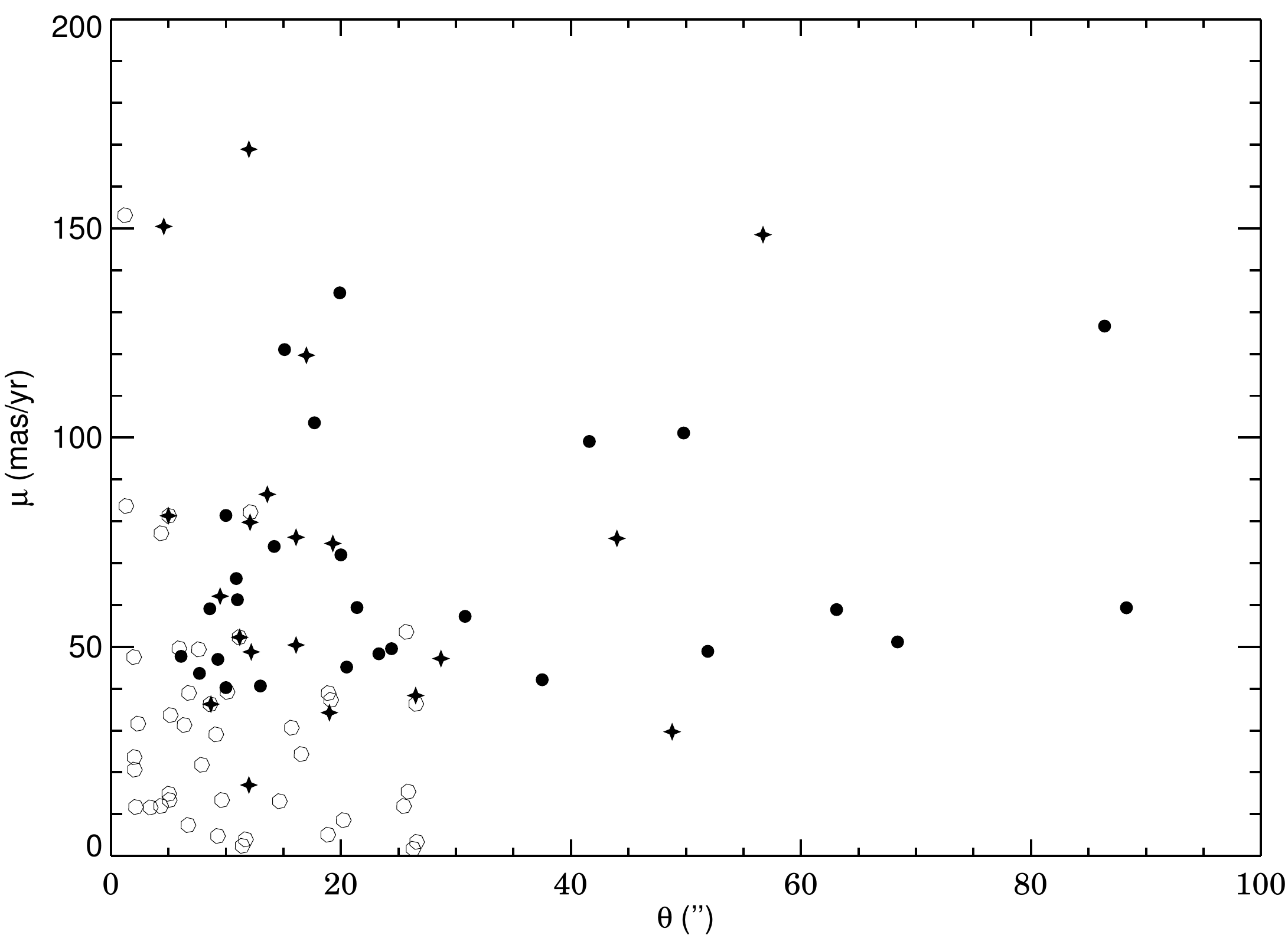}
\caption{Total proper motion ($\mu$) versus angular separation ($\theta$) for candidate DWDs identified in our search for common proper motion pairs (filled circles), in \citet[][open circles]{baxter14}, and in the literature (pluses). Although the \citet{baxter14} DWDs were also identified in SDSS, the overlap with our sample is small due to differences in the search techniques. Compared to other searches, we find systems with larger $\theta$ and $\mu$. }
\label {fig:mu_theta} 
\end{center}
\end{figure}

Figure \ref{fig:mu_theta} shows $\mu$ versus $\theta$ for our new DWDs, as well as for pairs from the literature, including the \citet{baxter14} DWDs. Our pairs are very different from those in the literature, and in particular from the DWDs identified by \citet{baxter14}, despite similar source catalogs \citep[SDSS DR9 here, DR7 for][]{baxter14}. 

These differences are largely due to the search methods employed: \citet{baxter14} initially used a less restrictive color-color region to select photometric WD candidates. These authors then searched for candidate wide DWDs by finding nearby pairs of blue objects without proper motion matching, similar to the astrometric approach described above. Compared to our candidates, these candidate DWDs have smaller $\theta$ and $\mu$ values. But the small overlap between the two samples argues for the value of both approaches.

We recover all of the new wide DWDs found in Paper I in this search. Another 21 wide DWDs from the literature fall within the SDSS DR9 footprint; we recover eight of these pairs. The reasons we fail to re-detect the other 13 DWDs are given in Table \ref{tab:nondetection}. Most are not recovered because one or both of the WDs are cool enough to have colors too red to fall within our WD photometric region. This suggests that our detection algorithm is insensitive to the wide DWDs that contain cooler WDs, but these WDs have such large \tauc\ that they are less useful for constraining the IFMR than the hotter WDs we do find. 

\section{Assembling a Spectroscopic Sample}\label{sec:spec}
\subsection{Observations and Reductions, and Spectra From the Literature}
Identifying a large number of DWDs is only the first step toward constraining the IFMR. While the WDs' effective temperatures (\Teff) can be derived from SDSS photometry to within a few 100 K, typical SDSS photometric errors are large enough to create uncertainties $\gtrsim$0.5 dex in the derived surface gravities (\logg). As a result, photometry alone is insufficient to determine the mass of a given WD with an accuracy better than $\approx$50\%. Furthermore, without a precise mass (and therefore radius), \tauc\ estimates must be based on \Teff\ alone and are uncertain by a factor of two or larger. 

We therefore engaged in a campaign to obtain spectra for WDs in DWDs. Our targets included the new systems identified in Paper I, those photometrically selected from DR9, and WDs from pairs in the literature that lacked spectroscopy.\footnote{The \citet{baxter14} systems were published too recently to be included in our spectroscopic campaign, although as we discuss below several of these DWDs were also in our sample.} Roughly 50 systems have $g \leq 19$~mag, making them ideal targets for the 3.5-m telescope at Apache Point Observatory (APO), NM.\footnote{The APO $3.5$-m telescope is owned and operated by the Astrophysical Research Consortium.} 

\LongTables
\begin{deluxetable}{lr@{: }lcc}
\tablecaption{DWDs in the DR9 Footprint not Recovered in this Search \label{tab:nondetection}}
\tablehead{
\colhead{System} & 
\multicolumn{2}{c}{Reason} &
\colhead{$(u-g)$} &
\colhead{$(g-r)$}
}
\startdata
LP 322-500A/B & A & low \Teff & 0.97 & 0.35 \\
LP 549-33/32 & B & low \Teff & 1.98 & 0.89 \\
GD 322 & B & low \Teff & 0.81 & 0.35 \\
LP 707-8/9 & B & low \Teff & 1.53 & 0.75 \\
LP 701-69/70 & B & low \Teff & 2.26 & 0.98 \\  
\multirow{2}{*}{LP 406-62/63} & A & low \Teff & 1.31 & 0.61 \\ 
 & B & low \Teff & 1.72 & 0.83 \\ 
\multirow{2}{*}{LP 647-33/34} & A & low \Teff & 1.62 & 0.68 \\ 
 & B & low \Teff & 1.65 & 0.76  \\
\multirow{2}{*}{LP 543-33/32} & A & low \Teff & 1.76 & 0.88 \\ 
 & B & low \Teff & 2.08 & 1.07 \\
\multirow{2}{*}{LP 096-66/65} & A & low \Teff & 1.63 & 0.93 \\ 
 & B & low \Teff & 1.13 & 0.53  \\
\multirow{2}{*}{LP 567-39/38} & A & low \Teff & 1.90 & 0.75 \\ 
 & B & low \Teff & 0.96 & 0.48  \\
\hline
G261-43 & A & saturated &  & \vspace{0.1cm} \\ 
J0926$+$1321 & B & low $\mu$\tablenotemark{a} &  & \vspace{0.1cm} \\ 
GD 559 & A,B & not stars\tablenotemark{b} &  & 
\enddata 
\tablecomments{Objects with low \Teff\ have $(u-g)$ and $(g-r)$ colors outside our photometric selection region for WDs.}
\tablenotetext{a}{J0926$+$1321 was identified by \citet{dobbie12} as a wide DWD based on these authors' $\mu$ calculations.}
\tablenotetext{b}{The SDSS pipeline classified both stars as galaxies.}
\end{deluxetable}

Over 13 half nights between 2012 Sep and 2013 Sep, we observed 34 pairs with the Dual Imaging Spectrograph in its high-resolution mode (R $\approx 2500$ at H$\beta$), which provides coverage from 3800 to 5000 \AA\ on the blue CCD. The slit was rotated so that spectra were taken simultaneously of both candidate WDs in each pair. The spectra were therefore not obtained at the parallactic angle. Under ideal conditions, objects with separations as small as 2\asec\ could be distinguished for reduction.

\clearpage
\begin{deluxetable*}{lllr@{ $\pm$ }lr@{ $\pm$ }lr@{ $\pm$ }lccc}
\tablecaption{Properties of Candidate and Confirmed Wide DWDs \label{tab:DWD_obs}}
\tablehead{
\colhead{Name} &
\colhead{$\alpha$} &
\colhead{$\delta$} &
\multicolumn{2}{c}{$g$} &
\multicolumn{2}{c}{$\mu_{\alpha}$} &
\multicolumn{2}{c}{$\mu_{\delta}$} &
\colhead{$\theta$} &
\colhead{Targeted for} &
\colhead{Source} \\
\colhead{} &
\colhead{} &
\colhead{} &
\multicolumn{2}{c}{(mag)} &
\multicolumn{2}{c}{(mas yr$^{-1}$)} &
\multicolumn{2}{c}{(mas yr$^{-1}$)} &
\colhead{(\asec)} &
\colhead{Spectroscopy?} &
\colhead{}
}
\startdata
\multirow{2}{*}{J0000$-$0308}  &  00:00:11.63  &  $-$03:08:31.9  &  20.11 & 0.02  &  42.43 & 3.25  &  $-$21.97 & 3.25  &  \multirow{2}{*}{6.1} & \multirow{2}{*}{N} & \multirow{2}{*}{1} \\
 &  00:00:12.04  &  $-$03:08:31.3  &  20.01 & 0.02  &  48.15 & 3.25  & $-$21.02 & 3.25  & & &\\
\multirow{2}{*}{J0000$-$1051}  &  00:00:22.53  &  $-$10:51:42.1  &  18.90 & 0.01  &  45.26 & 4.65  &  $-$23.81 & 4.65  &  \multirow{2}{*}{16.1} & \multirow{2}{*}{N} & \multirow{2}{*}{2} \\
 &  00:00:22.83  &  $-$10:51:26.6  &  20.21 & 0.02  &  42.37 & 4.08  & $-$23.57 & 4.08  & & &\\
\multirow{2}{*}{CDDS1 } & 00:01:42.84 & $+$25:15:06.1  &  17.79 & 0.02  & \multicolumn{2}{c}{}  & \multicolumn{2}{c}{} & \multirow{2}{*}{2.2} & \multirow{2}{*}{N}  & \multirow{2}{*}{3} \\
 & 00:01:42.79 & $+$25:15:04.0  &  18.70 & 0.16  & \multicolumn{2}{c}{}  & \multicolumn{2}{c}{} & & & \\
\multirow{2}{*}{J0002$+$0733}  &  00:02:15.33  &  $+$07:33:59.1  &  17.85 & 0.01  &  $-$103.4 & 2.77  &  $-$64.37 & 2.77  &  \multirow{2}{*}{15.1} & \multirow{2}{*}{Y} & \multirow{2}{*}{1}\\
 &  00:02:16.13  &  $+$07:33:49.9  &  18.07 & 0.01  &  $-$103.62 & 2.85  & $-$63.59 & 2.85  & & &\\
\multirow{2}{*}{J0029$+$0015}  &  00:29:25.29  &  $+$00:15:59.8  &  19.59 & 0.01  &  $-$27.87 & 3.62  &  $-$23.24 & 3.62  &  \multirow{2}{*}{8.7} & \multirow{2}{*}{N} & \multirow{2}{*}{2,3}\\
 &  00:29:25.62  &  $+$00:15:52.7  &  18.46 & 0.01  &  $-$28.33 & 3.13  & $-$22.58 & 3.13  && & \\
\multirow{2}{*}{J0030$+$1810}  &  00:30:51.75  &  $+$18:10:53.8  &  18.72 & 0.01  &    \multicolumn{2}{c}{}  & \multicolumn{2}{c}{}    &  \multirow{2}{*}{7.8} & \multirow{2}{*}{Y} & \multirow{2}{*}{1}\\
 &  00:30:51.80  &  $+$18:10:46.1  &  18.73 & 0.01  &   \multicolumn{2}{c}{}  & \multicolumn{2}{c}{}   & & &\\
\multirow{2}{*}{CDDS3 } & 00:52:12.26 &$+$13:53:02.0  &  17.71 & 0.03  & \multicolumn{2}{c}{}  & \multicolumn{2}{c}{}  & \multirow{2}{*}{6.8} & \multirow{2}{*}{N}  & \multirow{2}{*}{3} \\
 & 00:52:12.73 &$+$13:53:01.1  &  18.89 & 0.03  & \multicolumn{2}{c}{}  & \multicolumn{2}{c}{}  & & & \\
\multirow{2}{*}{LP 406-62/63}  &  01:04:56.33  &  $+$21:19:54.9  &  18.30 & 0.02  &  $-$210.33 & 2.51  &  $-$431.2 & 2.51  &  \multirow{2}{*}{28.0} & \multirow{2}{*}{N} & \multirow{2}{*}{4}\\
 &  01:04:57.81  &  $+$21:20:13.5  &  18.59 & 0.02  &  $-$209.48 & 2.62  & $-$436.89 & 2.62  & & &\\
\multirow{2}{*}{J0105$-$0741}  &  01:05:53.78  &  $-$07:41:22.2  &  20.51 & 0.03  &  $-$47.73 & 4.65  &  $-$66.24 & 4.65  &  \multirow{2}{*}{10.0} & \multirow{2}{*}{N} & \multirow{2}{*}{1}\\
 &  01:05:54.14  &  $-$07:41:30.8  &  18.57 & 0.01  &  $-$48.41 & 2.57  & $-$61.63 & 2.57  & & &\\
\multirow{2}{*}{J0107$+$0511}  &  01:07:17.20  &  $+$05:11:46.8  &  18.67 & 0.01  &  $-$59.61 & 2.86  &  $-$79.32 & 2.86  &  \multirow{2}{*}{41.6} & \multirow{2}{*}{N} & \multirow{2}{*}{1}\\
 &  01:07:19.89  &  $+$05:11:58.1  &  20.54 & 0.02  &  $-$49.86 & 3.48  & $-$73.05 & 3.48  & & &
\enddata
\tablerefs{1: This work; 2: Paper I; 3: \citet{baxter14}; 4: Literature (see Paper I for references).}
\tablecomments{The full table may be found in the online edition of the journal. The WDs in each pair are ordered by RA, regardless of the ``A'' or ``B'' designation that may exist in the literature.}
\end{deluxetable*}

All the spectra were trimmed, bias-corrected, cleaned of cosmic rays, flat-fielded, extracted, and dispersion-corrected using standard IRAF tasks.\footnote{IRAF is distributed by the National Optical Astronomy Observatories, which are operated by the Association of Universities for Research in Astronomy, Inc., under cooperative agreement with the National Science Foundation.} The spectra were flux calibrated using bright WD spectrophotometric standards in the IRAF database. To improve the final signal-to-noise ratio (S/N), spectra were co-added using the IRAF routine {\tt scombine}. Occasionally, the spectra to be combined were taken at different epochs, but the observing setup was identical across all observations. 

\begin{figure}
\begin{center}
\includegraphics[width=0.73\columnwidth,angle=90]{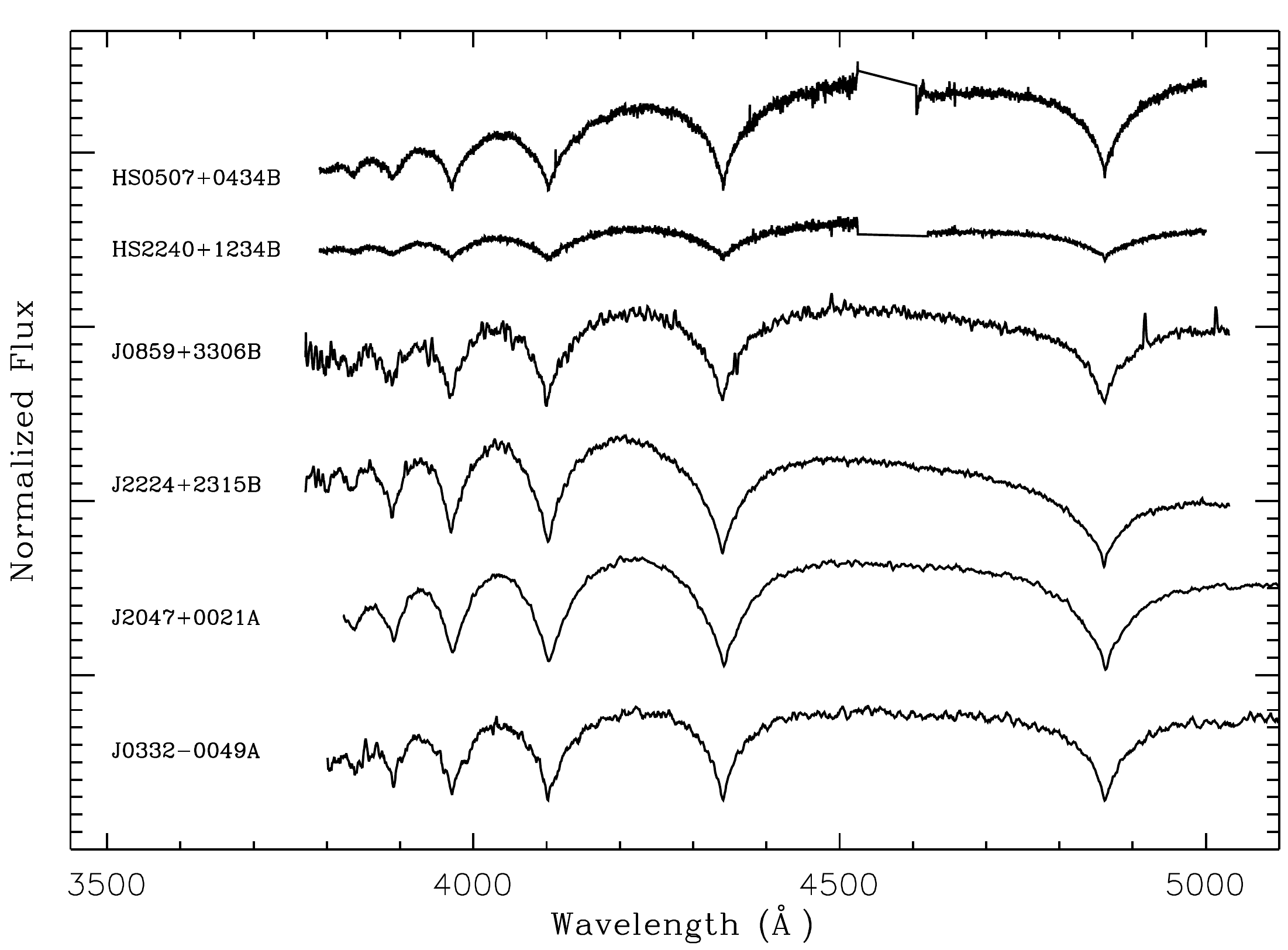}
\caption{Sample VLT, APO, and SDSS WD spectra. These DAs have \Teff\ of 10,000$-$15,000 K. The top two spectra were taken with the VLT (R $\approx$ 14,000) and are not flux calibrated. The middle two were taken at APO (R $\approx$ 2500) and the bottom two by SDSS (R $\approx$ 1800). These WDs range from $g\approx16$ (HS 2240$+$1234B) to $\approx$19 mag (J0859$+$3306B). The spectra here and in Figure~\ref{fig:non_DA_spec} have been smoothed using a boxcar average of width 5.}
\label {fig:sample_spec} 
\end{center}
\end{figure}

Wide DWDs identified in Paper I, as well as a few WDs from our photometric search in DR9, have at least one SDSS spectrum (R $\approx$ 1800), and we add these $\approx$30 SDSS spectra to our sample. Additionally, high-resolution Very Large Telescope (VLT; R $\approx$ 15,000) spectra for $\approx$10 WDs from the Supernova Progenitor Survey \citep{koester09} were provided by D.~Koester (priv.~communication). In total, we have 114 spectra for 97 WDs in wide DWDs; see Figure~\ref{fig:sample_spec} for sample spectra. The contamination by non-WDs is extremely low: only one of the 97 objects for which we obtained spectra is not a WD (J2124$-$1620A is an A star).

\subsection{Atmospheric Model Fits to Our Spectra}
\begin{figure}[h!]
\begin{center}
\includegraphics[width=\columnwidth]{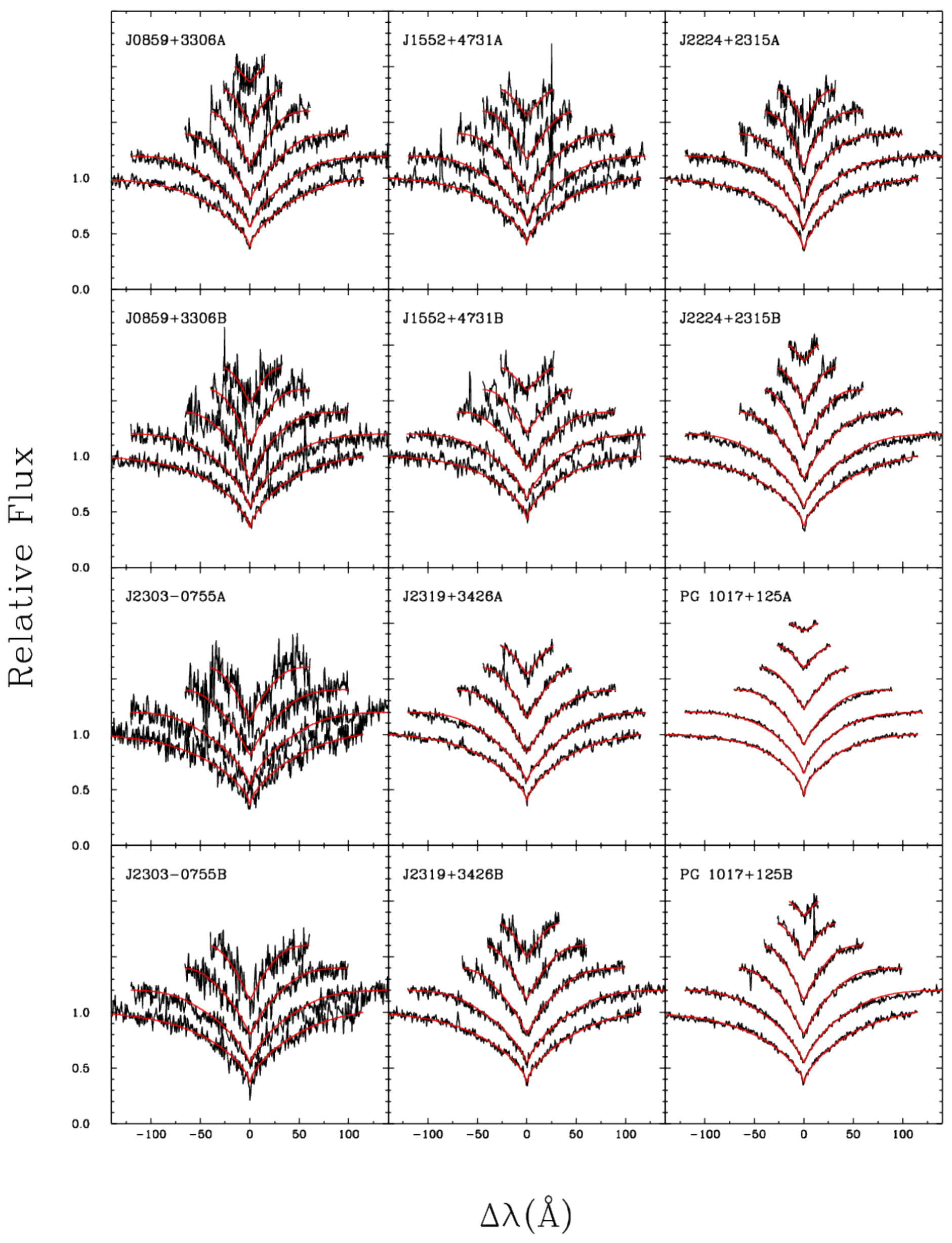}
\caption{Our model fits (red) to the APO spectra of six DA/DA pairs. The number of Balmer lines fit ranges from four to six, starting with H$\beta$ at the bottom. These fits are representative of the fits to all the WDs in our spectroscopic campaign.}
\label {fig:multipanel} 
\end{center}
\end{figure}

To obtain \Teff\ and \logg\ for these WDs, we used the spectroscopic technique developed by \citet{bergeron92} and described in \citet[][and references therein]{gianninas11}, which incorporates model atmospheres for WDs with $6.5 \leq$ \logg\ $\leq 9.5$. The observed and theoretical spectra are normalized to a continuum set to unity, and the observed  H$\beta$ to H8 lines are fit simultaneously to the synthetic spectra (see Figure~\ref{fig:multipanel}). The uncertainties in these quantities are a combination of the internal uncertainties, derived from the covariance matrix of the fitting functions, and external uncertainties of 1.2\% in \Teff\ and 0.038 dex in \logg, derived from multiple observations of the same object \citep[cf.][]{liebert05}.

These solutions are based on one-dimensional (1D) models using a standard mixing-length parameter ML2/$\alpha = 0.8$ \citep{tremblay10}. \citet{tremblay13} produced a new suite of WDs models and solved the radiation-hydrodynamics equations in three dimensions. These authors find that using this approach rather than mixing-length theory to approximate WDs with convective atmospheres leads to substantial differences in the derived masses for WDs with $7000 < $ \Teff\ $<$ 12,000~K. We applied the fitting formulas \citet{tremblay13} provide to the \Teff\ and \logg\ solutions for all of the WDs in our sample. In the relevant region of parameter space, these adjustments tend to shift our WD solutions to lower 

\noindent \Teff\ and \logg\ (or to older and less massive WDs) compared to mixing-length models. 

Next, we used the \citet{wood95} and \citet{fontaine01} models, depending on \Teff, to map our \Teff\ and \logg\ values to \tauc\ and masses ($M_{\rm WD}$) for each of our WDs. Our fits also provide distances to the WDs which are determined by comparing photometric magnitudes with absolute magnitudes from the spectroscopic solutions. The resulting quantities for the DA+DA DWDs with spectra are given in Table~\ref{tab:DWD_spec_quants}. 

\subsection{Our Spectroscopic Sample}
Our sample includes 27 DA/DA pairs. Table~\ref{tab:DWD_spec_quants} shows that lower S/N spectra result in larger \Teff\ and \logg\ uncertainties, and hence in the derived \tauc\ and $M_{\rm WD}$ \citep[for the dependence of these uncertainties on S/N, see figure 12,][]{gianninas05}. We therefore divided the sample into high-fidelity and low-fidelity pairs. We labeled systems with mass uncertainties $>$0.1 \Msun\ in at least one WD as low-fidelity. These pairs have spectra good enough to identify objects as DAs, but too poor to obtain accurate fits to model atmospheres.

As a further test, we considered the spectroscopic distances to each WD in these candidate DWDs. We designated systems with a distance difference $>$25\% as low-fidelity. These distances do not necessarily identify these systems as random alignments, but instead reflect the accuracy of the spectral fits. After identifying eight systems as low-fidelity, we are left with 19 high-fidelity pairs as a starting point for our analysis below.

\begin{figure}[t!]
\begin{center}
\includegraphics[width=1.7\columnwidth,angle=90]{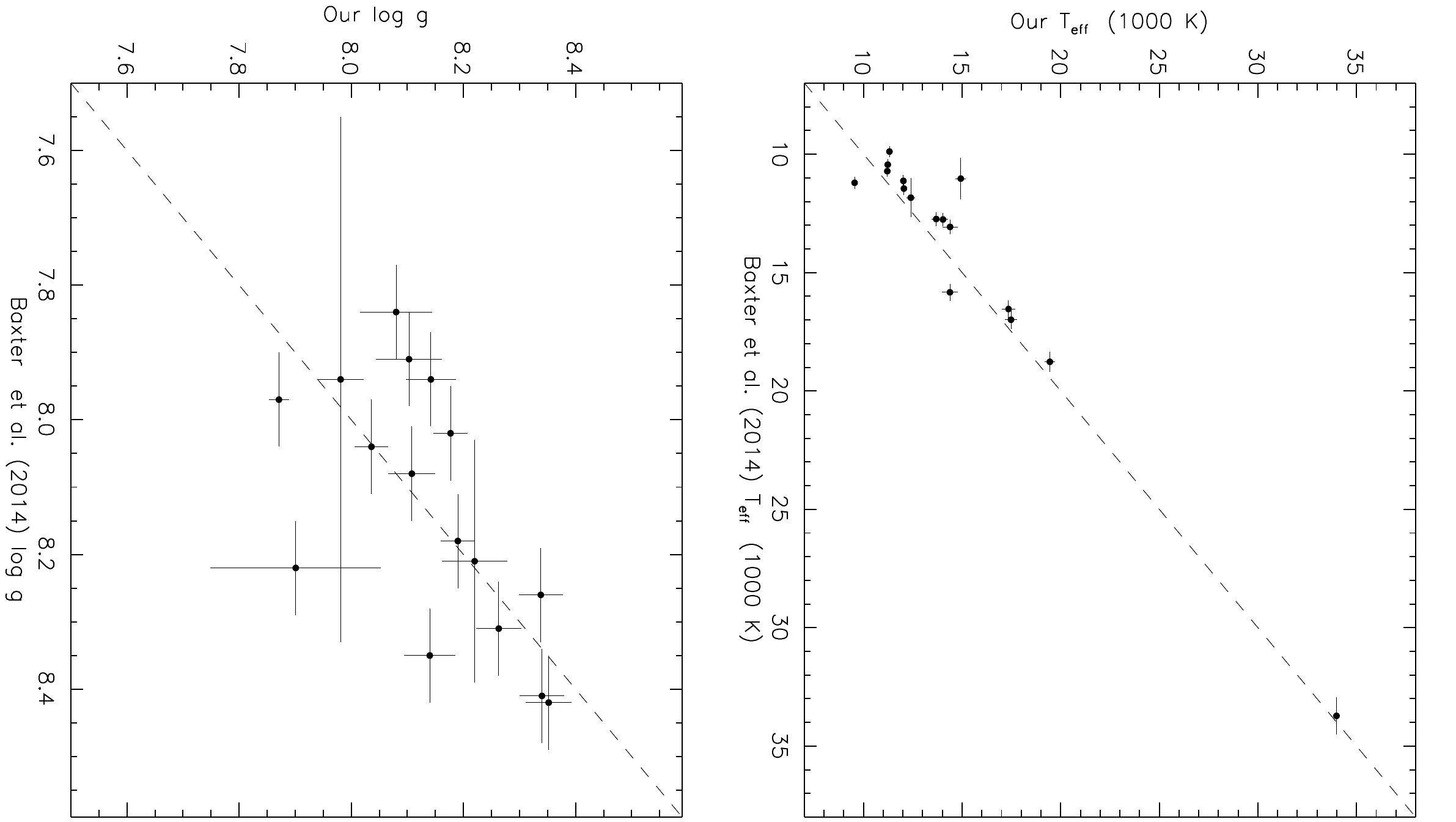}
\caption{\label{fig:baxter_compare} A comparison of \Teff\ and \logg\ values derived by \citet{baxter14} with the values from our spectroscopic fits to the eight systems contained in both spectroscopic data sets. The dashed line shows the 1:1 correspondence. While the \Teff\ values are broadly consistent, our values appear to be slightly higher in most cases. By contrast, the \logg\ values are well-matched for only half of the WDs and differ significantly for the other half, with no obvious trend in the differences.}
\end{center}
\end{figure}

Our spectroscopic campaign also uncovered a number of pairs that, while interesting, cannot be used to constrain the IFMR. These include the detection of two systems containing a helium-atmosphere (DB) WD \citep[one of which was already reported by][]{baxter14}. We also identify seven pairs that include a candidate magnetic (DAH) WD and six pairs with a (DC) WD too cool to show significant Balmer absorption features. Finally, we obtained spectra of the two known triple degenerate systems, G 21-15 and Gr 576/577, and identified PG 0901$+$140 as a candidate triple degenerate. These systems are discussed in greater detail in the Appendix. 

\clearpage
\begin{deluxetable*}{lcccr@{$\pm$}lr@{$\pm$}lr@{$\pm$}lr@{$\pm$}lr@{$\pm$}l}
\tablecaption{Fit Results for the DA/DA DWD Spectroscopic Sample}
\tablehead{
\colhead{Name} & 
\colhead{Telescope} & 
\colhead{\# of Fitted} &
\colhead{S/N\tablenotemark{a}} &
\multicolumn{2}{c}{\Teff} & 
\multicolumn{2}{c}{\logg} & 
\multicolumn{2}{c}{Distance} & 
\multicolumn{2}{c}{$M_{\rm WD}$} & 
\multicolumn{2}{c}{\tauc} \\
\colhead{} & 
\colhead{} & 
\colhead{Balmer Lines} &
\colhead{} &
\multicolumn{2}{c}{(K)} & 
\multicolumn{2}{c}{} & 
\multicolumn{2}{c}{(pc)} & 
\multicolumn{2}{c}{(\Msun)} & 
\multicolumn{2}{c}{(Myr)} 
}
\startdata
\cutinhead{High-Fidelity Systems}
HS 0507$+$0434A  &  VLT  & 6 & 25,22 & 21450 & 310  &  8.00 & 0.04  & 51 & 2 &  0.629 & 0.024  &  50 & 7  \\ 
HS 0507$+$0434B  &  VLT  & 6 & 19,12 & 12070 & 190  &  8.07 & 0.05  & 52 & 2 &  0.649 & 0.030  &  406 & 33  \\ 
HS 2240$+$1234A  &  VLT  & 6 & 1,11 & 15320 & 230  &  8.04 & 0.04  & 96 & 3 &  0.636 & 0.026  &  197 & 17  \\ 
HS 2240$+$1234B  &  VLT  & 5 & 1,10 & 14150 & 290  &  8.10 & 0.05  & 99 & 4 &  0.668 & 0.028  &  272 & 25  \\ 
J0332$-$0049A  &  SDSS  & 6 & 25,23 & 10940 & 180  &  8.10 & 0.06  &  168 & 7  &  0.661 & 0.036  &  544 & 52  \\ 
J0332$-$0049B  &  VLT  & 5 & 2,16 & 33990 & 490  &  7.87 & 0.05  & 177 & 6 &  0.594 & 0.022  &  6 & $<$1  \\ 
J0754$+$1239A  &  APO  & 5 & 17 & 14190 & 1070  &  8.24 & 0.12  &  258 & 23  &  0.755 & 0.074  &  335 & 95  \\ 
J0754$+$1239B  &  APO  & 5 & 15 & 13690 & 630  &  8.31 & 0.11  &  247 & 22  &  0.801 & 0.072  &  415 & 91  \\ 
J0827$-$0216A  &  APO  & 5 & 30 & 27310 & 450  &  8.49 & 0.06  &  260 & 13  &  0.933 & 0.035  &  68 & 11  \\ 
J0827$-$0216B  &  APO  & 5 & 24 & 27860 & 490  &  8.58 & 0.07  &  265 & 16  &  0.989 & 0.039  &  79 & 14  \\ 
J0859$+$3306A  &  APO  & 6 & 30 & 14930 & 340  &  7.98 & 0.06  &  232 & 10  &  0.602 & 0.034  &  194 & 23  \\ 
J0859$+$3306B  &  APO  & 5 & 21 & 12140 & 280  &  8.18 & 0.07  &  228 & 12  &  0.718 & 0.045  &  470 & 59  \\ 
J1231$+$5736A  &  APO  & 5 & 38 & 15360 & 290  &  8.01 & 0.05  &  217 & 18  &  0.618 & 0.031  &  185 & 19  \\ 
J1231$+$5736B  &  APO  & 5 & 22 & 11190 & 230  &  7.92 & 0.08  &  231 & 13  &  0.556 & 0.046  &  404 & 47  \\ 
J1257$+$1925A  &  SDSS  & 5 & 10 & 11750 & 470  &  7.91 & 0.16  &  456 & 51  &  0.552 & 0.089  &  350 & 83  \\ 
J1257$+$1925B  &  SDSS  & 5 & 57 & 47800 & 990  &  7.76 & 0.07  &  503 & 32  &  0.579 & 0.029  &  2 & $<$1  \\ 
J1313$+$2030A  &  APO  & 5 & 30 & 14390 & 450  &  8.34 & 0.06  &  156 & 8  &  0.823 & 0.037  &  382 & 49  \\ 
J1313$+$2030B  &  APO  & 6 & 40 & 14030 & 330  &  8.19 & 0.05  &  147 & 6  &  0.726 & 0.033  &  322 & 33  \\ 
J1552$+$4731A  &  APO  & 5 & 24 & 17350 & 410  &  8.10 & 0.07  &  382 & 20  &  0.679 & 0.044  &  148 & 23  \\ 
J1552$+$4731B  &  APO  & 5 & 27 & 19450 & 390  &  8.14 & 0.06  &  367 & 17  &  0.707 & 0.037  &  107 & 16  \\ 
J2222$-$0828A  &  APO  & 5 & 41 & 14380 & 440  &  8.18 & 0.05  &  92 & 4  &  0.719 & 0.033  &  295 & 35  \\ 
J2222$-$0828B  &  APO  & 5 & 29 & 11750 & 220  &  8.08 & 0.06  &  112 & 5  &  0.653 & 0.038  &  442 & 45  \\ 
HS 2220$+$2146A &  VLT  & 5 & 8,11 & 14270 & 270 & 8.15 & 0.04  & 79 & 2 & 0.702 & 0.022  &  289 & 22  \\
HS 2220$+$2146B &  VLT  & 5 & 10,16 & 18830 & 220 & 8.35 & 0.04 &  73 & 2  &  0.837 & 0.022  &  179 & 14  \\ 
J2224$+$2315A  &  APO  & 5 & 39 & 10930 & 180  &  8.16 & 0.06  &  137 & 6  &  0.702 & 0.036  &  604 & 61  \\ 
J2224$+$2315B  &  APO  & 6 & 47 & 13690 & 300  &  8.04 & 0.05  &  139 & 5  &  0.631 & 0.031  &  274 & 27  \\ 
J2303$-$0755A  &  APO  & 5 & 15 & 14400 & 960  &  8.27 & 0.10  &  194 & 16  &  0.777 & 0.065  &  339 & 84  \\ 
J2303$-$0755B  &  APO  & 5 & 18 & 13900 & 730  &  8.04 & 0.09  &  223 & 15  &  0.635 & 0.054  &  264 & 54  \\ 
J2319$+$3426A  &  APO  & 5 & 43 & 16320 & 270  &  8.12 & 0.05  &  162 & 6  &  0.684 & 0.030  &  183 & 18  \\ 
J2319$+$3426B  &  APO  & 5 & 33 & 14140 & 450  &  8.05 & 0.05  &  172 & 7  &  0.641 & 0.032  &  255 & 31  \\ 
LP 128-254  &  APO  & 6 & 35 & 14030 & 470  &  8.07 & 0.06  &  164 & 7  &  0.652 & 0.034  &  269 & 34  \\ 
LP 128-255  &  SDSS  & 5 & 25 & 11080 & 200  &  7.98 & 0.07  &  164 & 8  &  0.589 & 0.040  &  447 & 45  \\ 
LP 370-50  &  APO  & 5 & 30 & 7560 & 120  &  8.19 & 0.09  &  59 & 4  &  0.710 & 0.055  &  1668 & 324  \\ 
LP 370-51  &  APO  & 5 & 31 & 7210 & 120  &  8.14 & 0.10  &  61 & 4  &  0.681 & 0.061  &  1723 & 349  \\ 
PG 0901$+$140A\tablenotemark{b}  &  APO  & 6 & 67 & 9100 & 140  &  7.78 & 0.08  &  59 & 3  &  0.474 & 0.041  &  585 & 57  \\ 
PG 0901$+$140B  &  APO  & 6 & 39 & 8120 & 120  &  7.89 & 0.07  &  58 & 3  &  0.531 & 0.039  &  886 & 87  \\ 
PG 0922$+$162A  &  VLT  & 5 & 15,15 & 24480 & 350  &  8.28 & 0.04  &  158 & 6  &  0.797 & 0.028  &  59 & 9  \\ 
PG 0922$+$162B  &  VLT  & 4 & 20,8 & 26500 & 440  &  9.04 & 0.06  & 110 & 7 &  1.220 & 0.023  &  227 & 23  \\ 
PG 1017$+$125A  &  APO  & 6 & 90 & 22130 & 330  &  7.99 & 0.04  &  106 & 4  &  0.622 & 0.024  &  41 & 6  \\ 
PG 1017$+$125B  &  APO  & 6 & 50 & 13580 & 240  &  8.12 & 0.05  &  108 & 4  &  0.681 & 0.030  &  317 & 28  \\ 
\cutinhead{Low-Fidelity Systems}
J1002$+$3606A  &  APO  & 5 & 20 & 9720 & 160  &  8.42 & 0.10  &  152 & 12  &  0.863 & 0.063  &  1287 & 296  \\ 
J1002$+$3606B  &  SDSS  & 4 & 9 & 11650 & 580  &  8.26 & 0.21  &  226 & 38  &  0.767 & 0.135  &  594 & 217  \\ 
J1110$+$4517A  &  APO  & 6 & 23 & 13700 & 370  &  8.10 & 0.06  &  113 & 5  &  0.670 & 0.038  &  301 & 36  \\ 
J1110$+$4517B  &  APO  & 5 & 46 & 19000 & 300  &  8.12 & 0.05  &  152 & 6  &  0.692 & 0.029  &  111 & 13  \\ 
J1203$+$4948A  &  APO  & 6 & 25 & 11410 & 220  &  8.12 & 0.07  &  119 & 6  &  0.674 & 0.041  &  502 & 56  \\ 
J1203$+$4948B  &  APO  & 5 & 14 & 7250 & 150  &  8.06 & 0.21  &  118 & 18  &  0.631 & 0.129  &  1507 & 621  \\ 
J1309$+$5503A  &  SDSS  & 5 & 13 & 8120 & 160  &  7.94 & 0.18  &  136 & 16  &  0.560 & 0.100  &  951 & 234  \\ 
J1309$+$5503B  &  SDSS  & 5 & 22 & 8090 & 140  &  8.17 & 0.11  &  85 & 7  &  0.698 & 0.071  &  1345 & 277  \\ 
J1546$+$6159A  &  APO  & 5 & 20 & 10880 & 190  &  7.84 & 0.07  &  244 & 12  &  0.514 & 0.039  &  398 & 39  \\ 
J1546$+$6159B  &  APO  & 6 & 70 & 16510 & 270  &  8.06 & 0.05  &  143 & 6  &  0.649 & 0.029  &  159 & 16  \\ 
J1703$+$3304A\tablenotemark{c}  &  SDSS  & 5 & 12 & 9400 & 180  &  7.63 & 0.16  &  239 & 24  &  0.424 & 0.069  &  874 & 72  \\ 
J1703$+$3304B  &  SDSS  & 6 & 33 & 11030 & 190  &  8.19 & 0.06  &  156 & 7  &  0.718 & 0.037  &  614 & 64  \\ 
J0030$+$1810A  &  APO  & 5 & 11 & 14070 & 1050  &  8.42 & 0.18  &  222 & 33  &  0.870 & 0.115  &  458 & 167  \\ 
J0030$+$1810B  &  APO  & 5 & 5 & 15620 & 2930  &  8.43 & 0.35  &  244 & 77  &  0.880 & 0.214  &  351 & 306  \\ 
J1113$+$3238A  &  SDSS  & 4 & 10 & 6680 & 230  &  7.33 & 0.54  &  156 & 49  &  0.307 & 0.185  &  1474 & 466  \\ 
J1113$+$3238B  &  SDSS  & 4 & 8 & 7760 & 260  &  8.82 & 0.33  &  78 & 27  &  1.107 & 0.170  &  3646 & 233  \\ 
J2326$-$0023A  &  SDSS  & 4 & 9,9 & 7530 & 180  &  8.45 & 0.23  &  114 & 21  &  0.884 & 0.151  &  3028 & 968  \\ 
J2326$-$0023B  &  SDSS  & 6 & 28 & 10530 & 170  &  8.02 & 0.06  &  124 & 5  &  0.611 & 0.036  &  538 & 50  
\enddata 
\tablecomments{When multiple spectra were available, we list the fit to the DA's best spectrum, which we define as the fit that produces a distance that best matches its companion's. For previously identified systems, we label the WDs ``A'' and ``B'' as in the literature. We order newly identified systems by their RA as opposed to e.g., their relative brightnesses.}
\tablenotetext{a}{Entries with more than one listed S/N  indicate that multiple spectra were used to fit for the WD parameters.}
\tablenotetext{b}{Given its anomalously low mass, PG 0901$+$140A may be an unresolved triple system.}
\tablenotetext{c}{Despite its low mass, J1703$+$3304A (CDDS40-B) is unlikely to be an unresolved double degenerate (cf.~discussion in the Appendix), because the discrepant distances indicate a poor fit.}

\label{tab:DWD_spec_quants}
\end{deluxetable*}

\clearpage
\subsection{Comparison to the Baxter et al.~(2014) Fit Results}\label{baxt2}
\citet{baxter14} identified 53 DWDs, and 12 are included in our spectroscopic sample. \citet{baxter14} obtained spectra for eight of these pairs. In Figure~\ref{fig:baxter_compare}, we compare the \logg\ and \Teff\ values these authors derived to ours. Half the WDs have \Teff\ and \logg\ values in agreement; the other half have significantly different spectroscopic solutions, particularly in \logg. This is similar to what \citet{baxter14} found when comparing their spectroscopic results to the SDSS-derived results of \citet{kleinman13}: of the five \citet{kleinman13} WDs for which \citet{baxter14} have spectra, two have significantly differing \logg\ values. 

\begin{figure}
\begin{center}
\includegraphics[width=.98\columnwidth,angle=90]{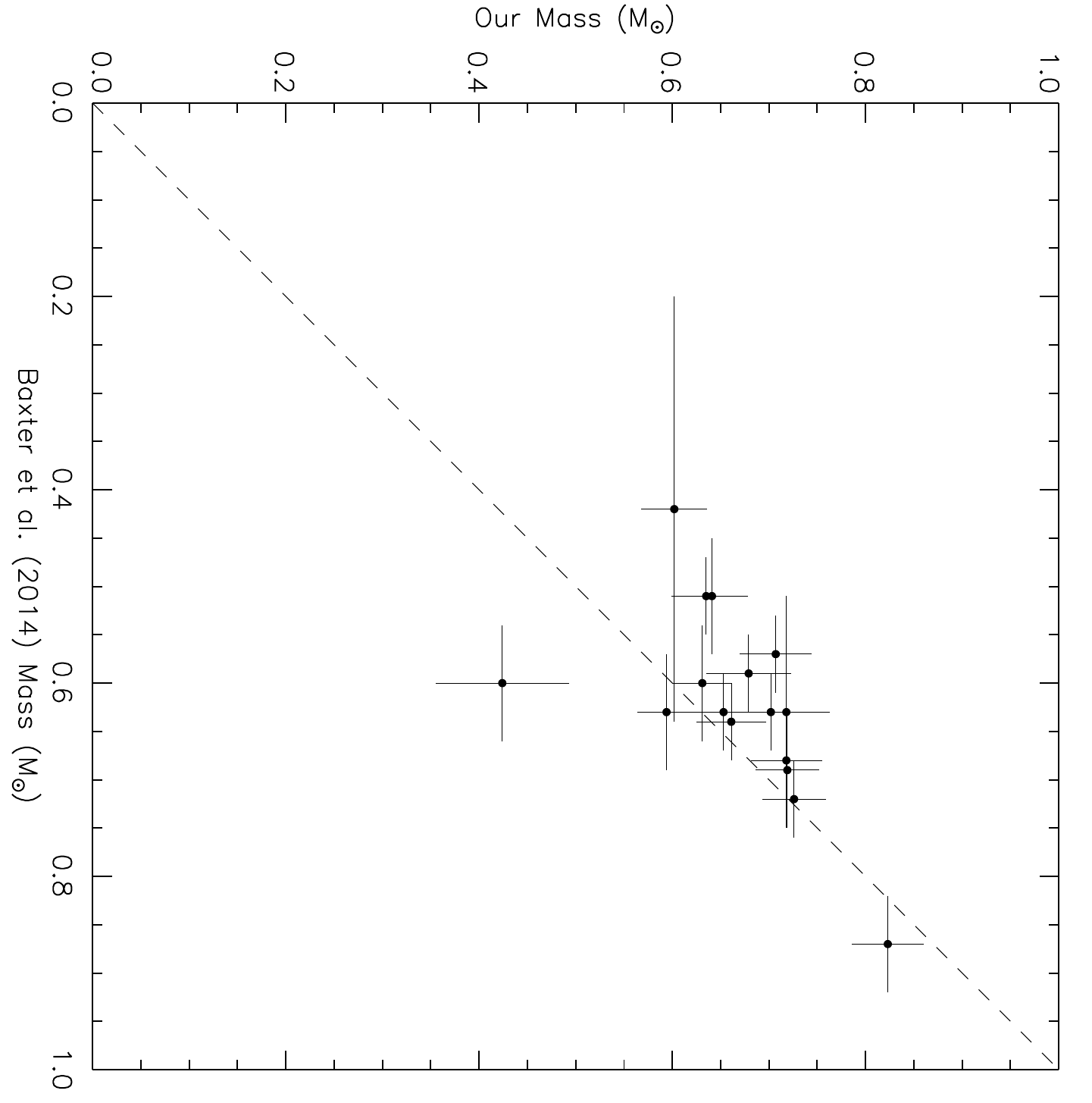}
\caption{\label{fig:baxter_mass} Comparison of the \Mwd\ of \citet{baxter14} and obtained from our spectra for 16 WDs in eight DWDs included in both spectroscopic samples. The dashed line shows the 1:1 correspondence. Our \Mwd\ estimates are systematically larger by $\approx$0.05~\Msun, presumably due to the use of spectra with different resolutions and fitting codes with different model atmospheres.}
\end{center}
\end{figure}

We show the WD masses derived from these spectral solutions for the 16 WDs in both spectroscopic samples in Figure~\ref{fig:baxter_mass}. We find that our $M_{\rm WD}$ estimates are systematically larger by $\approx$0.05 \Msun. We ascribe these differences to the combination of observations made with instruments with different resolutions and of spectral fitting done with different techniques and atmospheric models. 

As the \citet{baxter14} spectroscopic sample represents a significant addition to our own, we include the WDs in our analysis below. For systems in both the \citet{baxter14} spectroscopic sample and our own, we use the values derived from our spectra.

\subsection{Our DWD Sample for Constraining the IFMR} \label{sec:our_sample}
Standard stellar evolution theory predicts that the more massive WDs in DWDs are also the older WDs: more massive stars evolve faster, becoming more massive WDs with larger \tauc\ than their less massive companions. Of the 20 high-fidelity systems in Table~\ref{tab:DWD_spec_quants}, PG 0901$+$140 may be a triple system (see discussion in the Appendix) and five appear to host a more massive WD that is younger than its companion: J1231$+$5736, J1552$+$4731 (CDDS36), J2222$-$0828 (CDDS48), HS 2220$+$2146, and LP 128-254/255. We remove these and use the remaining 14 systems for our analysis.

We then consider the 10 \citet{baxter14} systems for which we have poor or no spectroscopic data. We remove five of these systems from our analysis. One of the WDs in CDDS16 has poor S/N; the higher-order Balmer lines are particularly noisy, which impacts the determination of \logg. \cite{baxter14} identified CDDS30 as a possible triple system. CDDS31 has a projected separation of $\approx$100 AU and could therefore have had previous mass-transfer episodes. The more massive WD in CDDS26 appears to be younger than its companion. Finally, SDSS spectra are available for both WDs in CDDS7, but the spectra are very poor. We add the remaining five pairs to our sample, so that we now have 19 DWDs with which to constrain the IFMR (see Table~\ref{tab:IFMR_sample}).

\section{Constraining the IFMR with DWDs}\label{sec:meth}
We begin by examining several of the basic assumptions that allow one to use DWDs to constrain the IFMR. These are that the two WDs are co-eval and have not been subjected to significant mass transfer during their lifetimes, that our DWD progenitors did not vary significantly in metallicity, and that the pre-WD lifetimes produced by stellar evolution codes for stars of a given mass are relatively insensitive to the parameters one uses in these calculations. We then revisit the \citet{finley97} result before developing a flexible, Bayesian method that takes into account measurement uncertainties and provides statistically rigorous constraints on the IFMR. 

\subsection{Examining the Underlying Assumptions}
\subsubsection{Co-evolution and Independence of Wide DWDs}
Binary star formation theory suggests that the collapse of gas clouds in multistellar systems occurs on a dynamical timescale \citep{shu87}; for typical binaries, this is $<$1 Myr. This expectation has been borne out by observations of binaries in the Taurus-Auriga cluster by \citet{kraus09}, which suggest that the age difference of the stars in binaries is even smaller than that of stars within an open cluster. Since the probability that field DWD progenitor systems formed through gravitational capture is extraordinarily small, we can safely assume the stars in a wide binary are born together through fragmentation \citep{boss88}. 

Wide binaries with small enough separations could have interacted in the past, potentially through wind-fed mass transfer. Indeed, below some critical (but still relatively large) separation, some amount of mass transfer is unavoidable, and may impact the system's evolution. For example, Mira, with a separation of $\approx$70 AU, shows mass accretion at a rate of $\approx$10$^{-10}$ \Msun~yr$^{-1}$ \citep{sokoloski10}, high enough to potentially induce periodic nova eruptions on the WD surface every $\approx$Myr, which would affect its derived $\tauc$.

Such mass accretion could be due to so-called wind Roche lobe overflow \citep{mohamed07,mohamed12}. These authors argue that when an AGB donor is

\clearpage
\begin{deluxetable*}{lr@{$\pm$}lr@{$\pm$}lr@{$\pm$}lr@{$\pm$}lccc}
\tablecaption{DA/DA DWDs Used to Constrain the IFMR}
\tablehead{
\colhead{Name} &
\multicolumn{2}{c}{$M_{\rm WD,1}$} &
\multicolumn{2}{c}{$M_{\rm WD,2}$} &
\multicolumn{2}{c}{$\tau_{\rm cool,1}$} &
\multicolumn{2}{c}{$\tau_{\rm cool,2}$} &
\colhead{Distance\tablenotemark{a}} &
\colhead{$\theta$} &
\colhead{$s$} \\
\colhead{} & 
\multicolumn{2}{c}{(M$_{\odot}$)} & 
\multicolumn{2}{c}{(M$_{\odot}$)} &
\multicolumn{2}{c}{(Myr)} &
\multicolumn{2}{c}{(Myr)} & 
\colhead{(pc)} & 
\colhead{(asec)} & 
\colhead{(AU)} 
}
\startdata
HS 0507$+$0434 & 0.649 & 0.030 & 0.629 & 0.024 & 406 & 33 & 50 & 7 & 51 & 18 & 919 \\
HS 2240$+$1234 & 0.668 & 0.028 & 0.636 & 0.026 & 272 & 25 & 197 & 17 & 97 & 12 & 1166 \\
J0332$-$0049 & 0.661 & 0.036 & 0.594 & 0.022 & 544 & 52 & 6 & $<1$ & 173 & 19 & 3290 \\
J0754$+$1239 & 0.801 & 0.072 & 0.755 & 0.074 & 415 & 91 & 335 & 95 & 252 & 2 & 504 \\
J0827$-$0216 & 0.989 & 0.039 & 0.933 & 0.035 & 79 & 14 & 68 & 11 & 262 & 3 & 787 \\
J0859$+$3306 & 0.718 & 0.045 & 0.602 & 0.034 & 470 & 59 & 194 & 23 & 231 & 9 & 2080 \\
J1257$+$1925 & 0.579 & 0.029 & 0.552 & 0.089 & 2 & $<1$ & 350 & 83 & 490 & 12 & 5888 \\
J1313$+$2030 & 0.823 & 0.037 & 0.726 & 0.033 & 382 & 49 & 322 & 33 & 150 & 6 & 901 \\
J2224$+$2315 & 0.702 & 0.036 & 0.631 & 0.031 & 604 & 61 & 274 & 27 & 138 & 3 & 415 \\
J2303$-$0755 & 0.777 & 0.065 & 0.635 & 0.054 & 339 & 84 & 264 & 54 & 209 & 8 & 1670 \\
J2319$+$3426 & 0.684 & 0.030 & 0.641 & 0.032 & 183 & 18 & 255 & 31 & 167 & 5 & 836 \\
LP 370-50/51 & 0.710 & 0.055 & 0.681 & 0.061 & 1670 & 320 & 1720 & 350 & 60 & 13 & 781 \\
PG 0922$+$162 & 1.220 & 0.023 & 0.797 & 0.028 & 227 & 23 & 59 & 9 & 158 & 5 & 791 \\
PG 1017$+$125 & 0.681 & 0.030 & 0.622 & 0.024 & 317 & 28 & 41 & 6 & 107 & 49 & 5250 \\
CDDS3 & 0.592 & 0.036 & 0.579 & 0.034 & 78 & 14 & 491 & 63 & 256 & 7 & 1792 \\
CDDS6 & 0.668 & 0.039 & 0.655 & 0.040 & 31 & 8 & 114 & 19 & 192 & 2 & 385 \\
CDDS9 & 0.598 & 0.058 & 0.535 & 0.070 & 142 & 32 & 376 & 69 & 506 & 12 & 6080 \\
CDDS14 & 0.644 & 0.041 & 0.590 & 0.045 & 428 & 54 & 598 & 91 & 211 & 5 & 1056 \\
CDDS40 & 0.694 & 0.043 & 0.634 & 0.040 & 876 & 129 & 530 & 70 & 187 & 11 & 1419
\enddata
\tablecomments{In this table WDs 1 and 2 are the more and less massive WDs in the pair, respectively.}
\tablenotetext{a}{This is the average of the distance to each WD.}
\label{tab:IFMR_sample}
\end{deluxetable*}

\noindent emitting a slow wind, a companion at tens of AU can channel a substantial fraction of the lost mass. \citet{abate13} suggest that wind Roche lobe overflow can occur when the companion is at a separation less than the AGB's dust formation radius, because at larger separations the increased opacity due to dust means that radiation pressure will quickly push the wind to escape the system. Observations suggest that this radius is also $\approx$tens of AU \citep{hofner09,karovicova13}. 

More work is needed to determine the spatial separation at which the effects of wind-fed mass transfer can be ignored. However, since the smallest projected binary separation of the wide DWDs in our sample is hundreds of AU (see Table~\ref{tab:IFMR_sample}), we expect mass accretion to be negligible, even in prior evolutionary states when the binaries may have had somewhat smaller separations.

\subsubsection{Metallicity of the DWD progenitors}\label{sec:metal}
In their analysis of WDs in the open cluster M37, \citet{kalirai05} found that their data were consistent with stars of a given $M_{\rm i}$ producing higher mass WDs than in previous studies. These authors suggested that the lower metallicity of M37 might result in less mass loss on the AGB and, therefore, more massive WDs. 

Theory supports this interpretation: \citet{renedo10} found that metal-poor stars undergo more thermal pulses on the AGB, resulting in more massive WDs. Variations in metallicity should result in WD masses varying by $\approx$0.1~\Msun\ for a given $M_{\rm i}$. However, for near-solar metallicities, the IFMR does not vary much \citep{marigo07, meng08, romero15}.

Fortunately, the WDs in our sample should have roughly similar, near-solar metallicities. We cannot directly test for the metallicity of a WD progenitor, but there are indications that most disk stars born in the past several Gyr likely have similar metallicities. For example, in a study of local F and G type stars, \citet{fuhrmann98} found that kinematically identified thin-disk stars have [Fe/H] and [Mg/H] metallicity indicators within 0.3 dex of solar.

As shown in Figure~\ref{fig:rpm}, our subset of candidate WDs with measured proper motions almost all have transverse velocities consistent with being in the Galactic disk. Furthermore, since we selected relatively hot WDs for the spectroscopic follow-up described in Section \ref{sec:spec}, all of the wide DWDs in our sample are relatively young. We therefore expect the progenitors of the WDs in our sample to have had a metallicity close to $z=0.02$.

\subsubsection{Robustness of the Pre-WD Lifetime Function}\label{sec:lifetime}
We obtain a pre-WD lifetime function, $\bs{F}$, by running models from {\tt MESA}. {\tt MESA} is a suite of modules that includes integrated equations-of-state tables, opacity tables, nuclear reaction networks, and elemental diffusion rates. The stellar evolution module, {\tt MESA star}, solves the stellar structure equations using a 1D, adaptive Lagrangian algorithm. Built with state-of-the-art prescriptions, {\tt MESA star} has been extensively tested and compared to observations and other stellar evolution codes \citep{paxton11,paxton13,paxton15}.

\begin{deluxetable}{ll}
\tablecaption{{\tt MESA} Model Parameters \label{tab:mesa}}
\tablehead{
\colhead{Parameter} & 
\colhead{Fiducial}\\
\colhead{} &
\colhead{Model}
}
\startdata
number of isotopes & 26 (up to Mg) \\ 
metallicity & $z=0.02$ \\ 
mixing length & $\alpha_{\rm MLT}=1.73$ \\ 
overshoot parameter & $f=0.014$ \\ 
photospheric model & {\tt simple\_photosphere} \\ 
RGB mass loss & Reimers, $\eta=0.5$ \\ 
AGB mass loss & Blocker, $\eta=0.5$ \\ 
opacity tables & \citet{grevesse93} \\ 
convection & Schwartzchild criterion \\ 
rotation & off
\enddata 
\end{deluxetable}

The parameters and prescriptions used in our fiducial model are given in Table \ref{tab:mesa}. Stars begin as gas clouds on the pre-main-sequence. Once fusion begins, the reaction network tracks the concentrations of isotopes. Convection is treated as an exponential diffusive process, with a diffusive constant and adjustable scale length based on the pressure scale height; semi-convection is not included in our fiducial model. Exponential mixing near the convective boundary due to overshooting is included \citep{freytag96,herwig97}. 

\begin{figure}[t!]
\begin{center}
\includegraphics[width=0.79\columnwidth,angle=90]{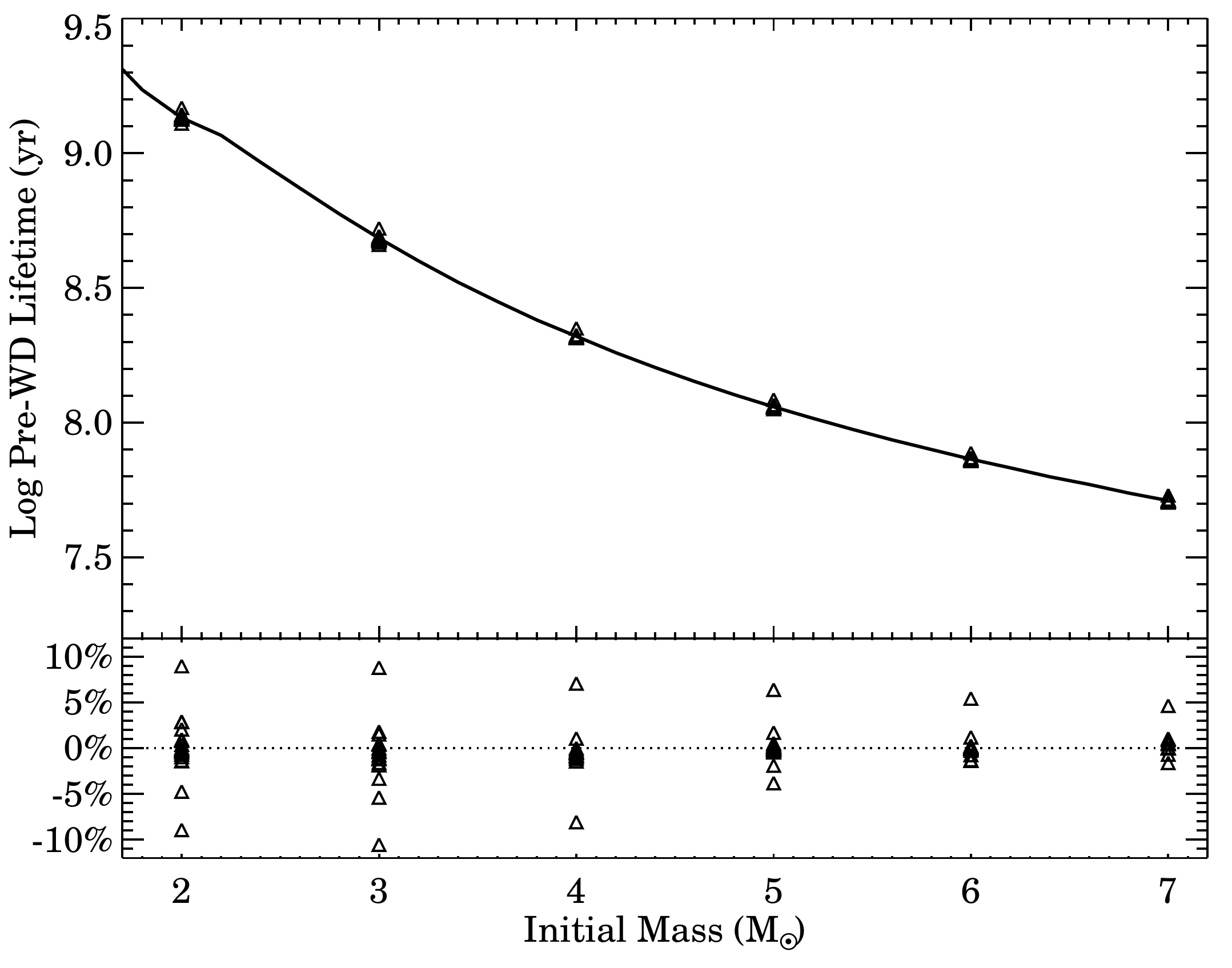}
\caption{\label{fig:1TP_age} Pre-WD lifetime (approximated by the age at the 1TP) as a function of initial mass. The solid line shows ages produced by our fiducial model, while the triangles show the results of {\tt MESA} models generated varying a number of prescriptions. The lower panel shows that the pre-WD lifetimes produced by these tests are robust, as they typically vary by less than 5\% even for very different models.}
\end{center}
\end{figure}

We stop our model at the first thermal pulse (1TP), the start of the thermally pulsing AGB (TP-AGB). We set the star's age here as its pre-WD lifetime, as evolution through the TP-AGB is quick ($10^5$ to $10^6$ yr, or $<$1\% of its lifetime), with higher mass stars evolving faster \citep{vassiliadis93}.

We run simulations for our fiducial model for $M_{\rm i} =$~0.6--8.0 \Msun\ in 0.2 \Msun\ steps. To determine the pre-WD lifetime for an arbitrary $M_{\rm i}$, we linearly interpolate over the values in this grid. The resulting pre-WD lifetime function is indicated by the solid line in Figure \ref{fig:1TP_age}.

In comparing {\tt MESA} to other stellar evolution codes, \citet{paxton11} found that the derived stellar lifetimes agreed within $\approx$5\%. However, \citet{paxton13} showed that the choice of parameters and prescriptions may have a stronger impact on the stellar lifetime. We therefore tested 19 {\tt MESA} models in addition to our fiducial model, varying the metallicity, atmospheric models, opacity tables, mixing length, and also included rotation and semi-convection. A detailed discussion of how each parameter affects the stellar lifetime is outside the scope of this work; we used these models to gauge the level of uncertainty in the stellar lifetimes.

The symbols in Figure~\ref{fig:1TP_age} show how the lifetimes we obtain vary from model to model for $2 \leq $ $M_{\rm i} \leq 7$~\Msun. The bottom panel shows that the majority of our 19 models produce lifetimes within a few percent of those obtained using our fiducial model. The largest differences in the lifetimes occur when changing the opacity tables (which are calculated based on different elemental abundances) and metallicities. Models run with the \citet{asplund09} opacity tables lead to lifetimes longer by $\approx$10\% than those from models run with the fiducial \citet{grevesse93} tables. Conversely, models with sub-solar metallicities lead to shorter lifetimes (e.g., by $\approx$10\% for $z=0.01$). In principle, these stellar evolution uncertainties should be included in a comprehensive model. However, as they are relatively minor and since we expect our DWD progenitors to have had similar, approximately solar metallicities, we ignore these uncertainties in our analysis below.

\subsection{Revisiting the \citet{finley97} Result}
\citet{finley97} constrained the IFMR using the wide DWD PG 0922$+$162. These authors compared the more massive WD ($\gapprox$1.10 \Msun) to similarly massive WDs in open clusters for which $M_{\rm i}$ had been published, thereby obtaining $M_{\rm i} = 6.5$$\pm$1.0~\Msun\ for this WD.\footnote{This is one of the difficulties in directly applying this method to other DWDs, since comparing the more massive WD in a DWD to WDs in open clusters cannot always be done for a generic DWD.} \citet{finley97} converted this mass into a pre-WD lifetime of 42--86 Myr, to which they added the \tauc\ of the massive WD to derive a system age of 320$\pm$32 Myr. These authors then used the less massive WD in PG 0922$+$162 ($M_{\rm WD} = 0.8$ \Msun) to constrain the IFMR: they derived a pre-WD lifetime for this WD of 231$\pm$34 Myr by subtracting its \tauc\ from the system age and obtained $M_{\rm i} = 3.8$$\pm$0.2~\Msun\ for its progenitor.

\begin{figure}[h!]
\begin{center}
\includegraphics[width=0.76\columnwidth,angle=90]{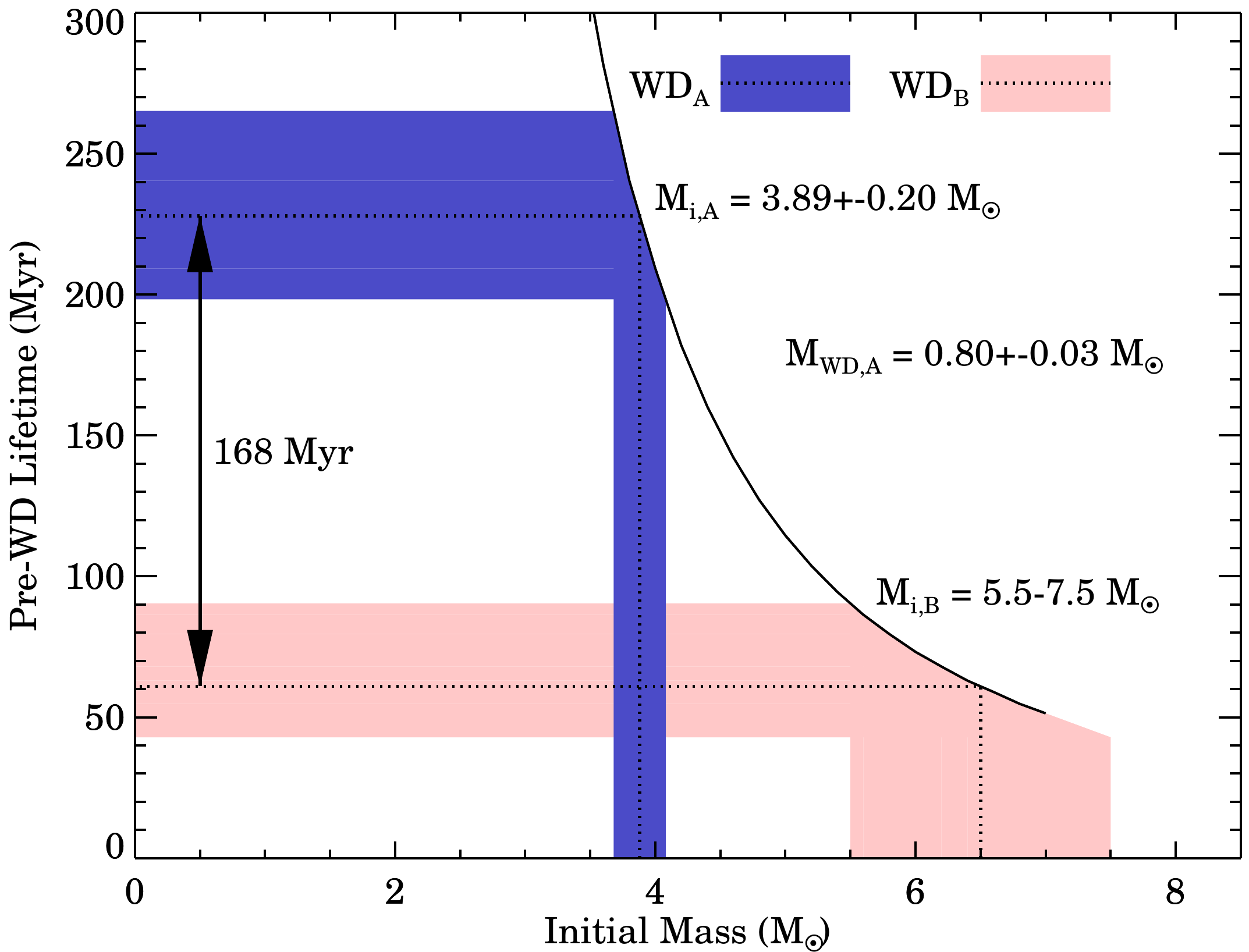}
\caption{\label{fig:PG0922} Converting $M_{\rm i}$ into pre-WD lifetimes. \citet{finley97} assigned the more massive WD$_{\rm B}$ in PG~0922$+$162 a $M_{\rm i} = 6.5$$\pm$1.0~\Msun, which corresponds to a pre-WD lifetime of 43--90 Myr (we use our {\tt mesa} stellar lifetime function here). Adding the \tauc\ difference between the two WDs gives a pre-WD lifetime of 198--265 Myr for the less massive WD$_{\rm A}$. We derive a corresponding $M_{\rm i} = 3.89$$\pm$0.20~\Msun\ for WD$_{\rm A}$; \citet{finley97} found 3.8$\pm$0.2 \Msun. Despite the relatively large uncertainty in the initial mass of WD$_{\rm B}$, the uncertainty in the initial mass of WD$_{\rm A}$ is small, particularly compared to typical uncertainties derived from other observational methods.}
\end{center}
\end{figure}

In Figure~\ref{fig:PG0922} we reproduce one of the key steps in this analysis, the conversion of pre-WD lifetimes into initial masses. Because the relation between lifetime and mass is steeper for longer-lived, lower-mass stars, even relatively large uncertainties in the assumed $M_{\rm i}$ for the more massive WD in PG 0922$+$162 results in stringent constraints on the $M_{\rm i}$ of the less massive WD. Indeed, the \citet{finley97} result is one of the most stringent constraints on the IFMR, and is one of the reasons why \citet{weidemann00} anchored his semi-empirical IFMR at $M_{\rm i} =4$~\Msun\ and $M_{\rm WD} = 0.8$~\Msun.

\subsection{A New Parametric Model for the IFMR} \label{sec:our_meth}
\begin{figure}[]
\begin{center}
\includegraphics[width=0.85\columnwidth]{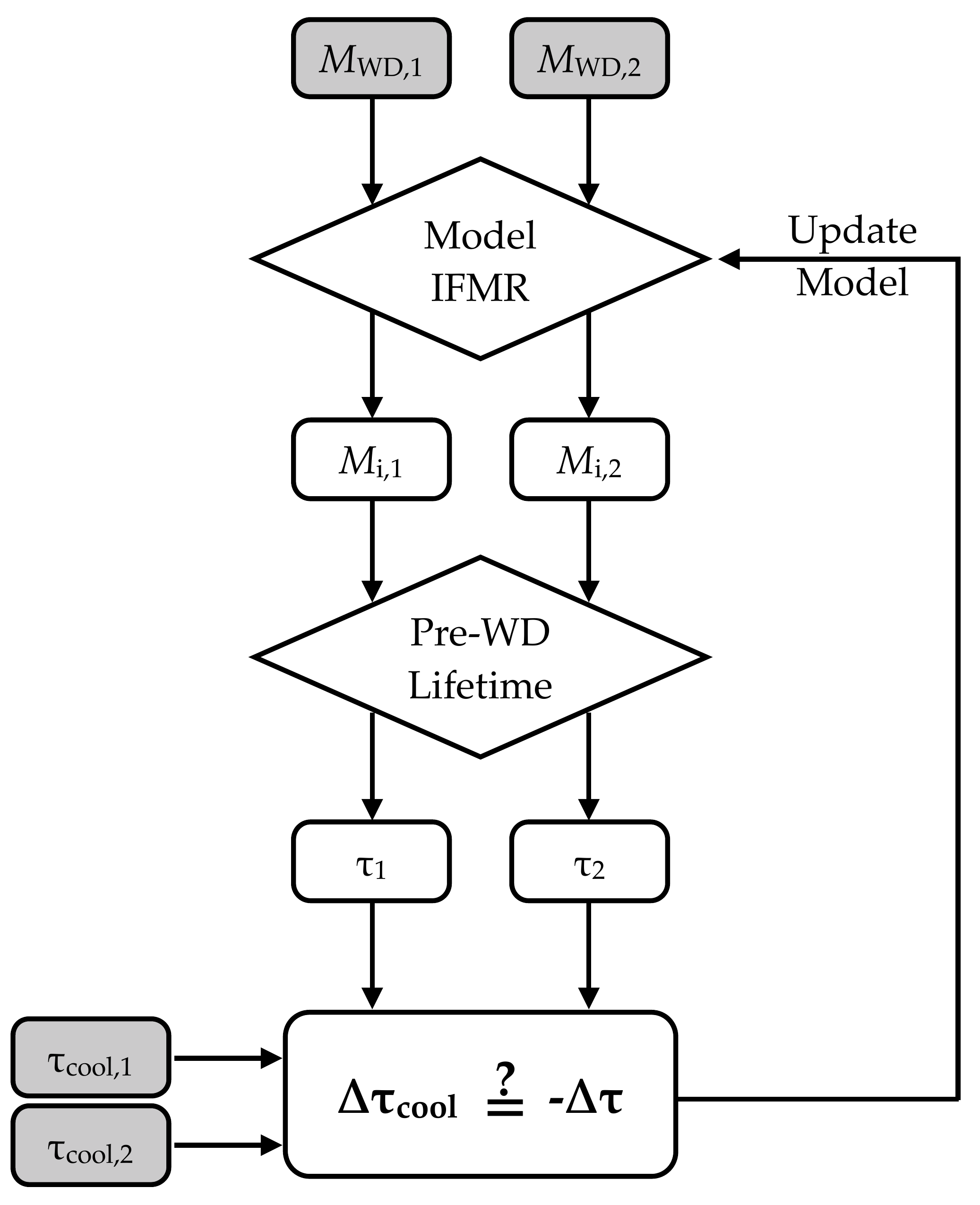}
\caption{\label{fig:DWD_diagram} A schematic summary of our method. Gray boxes indicate observed quantities. We convert each $M_{\rm WD}$ into a $M_{\rm i}$ using a candidate IFMR, and then find the corresponding pre-WD lifetime, $\tau$, using the function shown in Figure~\ref{fig:1TP_age}. The difference of these lifetimes should be equal to $\Delta \tauc$, which we obtain from the two observed $\tauc$. Our problem is then reduced to iterating over our model parameters to find the IFMR that makes this equality true for our set of wide DWDs. }\label{scheme}
\end{center}
\end{figure}

While the \cite{finley97} result indicates that DWDs can be powerful systems for constraining the IFMR, these authors' approach cannot be replicated for a generic set of wide DWDs such as the one we have assembled. We therefore develop a new method for constraining the IFMR with wide DWDs by constructing a parametric model for the IFMR. Our approach is presented in schematic form in Figure~\ref{scheme}.

We begin by considering the observables: the cooling ages, $\tauca$ and $\taucb$, and WD masses, $M_{\rm WD, 1}$ and $M_{\rm WD, 2}$. If the WDs are indeed co-eval and evolved independently, the difference in the cooling ages must be equal to the difference in the pre-WD lifetimes, $\ta$ and $\tb$. If WD$_1$ is the more massive, older WD and therefore had the shorter pre-WD lifetime, 
\begin{eqnarray}
\Delta \tauc &=& -\Delta \tau\ \label{eq:delta_eq_delta} \\
\tau_{\rm cool,1} - \tau_{\rm cool,2} &=& -(\ta - \tb). \label{eq:diff_eq_diff}
\end{eqnarray}

While $\tauca$ and $\taucb$ are observed, $\ta$ and $\tb$ are  obtained by a functional transformation from $M_{\rm WD}$:
\begin{eqnarray}
\tau &=& \bs{F} M_{\rm i} \\
&=& \bs{F}\bs{G}^{-1} M_{\rm WD}
\end{eqnarray}
where $\bs{F}$ is the pre-WD lifetime function, and we have applied the inverse IFMR, $\bs{G^{-1}}$ to obtain $M_{\rm i}$ from the observed WD masses. Combining these, we obtain:
\begin{equation}
\tauca - \taucb = \bs{F}\bs{G^{-1}}M_{\rm WD,2} - \bs{F}\bs{G^{-1}}M_{\rm WD,1}. \label{eq:delta_delta}
\end{equation}

Since $\bs{F}M_{\rm i}$ can be determined with accuracy from stellar evolution codes (see Section~\ref{sec:lifetime}), we now need to find the best $\bs{G^{-1}}$ that satisfies Equation~\ref{eq:delta_delta}. Below, we construct a likelihood function that evaluates the ability for any $\bs{G^{-1}}$ to account for the observations. We then define a parametric model for $\bs{G}$, and iterate over these model parameters, $\bs{\Theta}$, to find the highest likelihood set. As discussed below, because of the functional form of the IFMR, finding the best $\bs{G^{-1}}$ is directly equivalent to finding the best $\bs{G}$ and therefore constraining the IFMR.

\subsubsection{Bayesian Framework} \label{sec:bayes}
We calculate a likelihood function for a particular IFMR, using a Bayesian hierarchical model to account for measurement uncertainties. We begin with Bayes' rule:
\begin{equation}
P(\bs{\Theta} \given \bs{D}, I) = \frac{P(\bs{D} \given \bs{\Theta}, I) P(\bs{\Theta} \given I)}{P(\bs{D} \given I)},
\end{equation}
where $\bs{D}$ is the set of observed wide DWDs,\footnote{$\bs{D}$ refers to the set of wide DWDs, while $D$ refers to an individual system.} and $I$ represents our prior information and assumptions about the data and model. For instance, observational uncertainties are contained within $I$. $P(\bs{\Theta} \given \bs{D}, I)$ is the posterior probability we are looking for, $P(\bs{D} \given I)$ is a constant dependent only on the data, $P(\bs{\Theta} \given I)$ are the priors on our model, and $P(\bs{D} \given \bs{\Theta}, I)$ is the likelihood. The posterior probability over the set of data is a product over individual measurements:
\begin{equation}
P(\bs{D} \given \bs{\Theta}, I) = \prod_{ D \in \bs{D} } P(D \given \bs{\Theta}, I). \label{eq:prod_like}
\end{equation}

We now substitute in the individual observables:
\begin{equation}
P(D \given \bs{\Theta}, I) = P(M_1', M_2', \Delta \tauc' \given \bs{\Theta}, I).
\end{equation}

To construct our likelihood function, we first marginalize over $\ta$,  $\tb$, and $\Delta \tau$:
\begin{eqnarray}
P(D \given \bs{\Theta}, I) &=& \int_0^{\infty}\int_0^{\infty}\int_{-\infty}^{\infty} ~\der \ta ~\der \tb ~\der \Delta \tau \nonumber \\
& & \quad \times 
~P( \ta, \tb, \Delta \tau, M_1', M_2', \Delta \tauc' \given \bs{\Theta}, I).
\end{eqnarray}

Hereafter, primed quantities refer to observed values and unprimed quantities refer to true values. Because observations of each WD and of $\Delta \tauc$ are all independent, we can factor this probability:
\begin{eqnarray}
P(D \given \bs{\Theta}, I) &=& \int_0^{\infty} \int_0^{\infty}  ~\der \ta ~\der \tb 
~P(M_1' \given \ta, \bs{\Theta}, \sigma_{M_1}) \nonumber \\
& & \quad \times 
~P(M_2' \given \tb, \bs{\Theta}, \sigma_{M_2})
P(\ta \given \bs{\Theta})
P(\tb \given \bs{\Theta}) \nonumber \\
& & \quad \times 
~\int_{-\infty}^{\infty} ~\der \Delta \tau 
~P(\Delta \tauc' \given \Delta \tau, \sigma_{\Delta \tauc}) \nonumber \\
& & \quad \times ~P(\Delta \tau \given \ta, \tb),  \label{eq:split_1}
\end{eqnarray}
where here we also factored $I$ into individual observational uncertainties.
The conditional probability over $\Delta \tau$ is a delta function:
\begin{equation}
P(\Delta \tau \given \ta, \tb) = \delta(\Delta \tau - \ta + \tb),
\end{equation}
and the conditional probability over $\Delta \tauc'$ is a Gaussian distribution:
\begin{equation}
P(\Delta \tauc' \given \Delta \tau, \sigma_{\Delta \tauc}) = \mathcal{N} \left(-\Delta \tau \given \Delta \tauc', \sigma_{\Delta \tauc} \right),
\end{equation}
where the negative sign is from Equation \ref{eq:delta_eq_delta}. After reducing the innermost integral, Equation \ref{eq:split_1} simplifies to:
\begin{eqnarray}
P(D \given \bs{\Theta}, I) &=& \int_0^{\infty} \int_0^{\infty}  ~\der \ta ~\der \tb 
~P(M_1' \given \ta, \bs{\Theta}, \sigma_{M_1}) \nonumber \\
& & \quad \times 
~P(M_2' \given \tb, \bs{\Theta}, \sigma_{M_2})
P(\ta \given \bs{\Theta})
P(\tb \given \bs{\Theta}) \nonumber \\
& & \quad \times 
~\mathcal{N} \left[ (\ta-\tb) \given \Delta \tauc', \sigma_{\Delta \tauc} \right]. \label{eq:split_2}
\end{eqnarray}

Evaluating Equation \ref{eq:split_2} is unnecessarily computationally expensive, since only a small region of the domain has any contributing probability. We therefore use a Monte Carlo method to approximate the double integral as a single sum over randomly drawn $\ta$ and $\tb$. We make this approximation with a single rather than double sum because the first two terms in the integrand, the observed WD masses, are independent of each other:
\begin{eqnarray}
P(D \given \bs{\Theta}, I) &\approx& \frac{1}{N} \sum_k P(\ta_k \given \bs{\Theta})
P(\tb_k \given \bs{\Theta}) \nonumber \\
& & \quad \times 
~\mathcal{N} \left[ (\ta_k-\tb_k) \given \Delta \tauc', \sigma_{\Delta \tauc} \right], \label{eq:likelihood}
\end{eqnarray}
where there are $N$ random draws of $\ta_k$ and $\tb_k$ from the distributions:
\begin{eqnarray}
\tau_{1k} &\sim& P(\ta \given M_1', \bs{\Theta}, \sigma_{M_1}) = P(M_1' \given \ta, \bs{\Theta}, \sigma_{M_1}) \nonumber \\
\tau_{2k} &\sim& P(\tb \given M_2', \bs{\Theta}, \sigma_{M_2}) = P(M_2' \given \tb, \bs{\Theta}, \sigma_{M_2}). \label{eq:random_time}
\end{eqnarray}

Here, we have applied Bayes' theorem to make the equalities on the right side of these equations. To generate these random distributions:
\begin{eqnarray}
P(\ta \given M_1', \bs{\Theta}, \sigma_{M_1}) &=& P(\bs{G F^{-1}}\ta \given M_1', \sigma_{M_1}) \begin{vmatrix}
\frac{\der \bs{G F^{-1}}\ta}{\der \ta}
\end{vmatrix} \nonumber \\
P(\tb \given M_2', \bs{\Theta}, \sigma_{M_2}) &=& P( \bs{G F^{-1}}\tb \given M_2', \sigma_{M_2}) \begin{vmatrix}
\frac{\der \bs{G F^{-1}}\tb}{\der \tb}
\end{vmatrix}. \label{eq:random_draws}
\end{eqnarray}

We determine the derivatives numerically, and the observational uncertainties in WD mass are Gaussian:
\begin{eqnarray}
P(\bs{G F^{-1}}\ta \given M_1', \sigma_{M_1}) &=& \mathcal{N} (\bs{G F^{-1}}\ta \given M_1', \sigma_{M_1}) \nonumber \\
P(\bs{G F^{-1}}\tb \given M_2', \sigma_{M_2}) &=& \mathcal{N} (\bs{G F^{-1}}\tb \given M_2', \sigma_{M_2}). \label{eq:random_mass}
\end{eqnarray}

Equations \ref{eq:prod_like}, \ref{eq:likelihood}, \ref{eq:random_draws}, and \ref{eq:random_mass} define the model  likelihood. For a given model, finding the model parameters, $\bs{\Theta}$, implied by our set of wide DWDs involves maximizing the likelihood function in Equation~\ref{eq:likelihood}. 

We are interested in determining the precision of the constraints we can place on our model parameters, and therefore chose a Monte Carlo technique rather than a maximum-likelihood calculation. Specifically, we used the Markov Chain Monte Carlo algorithm {\tt emcee} \citep{foreman-mackey13}, which implements an affine invariant, ensemble sampler algorithm to search the parameter space \citep{goodman10}.

\subsubsection{Parametric Model} \label{sec:para_model}
We define our model by first including a parametrization that determines $P(\ta \given \bs{\Theta}$) and the corresponding quantity for the second WD. To keep the dimensionality of the problem low, our likelihood function (Equation~\ref{eq:likelihood}) marginalizes over all possible pre-WD lifetimes. 

Standard stellar evolution precludes the existence of WDs more massive than the Chandrasekhar mass ($\approx$1.35~\Msun). At the other end, the Galaxy is not old enough to produce isolated WDs less massive than $\lapprox$0.45 \Msun\ \citep{kilic07}, except in cases of extremely high metallicity. We therefore use a truncated Gaussian to describe the WD mass distribution, $0.45 ~\Msun < M_{\rm WD} < 1.35 ~\Msun$, which we can model using the mean mass $\mu_{\rm WD}$ and standard deviation $\sigma_{M_{\rm WD}}$ as parameters.

We set flat priors on $\mu_{\rm WD}$ and $\sigma_{M_{\rm WD}}$, limiting their ranges:
\begin{eqnarray}
\frac{\mu_{\rm WD}}{\Msun} &\in& [0.1,1.5] \nonumber \\
\frac{\sigma_{\rm WD}}{\Msun} &\in& [0.1,1.0]. \label{eq:prior_hyper}
\end{eqnarray}

More complex models may better represent the WD mass distribution, but for our purposes, this Gaussian model is sufficient. This allows us to add to our model two hyperparameters that weigh $\ta$ and $\tb$:
\begin{eqnarray}
P(\ta \given \bs{\Theta}) &=& \mathcal{N} (\bs{G F^{-1}}\ta \given \mu_{\rm WD}, \sigma_{M_{\rm WD}}) \\
P(\tb \given \bs{\Theta}) &=& \mathcal{N} (\bs{G F^{-1}}\tb \given \mu_{\rm WD}, \sigma_{M_{\rm WD}}). \label{eq:hyper}
\end{eqnarray}

Next, we choose a parametric form for the IFMR, $\bs{G}$. Observational IFMR constraints (such as those from open clusters) are traditionally fit with a linear relation \citep[e.g.,][]{williams09}. Here we opt for a more complex model based on our expectation that the IFMR has three distinct regimes: for $M_{\rm i}\  \lapprox\ 2$~\Msun, stars undergo a degenerate helium flash \citep{sweigart78}. For $2\ \lapprox\ M_{\rm i}\ \lapprox\ 4$~\Msun, stars will undergo stable, non-degenerate helium burning. Finally, second dredge-up becomes important for stars with $M_{\rm i}\ \gapprox\ 4$~\Msun\ \citep{dominguez99}.

Theoretical IFMRs from stellar evolution codes indicate roughly linear IFMRs for each of these regimes, with pivots at $M_{p,1} = 2$ and $M_{p,2} = 4$ \Msun. Assuming a continuous IFMR, we therefore construct a three-component, piecewise linear model with four free parameters: three separate slopes for each regime and a y-intercept that translates the relation vertically. 

\begin{figure}[h!]
\begin{center}
\includegraphics[width=0.90\columnwidth]{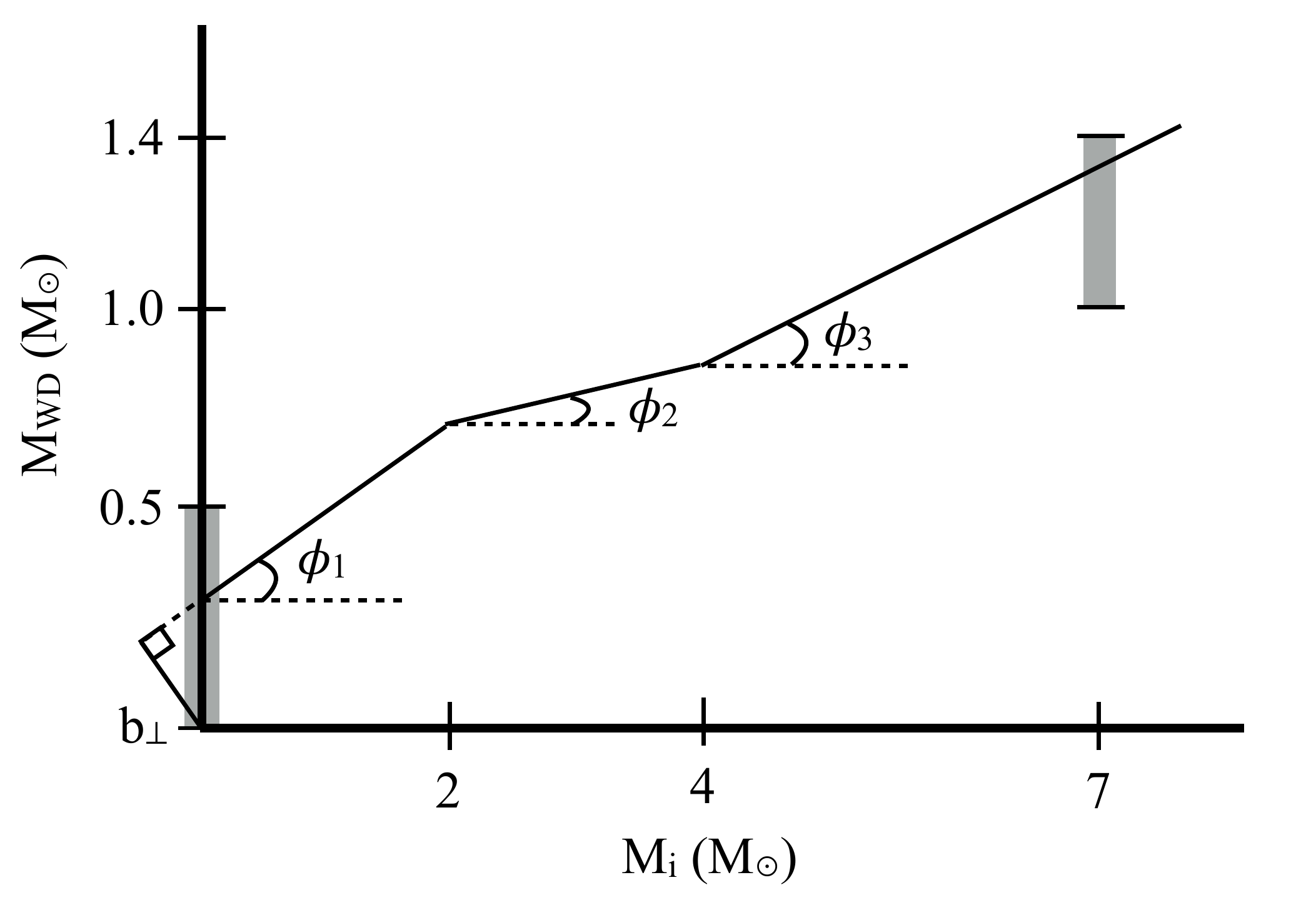}
\caption{\label{fig:model_params} The form of our IFMR ($\bs{G}$) is a three-component, piecewise linear model that we parameterize with the perpendicular distance of the line from the origin ($b_{\bot}$) and three horizontal angles ($\phi_1, \phi_2,$ and $\phi_3$). We also show model priors from Equation \ref{eq:prior_extras};  $\bs{G}$ must go through the gray regions at $M_{\rm i} =$ 0 and 7 $\Msun$. Our priors from Equation \ref{eq:prior_slopes} require that all three horizontal angles are positive and less than $\pi/2$ radians.
}
\end{center}
\end{figure}

Figure \ref{fig:model_params} demonstrates the parameterization of $\bs{G}$. We follow the recommendation of \citet{hogg10} to parametrize linear relations in terms of the angle each line makes with the horizontal and the perpendicular distance of the line from the origin ($\phi$ and $b_{\bot}$), instead of the slope and intercept ($m$ and $b$):
\begin{equation}
\bs{G} = \left\{ 
\begin{array} {ll}
{\rm tan}~\phi_1 ~(M_i) + b_1 & : M_{\rm i} < M_{p,1} \\
{\rm tan}~\phi_2 ~(M_i) + b_2 & : M_{p,1} < M_{\rm i} < M_{p,2} \\
{\rm tan}~\phi_3 ~(M_i) + b_3 & : M_{p,2} < M_{\rm i}. \\
\end{array}
\right.
\end{equation}

To ensure the model is continuous, we have: 
\begin{eqnarray}
 b_2 &=& ({\rm tan}~\phi_1 - {\rm tan}~\phi_2) ~M_{p,1} + b_1 \nonumber \\
 b_3 &=& ({\rm tan}~\phi_2 - {\rm tan}~\phi_3) ~M_{p,2} + b_2
\end{eqnarray}

Combined with our two WD mass distribution parameters, the piecewise function can then be expressed in terms of our four IFMR model parameters: 
\begin{eqnarray}
\bs{\Theta} &=& (\bs{\Theta_{\rm WD}}, \bs{\Theta_{\rm IFMR}}) \nonumber \\
\bs{\Theta} &=& (\mu_{\rm WD}, \sigma_{\rm WD}, b_{\bot}, \phi_1, \phi_2, \phi_3), \label{eq:model_params}
\end{eqnarray}
where $b_{\bot} = b_1 {\rm cos}~\phi_1$. We include priors on our model parameters to ensure that the model IFMR is an increasing function:
\begin{eqnarray}
\phi_1 &\in& [0, \pi/2] \nonumber \\
\phi_2 &\in& [0, \pi/2] \nonumber \\
\phi_3 &\in& [0, \pi/2]. \label{eq:prior_slopes}
\end{eqnarray}

Finally, we add a prior to our model so that $\bs{G}$ intercepts the y-axis between 0 and 0.5 \Msun, and a second prior that ensures that a 7.0 \Msun\ star produces a WD with a mass between 1.0 and 1.4 \Msun:
\begin{eqnarray}
0\ &<&\ \bs{G} (0.0 ~\Msun) < 0.5 ~\Msun \nonumber \\
1.0\ &<&\ \bs{G} (7.0 ~\Msun) < 1.4 ~\Msun. \label{eq:prior_extras}
\end{eqnarray}
These two constraints are shown by the gray regions in Figure \ref{fig:model_params}.

\subsection{Testing the Model with Mock Data} \label{sec:test}
We generate a set of mock observations using a test IFMR, then compare the derived constraints obtained from our model to the input IFMR. We first choose a test set of parameters for our IFMR and generate a mock sample of 20 wide DWDs. Table \ref{tab:mock_data} summarizes the distributions these mock data are drawn from.

The first WD in the pair is randomly assigned a mass from a truncated Gaussian distribution. The second WD's mass is then obtained based on a mass ratio, $q$, randomly generated from another Gaussian distribution. We determine $M_{\rm i}$ using the model IFMR, then the pre-WD lifetimes for each WD with the function shown in Figure \ref{fig:1TP_age}. 

$\tauc$ for the less-massive, slower-evolving WD is selected from a flat distribution in log space between 10 Myr and 1~Gyr. Following Equation~\ref{eq:diff_eq_diff}, we then assign the more massive WD the $\tauc$ of its companion plus the difference in pre-WD lifetimes. 

The WDs are then ``observed.'' Masses and cooling ages are randomly selected from Gaussian distributions centered on the true (mock) values, with standard deviations of 0.03 \Msun\ in WD mass and 10\% in $\tauc$ (these uncertainties are typical of WD spectral fits). Finally, the observed WD masses and $\tauc$ and their associated uncertainties are used as inputs for our Bayesian model. 

\begin{figure}[h!]
\begin{center}
\includegraphics[width=1.0\columnwidth]{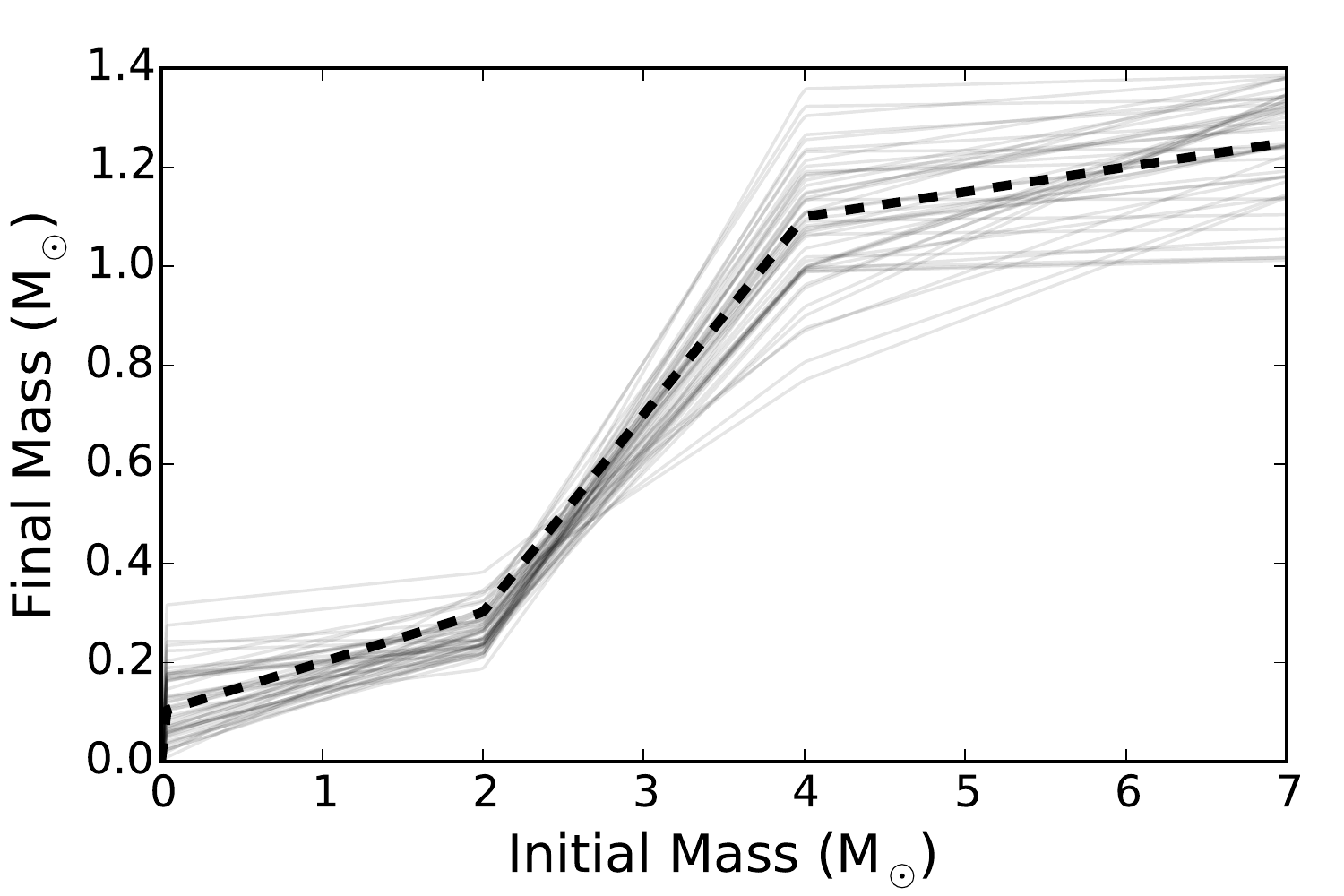}
\caption{\label{fig:mock_data_samples}   The input mock IFMR is shown by the black dashed line, while samples from the posterior are semi-transparent in gray. The posterior samples are evenly distributed around the input distribution.
}
\end{center}
\end{figure}

We use 32 separate chains in {\tt emcee}, running for 10,500 steps, the first 500 steps are a ``burn-in,'' which we throw away then check to make sure the chains have converged. Figure \ref{fig:mock_data_samples} shows our input model (black, dashed) and 50 randomly drawn samples (gray) from the posterior distribution of model parameters. Importantly, the posterior samples are evenly distributed around the input IFMR. The spread in the posterior samples indicate the constraints these mock systems place on the IFMR. 

\begin{deluxetable}{lll}
\tablecaption{Mock Data Parameters \label{tab:mock_data}}
\tablehead{
\colhead{Parameter} & 
\colhead{Input Distribution} &
\colhead{Range}
}
\startdata
$M_1$ & $\mathcal{N}(\mu = 0.75 ~\Msun, \sigma = 0.15 ~\Msun)$ & (0.6, 1.2) \\
$q$ & $\mathcal{N}(\mu = 1.0, \sigma = 0.15)$ & (0.45, 1.0) \\
log $\tau_2$ & $\mathcal{U}$ & (7.0, 9.0) \\
\cutinhead{Mock IFMR Parameters}
$\bs{\Theta_{\rm IFMR}}$ & (0.1, 0.1, 0.38, 0.05) &
\enddata 
\tablecomments{$\mathcal{N}$ is a Gaussian distribution with mean $\mu$ and standard deviation $\sigma$. $\mathcal{U}$ is the uniform distribution. The range dictates where the distributions are truncated. The model parameters, $\bs{\Theta_{\rm IFMR}}$ are defined in Equation~\ref{eq:model_params}. Observational uncertainties of 0.03 \Msun\ are assigned to $M_{\rm WD}$ and 10\% to  \tauc.}
\end{deluxetable}

\begin{figure}[h!]
\begin{center}
\includegraphics[width=1.00\columnwidth]{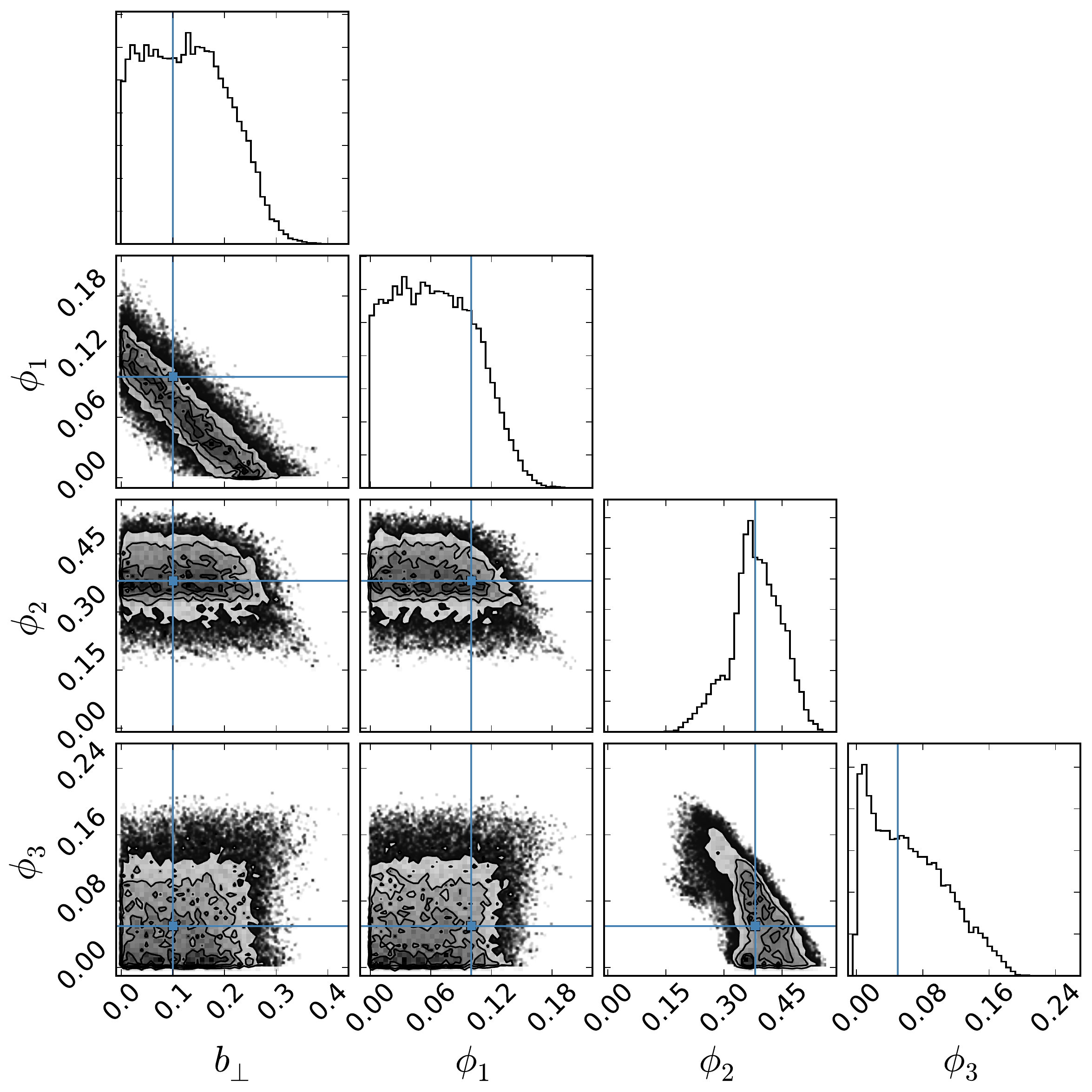}
\caption{\label{fig:mock_data_triangle}   Covariances between the four parameters in $\bs{\Theta_{\rm IFMR}}$ in our model for our mock data set. Lines indicate the input values for our parameters. Our posterior distributions are centered around the input parameters, indicating that our model converged to the correct solutions.}
\end{center}
\end{figure}

\begin{figure}[h!]
\begin{center}
\includegraphics[width=0.90\columnwidth]{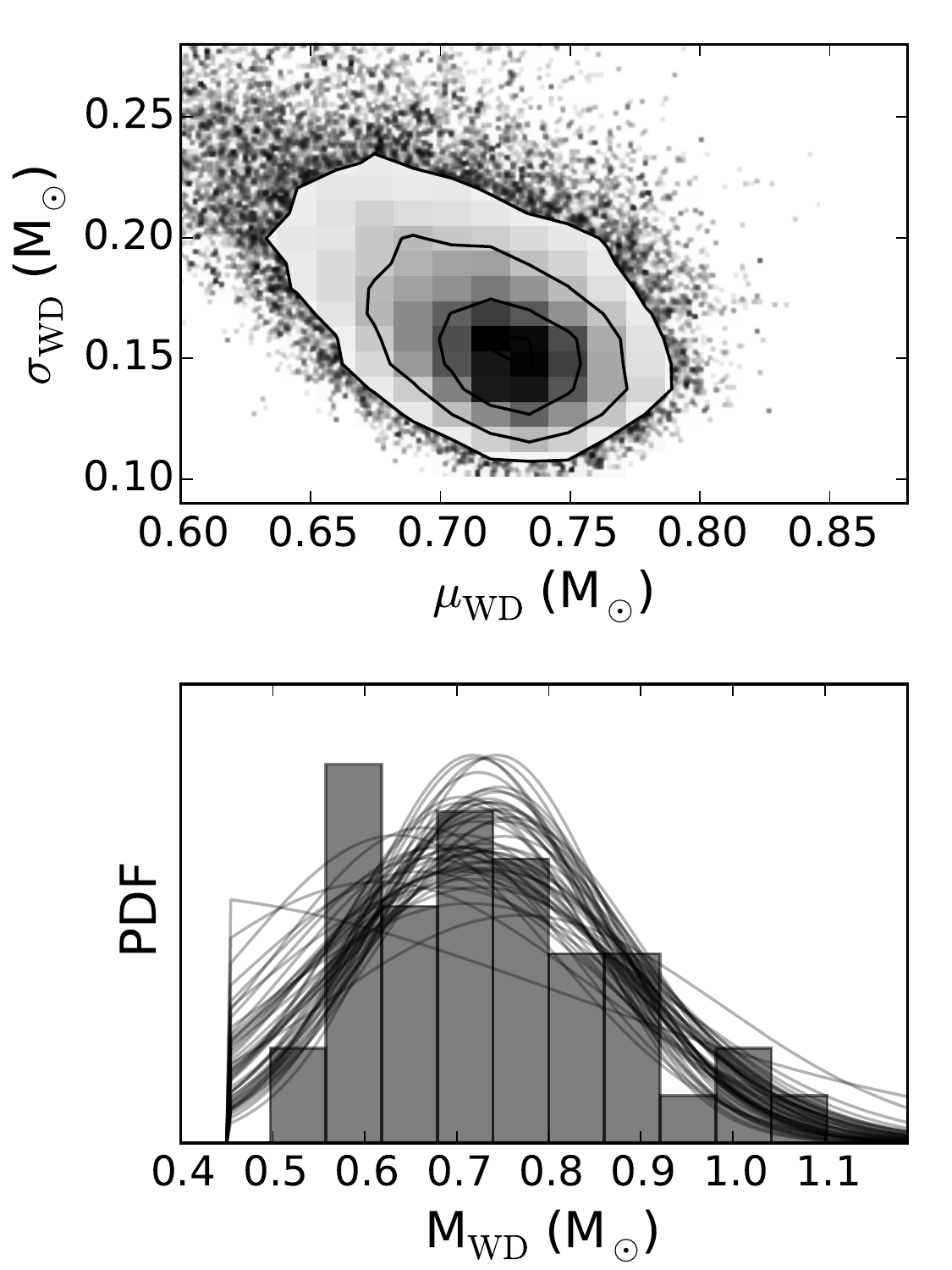}
\caption{\label{fig:mock_mass_dist} The top panel shows the covariance between the WD mass distribution model parameters $\mu_{\rm WD}$ and $\sigma_{\rm WD}$ for our mock data set. The bottom panel shows a normalized histogram of the WD masses in our test sample. Lines show samples from the posterior distribution of model parameters. }
\end{center}
\end{figure}

Figure \ref{fig:mock_data_triangle} shows the covariances between the four IFMR parameters in our model. Our model, for this particular combination of input parameters and mock wide DWDs, is able to constrain only certain regions of phase space. Figure \ref{fig:mock_data_samples} shows why; since there are no $<$0.5~\Msun\ WDs in the sample, our model is insensitive to the exact form of the first component in our piecewise linear model. Similarly, there are few $>$1.0~\Msun\ WDs in our sample, and the uncertainties in $\tauc$ estimates for these are of order the differences in the pre-WD lifetimes (however, these data can constrain $\phi_2$, the second piecewise linear component of our model). For the set of mock wide DWDs shown here, the constraints are most stringent between roughly 2 and 3 \Msun.

The top panel of Figure \ref{fig:mock_mass_dist} shows the covariance between the two model parameters describing the WD mass distribution. Although our model is too simple to exactly reproduce the input WD masses, the bottom panel shows that posterior samples from our model (lines) approximate the input WD mass distribution (histogram).

\subsection{Applying the Model to Wide DWDs}
\begin{figure}[h!]
\begin{center}
\includegraphics[width=0.99\columnwidth]{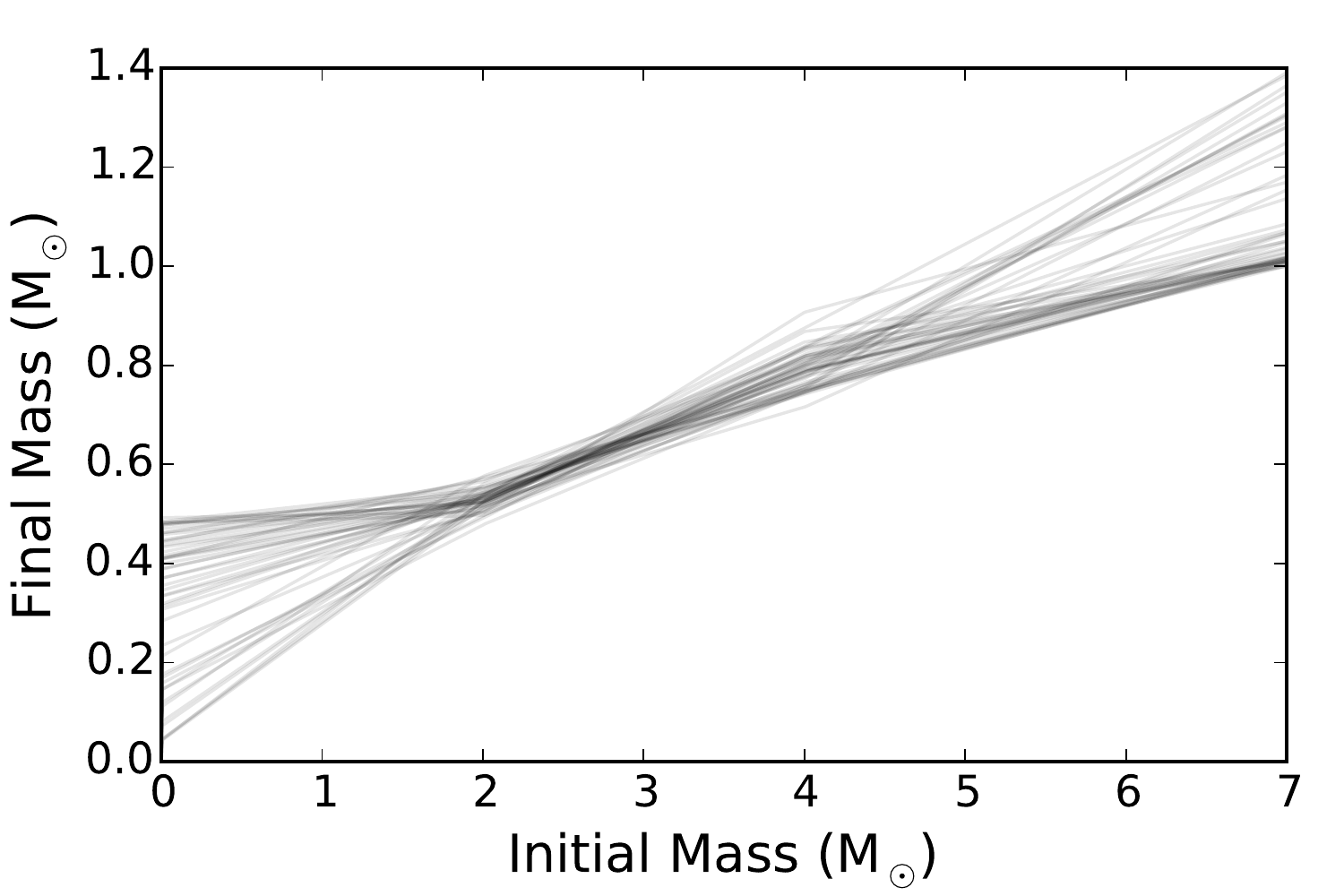}
\caption{\label{fig:DWD_samples} Samples from the posterior distribution for our fiducial model are semi-transparent in gray. The model converges for the second linear component ($M_i =$ 2--4~\Msun), but diverges outside of this mass range.}
\end{center}
\end{figure}

We applied our parametric model to our sample of wide DWDs using \texttt{emcee}. We check the 32 separate Markov chains to make sure they have converged. Here again, we throw away the first 500 steps and run the model for another 50000 steps. We make samples from our posterior distribution publicly available for download.\footnote{http://dx.doi.org/10.6084/m9.figshare.1572148}  Figure~\ref{fig:DWD_samples} shows selected samples from the posterior distribution of model parameters. The posterior samples converge between 2 and 4 \Msun. 

Reassuringly, this mass range corresponds to $\approx$0.5--0.8~\Msun\ WDs, roughly the masses of the WDs in our sample (see Table~\ref{tab:IFMR_sample}). Smaller $M_{\rm i}$ produce WDs too low-mass to be found in our sample, and our method is not as sensitive to the small differences in the pre-WD lifetimes of more massive stars. 

\begin{figure*}
\begin{center}
\includegraphics[width=0.92\textwidth]{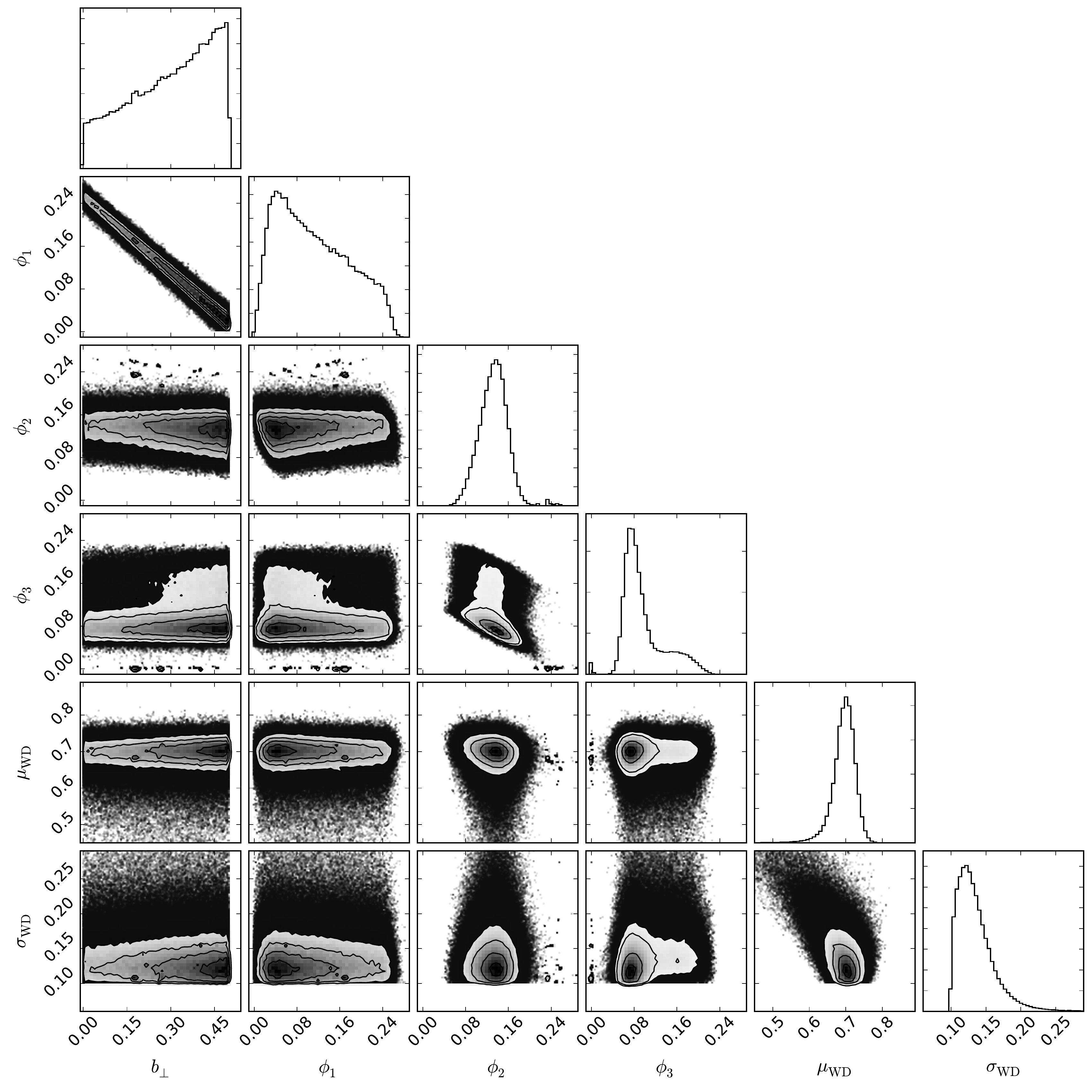}
\caption{\label{fig:DWD_triangle}  Covariances of the model parameters for our model applied to our sample of wide DWDs. }
\end{center}
\end{figure*}

The covariances between the different model parameters are shown in Figure~\ref{fig:DWD_triangle}. These confirm that the parameters have converged. In particular, the slope of the second piecewise linear component of our model, $\phi_2$, is well constrained, and the covariance between $b_{\bot}$ and $\phi_1$ indicates that the model is well constrained near the pivot point at $M_{\rm i} = 2~\Msun$.

\subsection{A Test of the Number of Model Parameters} \label{sec:discuss_model}
To test the dependence of our results on our model, we extend our model by allowing the pivot points to vary. We then have a model with eight parameters: $\mu_{\rm WD}$, $\sigma_{\rm WD}$, $\phi_1$, $\phi_2$, $\phi_3$, $b_{\bot}$, $M_{p,1}$, and $M_{p,2}$. We keep the same priors on the first six parameters in Equations \ref{eq:prior_hyper}, \ref{eq:prior_slopes}, and \ref{eq:prior_extras}, and add flat priors on the pivot masses:
\begin{eqnarray}
M_{p,1} &\in& [1.5, 2.5] \nonumber \\
M_{p,2} &\in& [3.5, 4.5]. \label{eq:prior_pivot}
\end{eqnarray}

\begin{figure}[h!]
\begin{center}
\includegraphics[width=1.05\columnwidth]{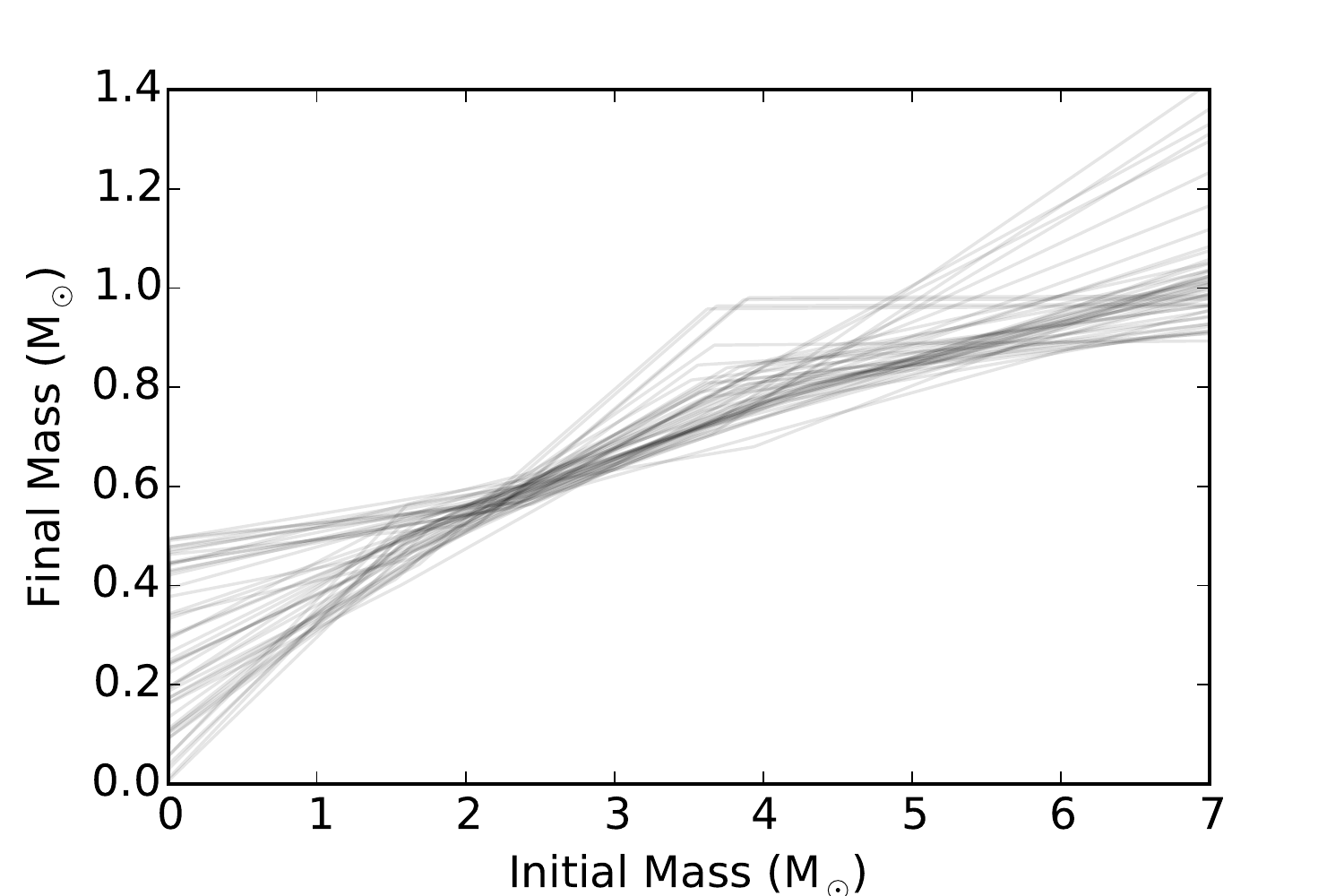}
\caption{\label{fig:DWD_samples_var_pivot} Samples from the posterior distribution for our eight-parameter model in which the pivot points for our three-component, piecewise model are allowed to vary. The results are consistent with those from the six-parameter model: our DWDs constrain the IFMR between $\approx$2--4 \Msun. }
\end{center}
\end{figure}

Figure~\ref{fig:DWD_samples_var_pivot} shows the results when we apply this eight-parameter model. Even with this more flexible model, the posterior distribution converges to a narrow distribution in $M_{\rm WD}$ for $M_{\rm i} =$ 2--4~\Msun, suggesting that this convergence is not a result of our original choice to fix the pivot points.

In model parameter space, however, the masses at which the IFMR pivots do not converge. This can be seen in Figure~\ref{fig:DWD_samples_var_pivot}: the pivot points vary across the whole allowed range. Perhaps with a future, larger data set, this model will be able to constrain $M_{p,1}$ and $M_{p,2}$, but the current sample of wide DWDs cannot. 

Because our original six-parameter model reasonably represents the constraints from our DWD sample and the Markov chains converge to a region in parameter space, we choose that model as our fiducial model for the remainder of our analysis. 

\section{Discussion} \label{sec:dis}

\subsection{Comparison to Theoretical Predictions} \label{sec:theory_IFMR}

While stellar evolution codes still disagree significantly on the functional form of the IFMR, major divergences typically occur after the first thermal pulse, and codes general produce similar core masses at the 1TP \citep{kalirai14}. Furthermore, one of the primary expectations from theory is that the core mass will not diminish on the TP-AGB \citep[e.g.,][]{karakas02}. This provides a significant sanity check for any observationally derived IFMR: for a given $M_{\rm i}$, the resulting WD should be at least as massive as the 1TP core mass. 

We compared samples from the posterior distribution from our fiducial model to the 1TP core masses obtained from the theoretical relations of \citet[][]{dominguez99}, \citet[][]{weiss09}, and \citet[][]{kalirai14}. Reassuringly, between 2 and 4 \Msun, our fiducial model constrains the IFMR to be above these relations.

Evolving from the 1TP to produce a WD is computationally challenging. The pulses that give the TP-AGB phase its name are due to thermal instabilities in the helium-burning shell of the AGB star \citep{schwarzschild65}. These pulses are suspected to form a temporary convective zone within the He inter-shell region. This leads to the so-called third dredge-up, in which the convective envelope extends into the inter-shell region, mixing heavier elements into the outer envelope \citep{iben75}. This sequence of successive dredge-up events and overshooting naturally explain the observed C abundances in AGB stars \citep[cf. discussion in][]{herwig00}.

Although convective overshooting has been seen in three-dimensional numerical simulations, such calculations are currently only possible for small regions of a stellar atmosphere over small timescales \citep[e.g.,][]{freytag12}. For now, stellar evolution predictions are typically limited by the ability of 1D models, which rely on some form of mixing-length theory \citep{erica1958} combined with a prescription for convective overshooting \citep[cf.][and references therein]{herwig00}, to calculate accurately the energy transport during these pulses. This can cause problems. For example, 1D approximations can lead to large radiation pressures developing at the base of the convective envelope, which can in turn lead to unphysically large (supersonic) radial velocities. One way to deal with this involves decreasing the opacity profiles by hand in these regions \citep[e.g.,][]{renedo10,pignatari13}.

\begin{figure}[h!]
\begin{center}
\includegraphics[width=0.99\columnwidth]{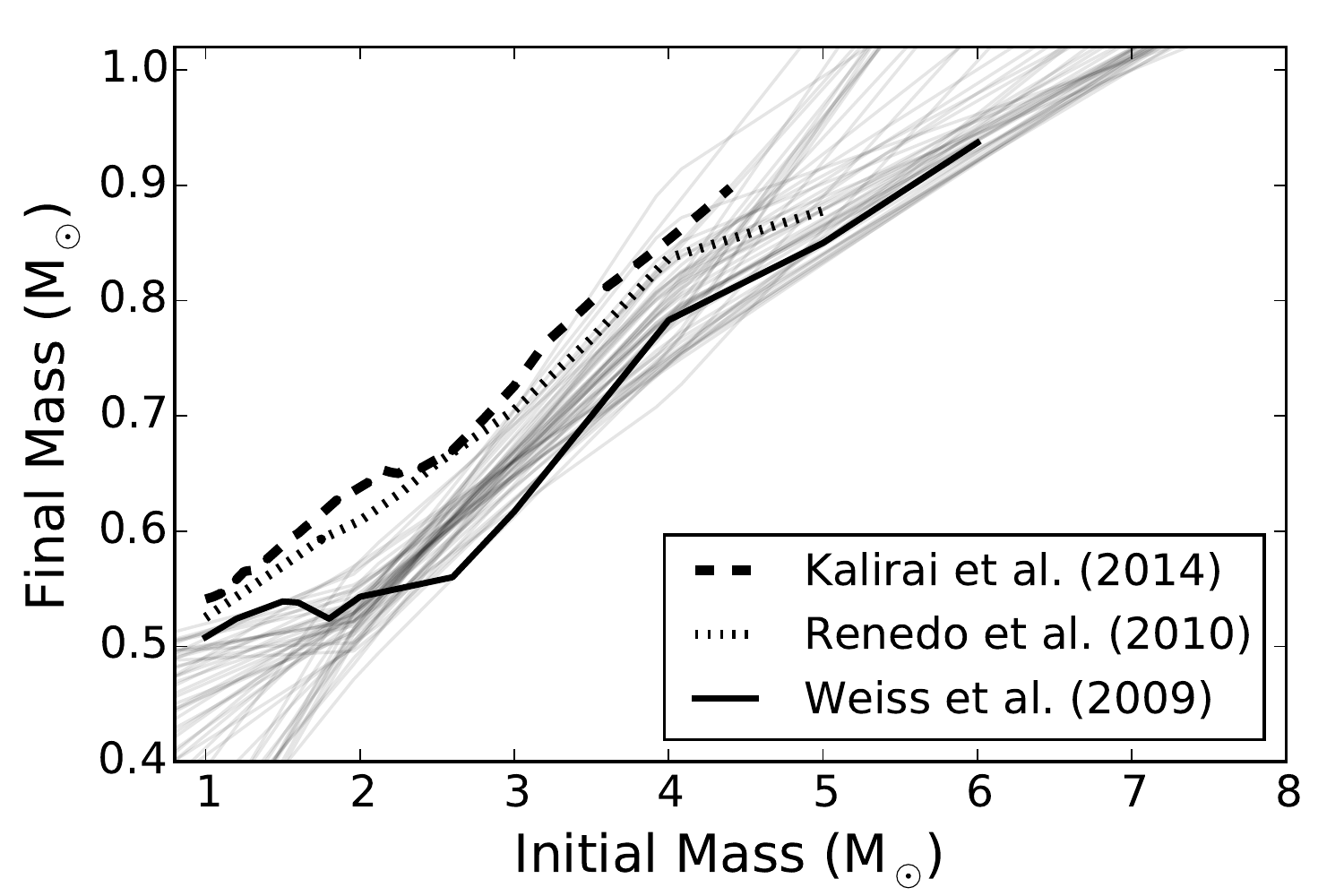}
\caption{\label{fig:DWD_theory} A comparison between samples from the posterior distribution from our fiducial model (gray lines) and theoretical IFMRs obtained from three separate stellar evolution codes. The differences between the codes are due primarily to treatment of the dredge-up and wind mass loss. }
\end{center}
\end{figure}

With these caveats in mind, we present in Figure~\ref{fig:DWD_theory} three IFMRs produced by different stellar evolution codes; the major differences between them stem from the treatments of dredge-up and wind mass loss. \citet{renedo10} argue, based on the theoretical results of \citet{canuto98} and observations of $s$-process abundances by \citet{lugaro03}, that the third dredge-up should be suppressed. In this scenario, the He core grows continuously, its mass only limited by the reduction of the envelope due to stellar winds, as described using the prescription of \citet{vassiliadis93}.

The \citet{kalirai14} IFMR uses the AGB models of \citet{marigo13a}, which improve the treatment of the onset of the third dredge-up and the calculation of opacities. However, the resulting IFMR is still dependent upon a prescription for the dredge-up efficiency   \citep[\citet{kalirai14} use the prescription of][]{karakas02} and the wind mass loss. \citet{kalirai14} calibrate their prescriptions by defining ignorance parameters and then finding the best fit ignorance parameters to match the observed open cluster constraints.

Finally, \citet{weiss09} use the exponential overshooting prescription of \citet{herwig97}. On the AGB, these authors use the wind mass loss prescription of \citet{wachter02} for C-rich AGB stars and the prescription of \citet{vLoon05} for O-rich AGB stars. The \citet{weiss09} simulations show little core growth on the TP-AGB, resulting in less massive WDs for a given $M_{\rm i}$ compared with the models produced by \citet{renedo10} and \citet{kalirai14}.

Figure~\ref{fig:DWD_theory} includes our posterior samples with these three theoretical IFMRs. For $M_{\rm i} =$~2--4 \Msun, our model converges to a region of parameter space that suggests that the IFMR may lie between the higher final-mass models of \citet{renedo10} and \citet{kalirai14} and the lower final-mass model of \citet{weiss09}. We may not yet be able to place stringent constraints on wind-mass-loss models and dredge-up efficiencies with the available data, but Figure~\ref{fig:DWD_theory} suggests that DWDs may eventually provide important new observational tests for stellar evolution codes.

\begin{figure}[h!]
\begin{center}
\includegraphics[width=0.99\columnwidth]{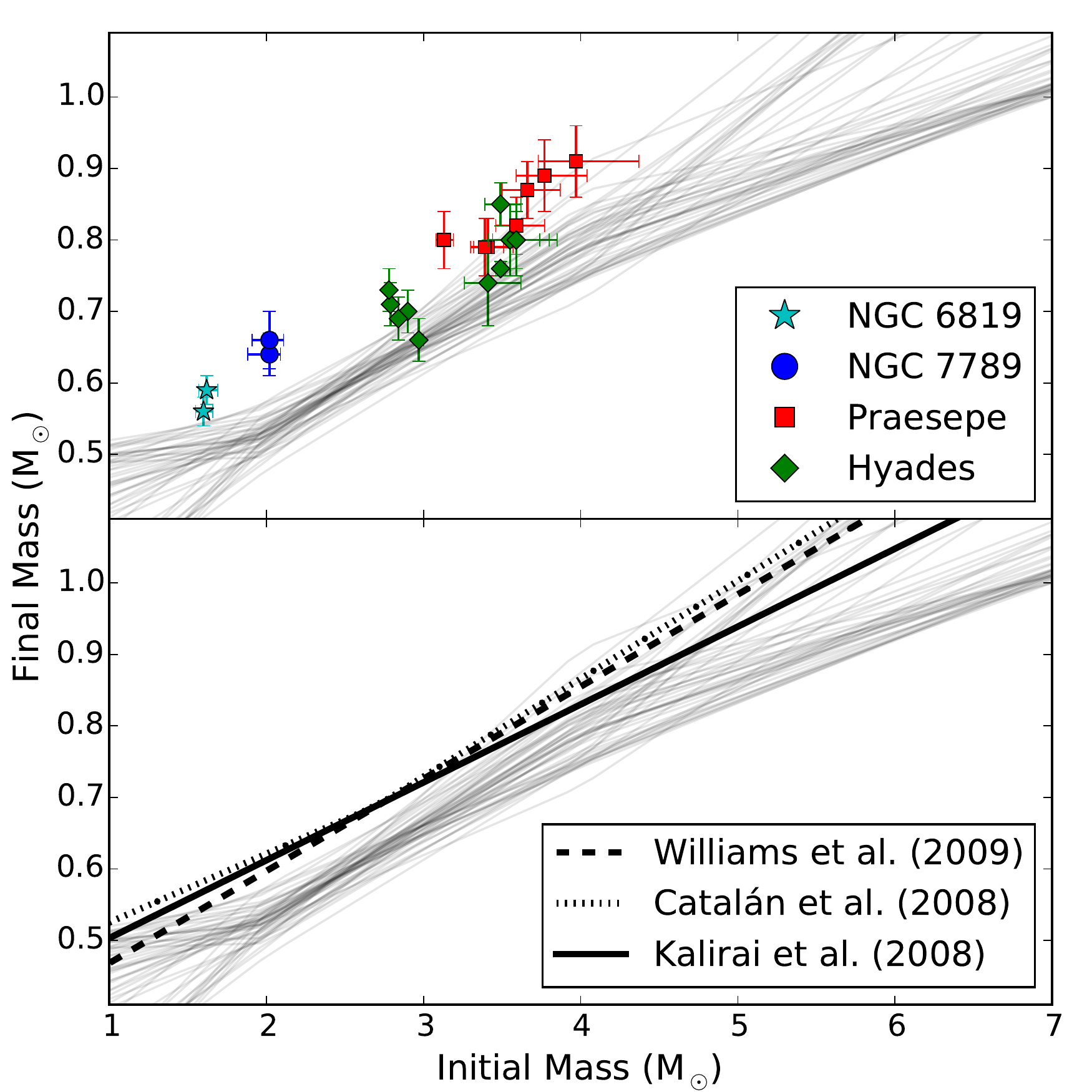}
\caption{\label{fig:DWD_obs} The top panel shows samples from the posterior distribution for our model (gray lines) and the observational constraints generated by WDs in four older open clusters \cite[from][]{kalirai14}. The bottom panel compares the same posterior samples to the semi-empirical linear fits of \citet{catalan08b}, \citet{kalirai08}, and \citet{williams09}.}
\end{center}
\end{figure}

\subsection{Comparison to Other Observational Constraints} \label{sec:IFMR_obs}
The most commonly used method for constraining the IFMR is that pioneered by \citet{sweeney76}. Here, the WDs are members of open clusters for which reliable ages can be obtained from e.g., isochrone fitting. There are at least two important limitations to consider when discussing this method. The first is that even for the nearby, well-studied open clusters that are generally used in these studies, significant disagreements about the stars' ages are not unusual. Furthermore, the techniques employed to obtain these ages differ, so that the constraints on the IFMR depend on different sets of systematic effects.

The second is that most easily accessible open clusters are young, with ages $<<$1 Gyr, so that only the most massive stars have evolved off the main-sequence. There are four notable exceptions: NGC 6819, NGC 7789, the Hyades, and Praesepe are all old enough for their $\approx$2--4 \Msun\ members to have evolved into WDs. \citet{kalirai14} recently re-analyzed the 18 WDs identified in these four clusters using improved WD atmospheric models. These data are shown in the top panel of Figure~\ref{fig:DWD_obs}. Our results are consistent with those derived from WDs in the Hyades, but the initial masses for WDs in NGC 7789 and Praesepe are systematically lower than our posterior samples.\footnote{We ignore the WDs in NGC 6819, since these evolved from $<$2~\Msun\ stars, outside the mass range to which we are sensitive.}

Various authors have pointed out that metallicity could lead to a natural spread in the IFMR \citep[e.g.,][]{kalirai05}. However, the four clusters hosting the WDs in Figure~\ref{fig:DWD_obs} all have near-solar metallicities,\footnote{
NGC 6819: [Fe/H]$ = -0.02$$\pm$0.02 \citep{lee-brown15}; NGC 7789: $+0.03$$\pm$0.07 \citep{overbeek15}; 
Praesepe: $+0.16$$\pm$0.05 \citep{carrera11}; 
Hyades: $+0.11$$\pm$0.01 \citep{carrera11}.}
while we also expect the DWDs in our sample to have roughly solar metallicity (see Section~\ref{sec:metal}). While \citet{meng08} and \citet{romero15} showed that metallicity-dependent mass-loss rates can result in WDs with masses varying by as much as 0.2 \Msun, the range is principally due to cases of extremely low or high metallicities. For metallicities near solar, variations in $M_{\rm WD} \lesssim 0.05$ \Msun\ (see figure 1 in Meng et al.~2008 and figure 4 in Romero et al.~2015). Metallicity differences are therefore unlikely to explain the discrepancy between our constraints and the WDs in these open clusters.

Instead, the difference could stem from uncertainties in the cluster ages. \citet{kalirai14} followed \citet{claver01} in adopting the {\it Hipparchos}-derived Hyades age of 625 Myr \citep{perryman98} for Praesepe.\footnote{For a recent discussion of the evidence that the two clusters are indeed the same age, see \citet{Douglas2014}.} By contrast, when fitting metallicity-specific isochrones to Praesepe, \citet{salaris09} found an age ranging from $\approx$450--650 Myr. The inclusion of convective overshooting, which extends stellar lifetimes, accounts for the difference. If, instead of 625 Myr, the Hyades and Praesepe are 550 Myr, the corresponding data points in Figure~\ref{fig:DWD_obs} would all shift right, toward higher $M_{\rm i}$, by $\approx$0.2--0.3~\Msun. These constraints would then be consistent with samples from our posterior distribution.\footnote{\citet{brandt15} recently argued that if stellar models accounting for rotation are used in the isochrone fitting, both clusters are $\approx$800 Myr, illustrating just how much uncertainty remains about the age of two of the nearest, best-studied open clusters.} 

The discrepancy with the two WDs in NGC 7789 could also be due to an inaccurate cluster age. \citet{kalirai14} adopt the age of 1.4 Gyr  obtained by \citet{kalirai08} from isochrone fits. However, isochrone-derived ages for this cluster vary noticeably in the literature, ranging from 1.1 Gyr \citep{mazzei88} to 1.6 Gyr \citep{gim98}. A modestly younger age of 1.2 Gyr for NGC 7789 increases the derived $M_{\rm i}$ by $\approx$0.2~\Msun, enough that these data also become consistent with samples from our posterior distribution. 

There are a number of other possibilities for the differences we see in Figure~\ref{fig:DWD_obs}. Our piecewise-linear model might be inappropriate because there is significant structure in the IFMR for $M_{\rm i} =$~2--4 \Msun. Some of the WDs in either the open cluster or wide DWD data may have unresolved binary companions that affected their evolution, and therefore should not be used to constrain the IFMR. To identify the source(s) of this discrepancy, progress needs to be made with both methods: we need more complex models than that presented here for the IFMR, combined with larger sets of wide DWDs, as well as new WDs in older open clusters. 

The bottom panel of Figure~\ref{fig:DWD_obs} shows that our posterior samples lie significantly below the semi-empirical IFMRs of \citet{catalan08b}, \citet{kalirai08}, and \citet{williams09}. These authors use somewhat different WD samples and spectroscopic solutions in generating their linear fits, but for $M_{\rm i} =$~2--4~\Msun, the constraints are obtained predominantly with the WDs shown in the top panel of Figure~\ref{fig:DWD_obs}.

Considering the uncertainties in the ages of the clusters hosting these WDs, we suggest that these semi-empirical relations be used with caution for $M_{\rm i}=$~2--4~\Msun. Samples from the posterior distribution from our model may be more accurate within this $M_{\rm i}$ mass range.

\section{Conclusions} \label{sec:concl}
In an effort to provide new, independent constraints on the IFMR, we began by conducting a comprehensive search for wide DWDs in the SDSS DR9 photometric catalog. Using two separate methods, we identified 65 new candidate systems. By combining these pairs with those already in the literature, we assembled a sample of 142 candidate and confirmed wide DWDs. 

To confirm the WD nature of the stars in these pairs, and to obtain accurate mass and \tauc\ measurements for them, we engaged in a spectroscopic campaign using the 3.5-m APO telescope (and also collected spectra from the literature). Our targets included new systems identified in \citet{andrews12}, those photometrically selected from DR9, and literature pairs that lacked spectroscopy. The contamination by non-WDs was extremely low: only one of the 97 objects for which we obtained spectra was not a WD. Fitting WD model atmospheres to our spectra gave us \logg\ and \Teff\ values; these were then converted to $M_{\rm WD}$ and \tauc\ for each star.

In addition to 27 DA/DA pairs, our campaign identified a number of interesting systems that cannot be used to constrain the IFMR.  We confirmed the nature of the second known DA/DB system, SDSS J0849$+$4712 (CDDS15), and identified SDSS J2355$+$1708 as the third DA/DB DWD. DBs may evolve from  H-deficient post-AGB stars \citep{althaus05}, and it is unclear whether the same IFMR applies to these stars as to DA WDs and their progenitors. We found four new DC WDs with DA companions (we also confirm the nature of another previously identified DA/DC system); the lack of absorption features makes it impossible to determine the mass of the DCs from spectra alone. We identify four new candidate DA/DAH pairs, and confirm the nature of three previously known systems; the DAH spectra cannot be fit with DA atmospheric models, and in general require higher-resolution and/or better S/N spectra to confirm their nature. 

Finally, we identify two candidate triple systems: PG 0901$+$140 (based on the anomalously low mass of one of its components, this system could be a triple degenerate) and J2047$+$0022B (which was previously identified by \citet{silvestri06} as an unresolved DQ+K7 binary). Because of the potential for mass transfer in the unresolved pair, candidate triple systems such as these had to be excluded from our efforts to constrain the IFMR.

We divided our double DAs into high- and low-fidelity pairs, labeling the DWDs with mass uncertainties $>$0.1~\Msun\ in at least one WD as low-fidelity. These pairs have spectra good enough to identify objects as DAs, but too poor to obtain accurate fits to model atmospheres. We also considered the spectroscopic distances to each WD in these candidate DWDs, and designated pairs with a distance difference $>$25\% as low-fidelity. Finally, we removed from our high-fidelity sample DWDs for which the more massive WD appeared to have a shorter \tauc\ than its less-massive companion, as this goes against standard expectations from stellar evolution. Combining our high-fidelity pairs with DWDs from \citet{baxter14} for which we lacked spectroscopic data, we obtained a sample of 19 DWDs with which to constrain the IFMR.

Because the members of wide binaries are co-eval and evolve independently (i.e., without mass transfer affecting their evolutionary pathways), the age of each WD in a DWD is the same and is the sum of each WD's \tauc\ and pre-WD lifetime. Using this as a starting point, we developed a hierarchical Bayesian framework that tests the likelihood that any particular IFMR accounts for the observed masses and cooling ages, and corresponding uncertainties, of a wide DWD. We then constructed a parametrized form of the IFMR, choosing a continuous, piecewise-linear function with pivot points at 2 and 4 \Msun. We also included a model for the underlying WD mass distribution, which we varied simultaneously with the IFMR parameters. 

We first tested our model on a set of mock wide DWDs and successfully recovered our input IFMR. We then applied our model to our 19 high-fidelity DWDs, using a Markov Chain Monte Carlo approach to find the region of parameter space implied by the data. We make samples from our posterior distribution of parameters available for download. The resulting constraints are comparable to previous constraints on the IFMR obtained using WDs in open clusters. However, our results produce larger $M_{\rm i}$ than previous observations (or alternatively, we found that stars of a given $M_{\rm i}$ produce less massive WDs). Importantly, our constraints are most sensitive to $M_{\rm i} =$~2--4~\Msun, a regime that the open cluster data have difficulty testing. We found no improvement when we tested an expanded model in which the pivot points vary; our original model with fixed pivot points is sufficient to describe our current set of DWDs.

Our model can be expanded to include other constraints, including those from open cluster WDs, globular cluster WD cooling tracks, and Sirius-like binaries. Because each of these methods has their own associated systematic uncertainties, including these in a statistically responsible way is not straightforward, however. 

Our three-component, piecewise linear model is only an approximation for the true, physical IFMR. Ultimately, we would like to be able to constrain physically meaningful stellar evolution parameters. \citet{kalirai14} recently performed such an analysis using the WDs presented in Figure~\ref{fig:DWD_obs}. Our method here could similarly be expanded to constrain stellar evolution directly. For example, the form of the IFMR within the $M_{\rm i} =$~2--4 \Msun\ range is sensitive to physics on the TP-AGB. DWD data may be precise enough to place important constraints on uncertain dredge-up and overshooting physics. We leave this to future work.

In our search for new DWDs, we relied on SDSS photometry combined with proper motion measurements to identify WD candidates, while masses and \tauc\ were derived from fits to WD template spectra. With precision astrometry from the {\it Gaia} space telescope, identifying WDs and matching them with common proper motion companions will be significantly easier. Furthermore, as pointed out by  \citet{carrasco14}, data from the BP and RP spectrophotometers, combined with distance measurements and a mass-radius relation, will provide $\Teff$ and $M_{\rm WD}$ for every WD identified. These authors estimate {\it Gaia} will find some 250,000 to 500,000 WDs. We do not know the space density of wide DWDs in the Galaxy, but it is hard to escape the conclusion that {\it Gaia} will identify hundreds to thousands of new wide DWDs. With a measured $\Teff$ and $M_{\rm WD}$ for the WDs in each of these pairs, these wide DWDs will potentially revolutionize our understanding of the IFMR.

\acknowledgments We thank the referees for their comments which led to improvements in the manuscript. We thank the observing specialists for their help with the APO observations. We thank P.~Bergeron for fitting the DB WDs in our sample and D.~Koester for providing us with VLT spectra of several of the WDs discussed here. We thank Lars Bildsten, Silvia Catal\'an, Falk Herwig, David Hogg, Marcelo Miller Bertolami, Rodolfo Montez, Jr., and Adrian Price-Whelan for stimulating and helpful discussions. M.A.A., M.K., and A.G.~gratefully acknowledge the support of the NSF and NASA under grants AST-1255419, AST-1312678, and NNX14AF65G, respectively.

This work made use of the {\it yeti} cluster at the Columbia University Shared Research Computing Facility, which is supported by NIH Research Facility Improvement Grant 1G20RR030893-01 and matching funds from New York State/the Empire State Development's Division of Science, Technology and Innovation (NYSTAR) under Contract C090171.

Funding for SDSS-III has been provided by the Alfred P.~Sloan Foundation, the Participating Institutions, the National Science Foundation, and the U.S. Department of Energy Office of Science. The SDSS-III web site is http://www.sdss3.org/.

SDSS-III is managed by the Astrophysical Research Consortium for the Participating Institutions of the SDSS-III Collaboration including the University of Arizona, the Brazilian Participation Group, Brookhaven National Laboratory, University of Cambridge, Carnegie Mellon University, University of Florida, the French Participation Group, the German Participation Group, Harvard University, the Instituto de Astrof\'isica de Canarias, the Michigan State/Notre Dame/JINA Participation Group, Johns Hopkins University, Lawrence Berkeley National Laboratory, Max Planck Institute for Astrophysics, Max Planck Institute for Extraterrestrial Physics, New Mexico State University, New York University, Ohio State University, Pennsylvania State University, University of Portsmouth, Princeton University, the Spanish Participation Group, University of Tokyo, University of Utah, Vanderbilt University, University of Virginia, University of Washington, and Yale University.

\appendix

\label{sec:appendix}

Our spectroscopic campaign uncovered a number of interesting DWDs that could not be used to constrain the IFMR. These include two systems in which a DA is paired with a DB WD, which is a WD lacking an optically thick H layer and with \Teff\ \gapprox\ 12,000 K. SDSS J0849$+$4712 (CDDS15) was previously identified by \citet{baxter14} as the second known DA/DB system after L 151-81 A/B \citep{oswalt88}. Here we report J2355$+$1708 as the third identified wide DA/DB DWD. 

\begin{figure}[]
\begin{center}
\includegraphics[width=1.0\columnwidth,angle=90]{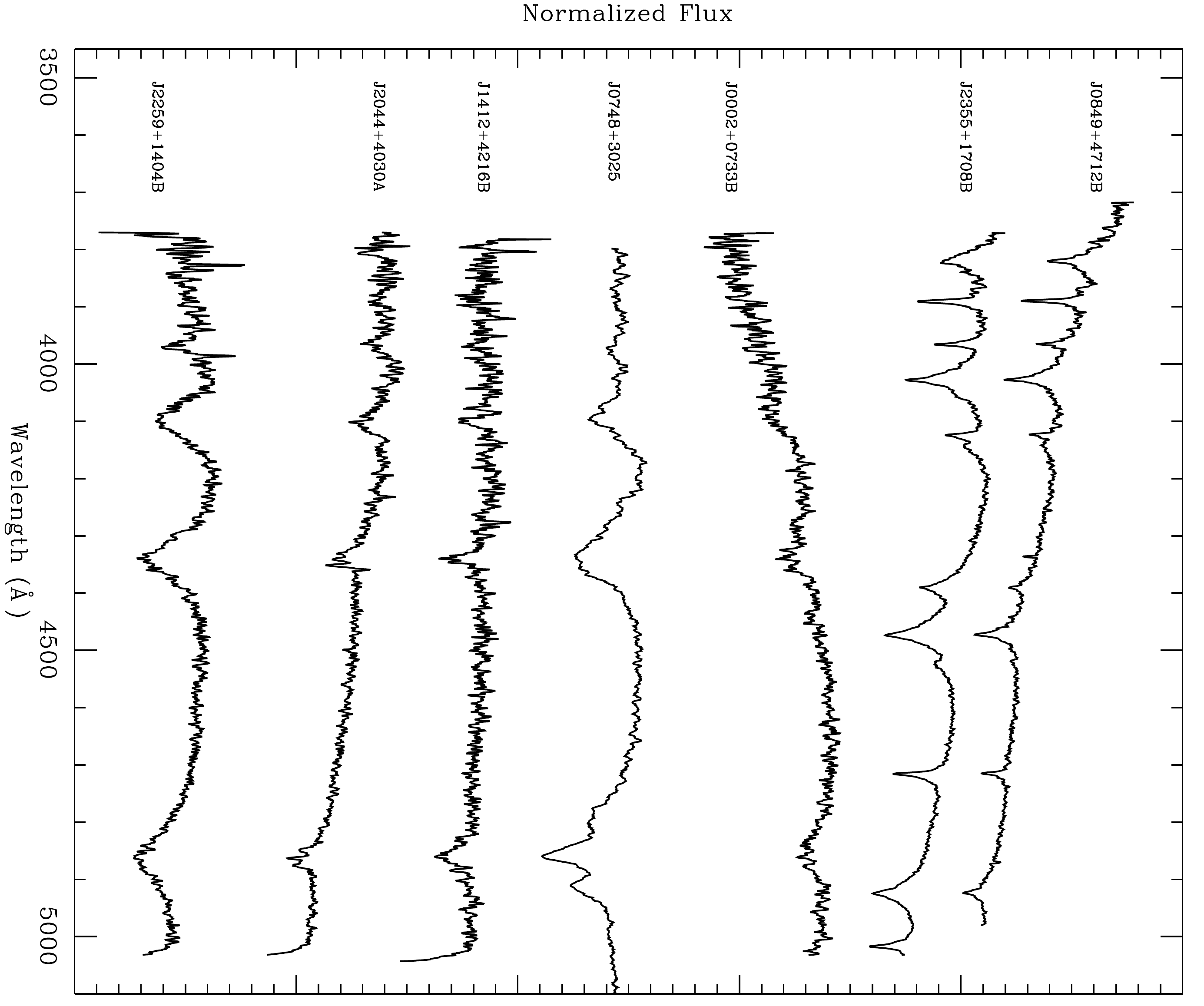}    
\caption{The two top spectra are for the DBs in our sample: J0849$+$4712B (CDDS15-A) and J2355$+$1708B. The other spectra are for five DAHs: J0002$+$0733B, J0748$+$3025, J1412$+$4216B, J2044$+$4030A, and J2259$+$1404B. We do not include the SDSS spectra for PG 1258$+$593B and J1314$+$1732A, which show Zeeman splitting in the H$\alpha$ line, but not in the higher order Balmer lines.}
\label{fig:non_DA_spec}
\end{center}
\end{figure}

The DBs in these systems, SDSS J0849$+$4712B (CDDS15-A) and J2355$+$1708B, are shown in Figure~\ref{fig:non_DA_spec}. The fit results for these stars in Table~\ref{tab:DWD_spec_quants} were provided by P.~Bergeron (priv.~communication). DBs probably form through a different channel than DAs. They may evolve from H-deficient post-AGB stars \citep{althaus05}, and it is unclear whether the same IFMR applies to these stars.

As DBs cool to $\Teff\ \lapprox\ 12,000$ K, they can no longer ionize He. They are then known as DC WDs and characterized by their featureless blackbody spectra. The lack of absorption features makes it impossible to determine the mass of these DC WDs from spectra alone. We identify four DC WDs in our sample: LP 549-32, J0029$+$0015A (CDDS2), J0344$+$1510A, and J1544$+$2344B. GD 559 was previously identified as a DA/DC wide pair by \citet{farihi05}. Due to their cool temperatures, the spectra of these objects generally have a low S/N.

Magnetic DAs (DAHs) are identified by Zeeman splitting in the Balmer lines. We have identified four new candidate DAHs: J0002$+$0733B, J1314$+$1732A, J1412$+$4216B, and J2044$+$4030A. We also confirm the DAH nature of three previously identified DAH WDs in wide pairs: PG 1258+593B \citep{girven10}, J0748$+$3025 \citep[CDDS11][]{dobbie13,baxter14} J2259$+$1404B \citep[CDDS52][]{baxter14}. Spectra for several of these are shown in Figure~\ref{fig:non_DA_spec}. Since DA atmospheric models cannot be used here, we only include spectroscopic fits to the DAHs' companions in Table~\ref{tab:non_DA_spec}.

We label the DA/DC and DA/DAH systems as candidates in Table~\ref{tab:non_DA_spec} because higher-resolution and/or better S/N spectra are required to confirm the nature of the non-DAs in these pairs.

The ``B'' component of the wide DWD J2047$+$0021B was identified by \citet{silvestri06} as a carbon atmosphere WD (DQ) with a K7 companion. This would make it the second such triple system composed of two WDs and a K star, after CDDS30 \citep{baxter14}. We list J2047$+$0021 as a candidate triple system in Table \ref{tab:non_DA_spec}.

Finally, while it is possible to form single WDs with masses as low as 0.45 \Msun\ \citep{kilic07}, observations of WDs with $M_{\rm WD} <0.45$~\Msun\ indicate a binary fraction $\gtrsim$80\% \citep{brown11}, suggesting that the low-mass WDs in several of our systems are in fact unresolved binaries. These systems are listed as candidate triple systems in Table~\ref{tab:non_DA_spec}. This list includes the two known systems, G 21-15 \citep{farihi05} and Gr 576/577 \citep{maxted00}, which are hierarchical triples composed of a close pair of WDs with a degenerate tertiary body in a larger orbit. In both systems, our spectroscopic fits return $\approx$0.45~\Msun\ for the unresolved close pair. Additionally, PG 0901$+$140A (with $M_{\rm WD} = 0.47 \pm 0.04$~\Msun) may be an unresolved degenerate pair. Because of the potential for mass transfer in the unresolved pairs, these systems cannot be used to constrain the IFMR and were excluded from our analysis. We note here that our spectroscopic solutions for PG 0901$+$140 differ substantially from those of \citet{farihi05}, who find that both WDs in the system are approximately $0.8$~\Msun.

For completeness, we also include in Table~\ref{tab:non_DA_spec} those systems for which we had only one spectrum and our one spectroscopic contaminant, the DA/A-star pair SDSS J2124$-$1620.

\setlength{\baselineskip}{0.6\baselineskip}
\bibliography{references}

\begin{thebibliography}{103}
\expandafter\ifx\csname natexlab\endcsname\relax\def\natexlab#1{#1}\fi

\bibitem[{{Abate} {et~al.}(2013){Abate}, {Pols}, {Izzard}, {Mohamed}, \& {de
  Mink}}]{abate13}
{Abate}, C., {Pols}, O.~R., {Izzard}, R.~G., {Mohamed}, S.~S., \& {de Mink},
  S.~E. 2013, \aap, 552, A26

\bibitem[{{Abazajian} {et~al.}(2009){Abazajian}, {Adelman-McCarthy},
  {Ag{\"u}eros}, {Allam}, {Allende Prieto}, {An}, {Anderson}, {Anderson},
  {Annis}, {Bahcall}, \& et~al.}]{DR7paper}
{Abazajian}, K.~N. {et~al.} 2009, \apjs, 182, 543

\bibitem[{{Ahn} {et~al.}(2012){Ahn}, {Alexandroff}, {Allende Prieto},
  {Anderson}, {Anderton}, {Andrews}, {Aubourg}, {Bailey}, {Balbinot}, {Barnes},
  \& et~al.}]{DR9paper}
{Ahn}, C.~P. {et~al.} 2012, \apjs, 203, 21

\bibitem[{{Althaus} {et~al.}(2005){Althaus}, {Serenelli}, {Panei},
  {C{\'o}rsico}, {Garc{\'{\i}}a-Berro}, \& {Sc{\'o}ccola}}]{althaus05}
{Althaus}, L.~G., {Serenelli}, A.~M., {Panei}, J.~A., {C{\'o}rsico}, A.~H.,
  {Garc{\'{\i}}a-Berro}, E., \& {Sc{\'o}ccola}, C.~G. 2005, \aap, 435, 631

\bibitem[{{Andrews} {et~al.}(2012){Andrews}, {Ag{\"u}eros}, {Belczynski},
  {Dhital}, {Kleinman}, \& {West}}]{andrews12}
{Andrews}, J.~J., {Ag{\"u}eros}, M.~A., {Belczynski}, K., {Dhital}, S.,
  {Kleinman}, S.~J., \& {West}, A.~A. 2012, \apj, 757, 170

\bibitem[{{Asplund} {et~al.}(2009){Asplund}, {Grevesse}, {Sauval}, \&
  {Scott}}]{asplund09}
{Asplund}, M., {Grevesse}, N., {Sauval}, A.~J., \& {Scott}, P. 2009, \araa, 47,
  481

\bibitem[{{Baxter} {et~al.}(2014){Baxter}, {Dobbie}, {Parker}, {Casewell},
  {Lodieu}, {Burleigh}, {Lawrie}, {K{\"u}lebi}, {Koester}, \&
  {Holland}}]{baxter14}
{Baxter}, R.~B. {et~al.} 2014, \mnras, 440, 3184

\bibitem[{{Bergeron} {et~al.}(1992){Bergeron}, {Saffer}, \&
  {Liebert}}]{bergeron92}
{Bergeron}, P., {Saffer}, R.~A., \& {Liebert}, J. 1992, \apj, 394, 228

\bibitem[{{B{\"o}hm-Vitense}(1958)}]{erica1958}
{B{\"o}hm-Vitense}, E. 1958, \zap, 46, 108

\bibitem[{{Boss}(1988)}]{boss88}
{Boss}, A.~P. 1988, Comments on Astrophysics, 12, 169

\bibitem[{{Brandt} \& {Huang}(2015)}]{brandt15}
{Brandt}, T.~D., \& {Huang}, C.~X. 2015, \apj, 807, 24

\bibitem[{{Brown} {et~al.}(2011){Brown}, {Kilic}, {Brown}, \&
  {Kenyon}}]{brown11}
{Brown}, J.~M., {Kilic}, M., {Brown}, W.~R., \& {Kenyon}, S.~J. 2011, \apj,
  730, 67

\bibitem[{{Canuto}(1998)}]{canuto98}
{Canuto}, V.~M. 1998, \apjl, 508, L103

\bibitem[{{Carrasco} {et~al.}(2014){Carrasco}, {Catal{\'a}n}, {Jordi},
  {Tremblay}, {Napiwotzki}, {Luri}, {Robin}, \& {Kowalski}}]{carrasco14}
{Carrasco}, J.~M., {Catal{\'a}n}, S., {Jordi}, C., {Tremblay}, P.-E.,
  {Napiwotzki}, R., {Luri}, X., {Robin}, A.~C., \& {Kowalski}, P.~M. 2014,
  \aap, 565, A11

\bibitem[{{Carrera} \& {Pancino}(2011)}]{carrera11}
{Carrera}, R., \& {Pancino}, E. 2011, \aap, 535, A30

\bibitem[{{Casali} {et~al.}(2007){Casali}, {Adamson}, {Alves de Oliveira},
  {Almaini}, {Burch}, {Chuter}, {Elliot}, {Folger}, {Foucaud}, {Hambly},
  {Hastie}, {Henry}, {Hirst}, {Irwin}, {Ives}, {Lawrence}, {Laidlaw}, {Lee},
  {Lewis}, {Lunney}, {McLay}, {Montgomery}, {Pickup}, {Read}, {Rees}, {Robson},
  {Sekiguchi}, {Vick}, {Warren}, \& {Woodward}}]{casali07}
{Casali}, M. {et~al.} 2007, \aap, 467, 777

\bibitem[{{Catal{\'a}n}(2015)}]{catalan15}
{Catal{\'a}n}, S. 2015, in Astronomical Society of the Pacific Conference
  Series, Vol. 493, 19th European Workshop on White Dwarfs, ed. P.~{Dufour},
  P.~{Bergeron}, \& G.~{Fontaine}, 325

\bibitem[{{Catal{\'a}n} {et~al.}(2008){Catal{\'a}n}, {Isern},
  {Garc{\'{\i}}a-Berro}, \& {Ribas}}]{catalan08b}
{Catal{\'a}n}, S., {Isern}, J., {Garc{\'{\i}}a-Berro}, E., \& {Ribas}, I. 2008,
  \mnras, 387, 1693

\bibitem[{{Claver} {et~al.}(2001){Claver}, {Liebert}, {Bergeron}, \&
  {Koester}}]{claver01}
{Claver}, C.~F., {Liebert}, J., {Bergeron}, P., \& {Koester}, D. 2001, \apj,
  563, 987

\bibitem[{{Dhital} {et~al.}(2010){Dhital}, {West}, {Stassun}, \&
  {Bochanski}}]{dhital10}
{Dhital}, S., {West}, A.~A., {Stassun}, K.~G., \& {Bochanski}, J.~J. 2010, \aj,
  139, 2566

\bibitem[{{Dhital} {et~al.}(2015){Dhital}, {West}, {Stassun}, {Schluns}, \&
  {Massey}}]{dhital15}
{Dhital}, S., {West}, A.~A., {Stassun}, K.~G., {Schluns}, K.~J., \& {Massey},
  A.~P. 2015, \aj, 150, 57

\bibitem[{{Dobbie} {et~al.}(2012){Dobbie}, {Baxter}, {K{\"u}lebi}, {Parker},
  {Koester}, {Jordan}, {Lodieu}, \& {Euchner}}]{dobbie12}
{Dobbie}, P.~D., {Baxter}, R., {K{\"u}lebi}, B., {Parker}, Q.~A., {Koester},
  D., {Jordan}, S., {Lodieu}, N., \& {Euchner}, F. 2012, \mnras, 421, 202

\bibitem[{{Dobbie} {et~al.}(2013){Dobbie}, {K{\"u}lebi}, {Casewell},
  {Burleigh}, {Parker}, {Baxter}, {Lawrie}, {Jordan}, \& {Koester}}]{dobbie13}
{Dobbie}, P.~D. {et~al.} 2013, \mnras, 428, L16

\bibitem[{{Dominguez} {et~al.}(1999){Dominguez}, {Chieffi}, {Limongi}, \&
  {Straniero}}]{dominguez99}
{Dominguez}, I., {Chieffi}, A., {Limongi}, M., \& {Straniero}, O. 1999, \apj,
  524, 226

\bibitem[{{Douglas} {et~al.}(2014){Douglas}, {Ag{\"u}eros}, {Covey}, {Bowsher},
  {Bochanski}, {Cargile}, {Kraus}, {Law}, {Lemonias}, {Arce}, {Fierroz}, \&
  {Kundert}}]{Douglas2014}
{Douglas}, S.~T. {et~al.} 2014, \apj, 795, 161

\bibitem[{{Farihi} {et~al.}(2005){Farihi}, {Becklin}, \&
  {Zuckerman}}]{farihi05}
{Farihi}, J., {Becklin}, E.~E., \& {Zuckerman}, B. 2005, \apjs, 161, 394

\bibitem[{{Finley} \& {Koester}(1997)}]{finley97}
{Finley}, D.~S., \& {Koester}, D. 1997, \apjl, 489, L79

\bibitem[{{Fontaine} {et~al.}(2001){Fontaine}, {Brassard}, \&
  {Bergeron}}]{fontaine01}
{Fontaine}, G., {Brassard}, P., \& {Bergeron}, P. 2001, \pasp, 113, 409

\bibitem[{{Foreman-Mackey} {et~al.}(2013){Foreman-Mackey}, {Hogg}, {Lang}, \&
  {Goodman}}]{foreman-mackey13}
{Foreman-Mackey}, D., {Hogg}, D.~W., {Lang}, D., \& {Goodman}, J. 2013, \pasp,
  125, 306

\bibitem[{{Freytag} {et~al.}(1996){Freytag}, {Ludwig}, \&
  {Steffen}}]{freytag96}
{Freytag}, B., {Ludwig}, H.-G., \& {Steffen}, M. 1996, \aap, 313, 497

\bibitem[{{Freytag} {et~al.}(2012){Freytag}, {Steffen}, {Ludwig},
  {Wedemeyer-B{\"o}hm}, {Schaffenberger}, \& {Steiner}}]{freytag12}
{Freytag}, B., {Steffen}, M., {Ludwig}, H.-G., {Wedemeyer-B{\"o}hm}, S.,
  {Schaffenberger}, W., \& {Steiner}, O. 2012, Journal of Computational
  Physics, 231, 919

\bibitem[{{Fuhrmann}(1998)}]{fuhrmann98}
{Fuhrmann}, K. 1998, \aap, 338, 161

\bibitem[{{Gianninas} {et~al.}(2005){Gianninas}, {Bergeron}, \&
  {Fontaine}}]{gianninas05}
{Gianninas}, A., {Bergeron}, P., \& {Fontaine}, G. 2005, \apj, 631, 1100

\bibitem[{{Gianninas} {et~al.}(2011){Gianninas}, {Bergeron}, \&
  {Ruiz}}]{gianninas11}
{Gianninas}, A., {Bergeron}, P., \& {Ruiz}, M.~T. 2011, \apj, 743, 138

\bibitem[{{Gim} {et~al.}(1998){Gim}, {Vandenberg}, {Stetson}, {Hesser}, \&
  {Zurek}}]{gim98}
{Gim}, M., {Vandenberg}, D.~A., {Stetson}, P.~B., {Hesser}, J.~E., \& {Zurek},
  D.~R. 1998, \pasp, 110, 1318

\bibitem[{{Girven} {et~al.}(2010){Girven}, {G{\"a}nsicke}, {K{\"u}lebi},
  {Steeghs}, {Jordan}, {Marsh}, \& {Koester}}]{girven10}
{Girven}, J., {G{\"a}nsicke}, B.~T., {K{\"u}lebi}, B., {Steeghs}, D., {Jordan},
  S., {Marsh}, T.~R., \& {Koester}, D. 2010, \mnras, 404, 159

\bibitem[{{Girven} {et~al.}(2011){Girven}, {G{\"a}nsicke}, {Steeghs}, \&
  {Koester}}]{girven11}
{Girven}, J., {G{\"a}nsicke}, B.~T., {Steeghs}, D., \& {Koester}, D. 2011,
  \mnras, 417, 1210

\bibitem[{{Goodman} \& {Weare}(2010)}]{goodman10}
{Goodman}, J., \& {Weare}, J. 2010, Comm. App. Math. Comp. Sci., 5, 65

\bibitem[{{Green} {et~al.}(1986){Green}, {Schmidt}, \& {Liebert}}]{green86}
{Green}, R.~F., {Schmidt}, M., \& {Liebert}, J. 1986, \apjs, 61, 305

\bibitem[{{Greenstein} {et~al.}(1983){Greenstein}, {Dolez}, \&
  {Vauclair}}]{greenstein83}
{Greenstein}, J.~L., {Dolez}, N., \& {Vauclair}, G. 1983, \aap, 127, 25

\bibitem[{{Grevesse} \& {Noels}(1993)}]{grevesse93}
{Grevesse}, N., \& {Noels}, A. 1993, in Origin and Evolution of the Elements,
  ed. N.~{Prantzos}, E.~{Vangioni-Flam}, \& M.~{Casse}, 15--25

\bibitem[{{Hambly} {et~al.}(2008){Hambly}, {Collins}, {Cross}, {Mann}, {Read},
  {Sutorius}, {Bond}, {Bryant}, {Emerson}, {Lawrence}, {Rimoldini}, {Stewart},
  {Williams}, {Adamson}, {Hirst}, {Dye}, \& {Warren}}]{hambly08}
{Hambly}, N.~C. {et~al.} 2008, \mnras, 384, 637

\bibitem[{{Herwig}(2000)}]{herwig00}
{Herwig}, F. 2000, \aap, 360, 952

\bibitem[{{Herwig} {et~al.}(1997){Herwig}, {Bloecker}, {Schoenberner}, \& {El
  Eid}}]{herwig97}
{Herwig}, F., {Bloecker}, T., {Schoenberner}, D., \& {El Eid}, M. 1997, \aap,
  324, L81

\bibitem[{{Hewett} {et~al.}(2006){Hewett}, {Warren}, {Leggett}, \&
  {Hodgkin}}]{hewett06}
{Hewett}, P.~C., {Warren}, S.~J., {Leggett}, S.~K., \& {Hodgkin}, S.~T. 2006,
  \mnras, 367, 454

\bibitem[{{Hodgkin} {et~al.}(2009){Hodgkin}, {Irwin}, {Hewett}, \&
  {Warren}}]{hodgkin09}
{Hodgkin}, S.~T., {Irwin}, M.~J., {Hewett}, P.~C., \& {Warren}, S.~J. 2009,
  \mnras, 394, 675

\bibitem[{{H{\"o}fner}(2009)}]{hofner09}
{H{\"o}fner}, S. 2009, in Astronomical Society of the Pacific Conference
  Series, Vol. 414, Cosmic Dust - Near and Far, ed. T.~{Henning},
  E.~{Gr{\"u}n}, \& J.~{Steinacker}, 3

\bibitem[{{Hogg} {et~al.}(2010){Hogg}, {Bovy}, \& {Lang}}]{hogg10}
{Hogg}, D.~W., {Bovy}, J., \& {Lang}, D. 2010, ArXiv e-prints

\bibitem[{{Iben}(1975)}]{iben75}
{Iben}, Jr., I. 1975, \apj, 196, 525

\bibitem[{{Kalirai} {et~al.}(2008){Kalirai}, {Hansen}, {Kelson}, {Reitzel},
  {Rich}, \& {Richer}}]{kalirai08}
{Kalirai}, J.~S., {Hansen}, B.~M.~S., {Kelson}, D.~D., {Reitzel}, D.~B.,
  {Rich}, R.~M., \& {Richer}, H.~B. 2008, \apj, 676, 594

\bibitem[{{Kalirai} {et~al.}(2014){Kalirai}, {Marigo}, \&
  {Tremblay}}]{kalirai14}
{Kalirai}, J.~S., {Marigo}, P., \& {Tremblay}, P.-E. 2014, \apj, 782, 17

\bibitem[{{Kalirai} {et~al.}(2005){Kalirai}, {Richer}, {Reitzel}, {Hansen},
  {Rich}, {Fahlman}, {Gibson}, \& {von Hippel}}]{kalirai05}
{Kalirai}, J.~S., {Richer}, H.~B., {Reitzel}, D., {Hansen}, B.~M.~S., {Rich},
  R.~M., {Fahlman}, G.~G., {Gibson}, B.~K., \& {von Hippel}, T. 2005, \apjl,
  618, L123

\bibitem[{{Karakas} {et~al.}(2002){Karakas}, {Lattanzio}, \&
  {Pols}}]{karakas02}
{Karakas}, A.~I., {Lattanzio}, J.~C., \& {Pols}, O.~R. 2002, PASA, 19, 515

\bibitem[{{Karovicova} {et~al.}(2013){Karovicova}, {Wittkowski}, {Ohnaka},
  {Boboltz}, {Fossat}, \& {Scholz}}]{karovicova13}
{Karovicova}, I., {Wittkowski}, M., {Ohnaka}, K., {Boboltz}, D.~A., {Fossat},
  E., \& {Scholz}, M. 2013, \aap, 560, A75

\bibitem[{{Kepler} {et~al.}(2015){Kepler}, {Pelisoli}, {Koester}, {Ourique},
  {Kleinman}, {Romero}, {Nitta}, {Eisenstein}, {Costa}, {K{\"u}lebi}, {Jordan},
  {Dufour}, {Giommi}, \& {Rebassa-Mansergas}}]{kepler15}
{Kepler}, S.~O. {et~al.} 2015, \mnras, 446, 4078

\bibitem[{{Kilic} {et~al.}(2006){Kilic}, {Munn}, {Harris}, {Liebert}, {von
  Hippel}, {Williams}, {Metcalfe}, {Winget}, \& {Levine}}]{kilic06}
{Kilic}, M. {et~al.} 2006, \aj, 131, 582

\bibitem[{{Kilic} {et~al.}(2007){Kilic}, {Stanek}, \& {Pinsonneault}}]{kilic07}
{Kilic}, M., {Stanek}, K.~Z., \& {Pinsonneault}, M.~H. 2007, \apj, 671, 761

\bibitem[{{Kleinman} {et~al.}(2013){Kleinman}, {Kepler}, {Koester}, {Pelisoli},
  {Pe{\c c}anha}, {Nitta}, {Costa}, {Krzesinski}, {Dufour}, {Lachapelle},
  {Bergeron}, {Yip}, {Harris}, {Eisenstein}, {Althaus}, \&
  {C{\'o}rsico}}]{kleinman13}
{Kleinman}, S.~J. {et~al.} 2013, \apjs, 204, 5

\bibitem[{{Koester} {et~al.}(2009){Koester}, {Voss}, {Napiwotzki},
  {Christlieb}, {Homeier}, {Lisker}, {Reimers}, \& {Heber}}]{koester09}
{Koester}, D., {Voss}, B., {Napiwotzki}, R., {Christlieb}, N., {Homeier}, D.,
  {Lisker}, T., {Reimers}, D., \& {Heber}, U. 2009, \aap, 505, 441

\bibitem[{{Kraus} \& {Hillenbrand}(2009)}]{kraus09}
{Kraus}, A.~L., \& {Hillenbrand}, L.~A. 2009, \apj, 704, 531

\bibitem[{{Lawrence} {et~al.}(2007){Lawrence}, {Warren}, {Almaini}, {Edge},
  {Hambly}, {Jameson}, {Lucas}, {Casali}, {Adamson}, {Dye}, {Emerson},
  {Foucaud}, {Hewett}, {Hirst}, {Hodgkin}, {Irwin}, {Lodieu}, {McMahon},
  {Simpson}, {Smail}, {Mortlock}, \& {Folger}}]{lawrence07}
{Lawrence}, A. {et~al.} 2007, \mnras, 379, 1599

\bibitem[{{Lee-Brown} {et~al.}(2015){Lee-Brown}, {Anthony-Twarog},
  {Deliyannis}, {Rich}, \& {Twarog}}]{lee-brown15}
{Lee-Brown}, D.~B., {Anthony-Twarog}, B.~J., {Deliyannis}, C.~P., {Rich}, E.,
  \& {Twarog}, B.~A. 2015, \aj, 149, 121

\bibitem[{{Liebert} {et~al.}(2005){Liebert}, {Bergeron}, \&
  {Holberg}}]{liebert05}
{Liebert}, J., {Bergeron}, P., \& {Holberg}, J.~B. 2005, \apjs, 156, 47

\bibitem[{{Lugaro} {et~al.}(2003){Lugaro}, {Herwig}, {Lattanzio}, {Gallino}, \&
  {Straniero}}]{lugaro03}
{Lugaro}, M., {Herwig}, F., {Lattanzio}, J.~C., {Gallino}, R., \& {Straniero},
  O. 2003, \apj, 586, 1305

\bibitem[{{Marigo} {et~al.}(2013){Marigo}, {Bressan}, {Nanni}, {Girardi}, \&
  {Pumo}}]{marigo13a}
{Marigo}, P., {Bressan}, A., {Nanni}, A., {Girardi}, L., \& {Pumo}, M.~L. 2013,
  \mnras, 434, 488

\bibitem[{{Marigo} \& {Girardi}(2007)}]{marigo07}
{Marigo}, P., \& {Girardi}, L. 2007, \aap, 469, 239

\bibitem[{{Maxted} {et~al.}(2000){Maxted}, {Marsh}, {Moran}, \&
  {Han}}]{maxted00}
{Maxted}, P.~F.~L., {Marsh}, T.~R., {Moran}, C.~K.~J., \& {Han}, Z. 2000,
  \mnras, 314, 334

\bibitem[{{Mazzei} \& {Pigatto}(1988)}]{mazzei88}
{Mazzei}, P., \& {Pigatto}, L. 1988, \aap, 193, 148

\bibitem[{{Meng} {et~al.}(2008){Meng}, {Chen}, \& {Han}}]{meng08}
{Meng}, X., {Chen}, X., \& {Han}, Z. 2008, \aap, 487, 625

\bibitem[{{Mohamed} \& {Podsiadlowski}(2007)}]{mohamed07}
{Mohamed}, S., \& {Podsiadlowski}, P. 2007, in Astronomical Society of the
  Pacific Conference Series, Vol. 372, 15th European Workshop on White Dwarfs,
  ed. R.~{Napiwotzki} \& M.~R. {Burleigh}, 397

\bibitem[{{Mohamed} \& {Podsiadlowski}(2012)}]{mohamed12}
{Mohamed}, S., \& {Podsiadlowski}, P. 2012, Baltic Astronomy, 21, 88

\bibitem[{{Munn} {et~al.}(2004){Munn}, {Monet}, {Levine}, {Canzian}, {Pier},
  {Harris}, {Lupton}, {Ivezi{\'c}}, {Hindsley}, {Hennessy}, {Schneider}, \&
  {Brinkmann}}]{munn04}
{Munn}, J.~A. {et~al.} 2004, \aj, 127, 3034

\bibitem[{{Munn} {et~al.}(2008){Munn}, {Monet}, {Levine}, {Canzian}, {Pier},
  {Harris}, {Lupton}, {Ivezi{\'c}}, {Hindsley}, {Hennessy}, {Schneider}, \&
  {Brinkmann}}]{munn08}
---. 2008, \aj, 136, 895

\bibitem[{{Oswalt} {et~al.}(1988){Oswalt}, {Hintzen}, {Liebert}, \&
  {Sion}}]{oswalt88}
{Oswalt}, T.~D., {Hintzen}, P.~M., {Liebert}, J.~W., \& {Sion}, E.~M. 1988,
  \apjl, 333, L87

\bibitem[{{Overbeek} {et~al.}(2015){Overbeek}, {Friel}, {Jacobson}, {Johnson},
  {Pilachowski}, \& {M{\'e}sz{\'a}ros}}]{overbeek15}
{Overbeek}, J.~C., {Friel}, E.~D., {Jacobson}, H.~R., {Johnson}, C.~I.,
  {Pilachowski}, C.~A., \& {M{\'e}sz{\'a}ros}, S. 2015, \aj, 149, 15

\bibitem[{{Paxton} {et~al.}(2011){Paxton}, {Bildsten}, {Dotter}, {Herwig},
  {Lesaffre}, \& {Timmes}}]{paxton11}
{Paxton}, B., {Bildsten}, L., {Dotter}, A., {Herwig}, F., {Lesaffre}, P., \&
  {Timmes}, F. 2011, \apjs, 192, 3

\bibitem[{{Paxton} {et~al.}(2013){Paxton}, {Cantiello}, {Arras}, {Bildsten},
  {Brown}, {Dotter}, {Mankovich}, {Montgomery}, {Stello}, {Timmes}, \&
  {Townsend}}]{paxton13}
{Paxton}, B. {et~al.} 2013, \apjs, 208, 4

\bibitem[{{Paxton} {et~al.}(2015){Paxton}, {Marchant}, {Schwab}, {Bauer},
  {Bildsten}, {Cantiello}, {Dessart}, {Farmer}, {Hu}, {Langer}, {Townsend},
  {Townsley}, \& {Timmes}}]{paxton15}
---. 2015, ArXiv e-prints

\bibitem[{{Perryman} {et~al.}(1998){Perryman}, {Brown}, {Lebreton}, {Gomez},
  {Turon}, {Cayrel de Strobel}, {Mermilliod}, {Robichon}, {Kovalevsky}, \&
  {Crifo}}]{perryman98}
{Perryman}, M.~A.~C. {et~al.} 1998, \aap, 331, 81

\bibitem[{{Pignatari} {et~al.}(2013){Pignatari}, {Herwig}, {Hirschi},
  {Bennett}, {Rockefeller}, {Fryer}, {Timmes}, {Heger}, {Jones}, {Battino},
  {Ritter}, {Dotter}, {Trappitsch}, {Diehl}, {Frischknecht}, {Hungerford},
  {Magkotsios}, {Travaglio}, \& {Young}}]{pignatari13}
{Pignatari}, M. {et~al.} 2013, ArXiv e-prints

\bibitem[{{Renedo} {et~al.}(2010){Renedo}, {Althaus}, {Miller Bertolami},
  {Romero}, {C{\'o}rsico}, {Rohrmann}, \& {Garc{\'{\i}}a-Berro}}]{renedo10}
{Renedo}, I., {Althaus}, L.~G., {Miller Bertolami}, M.~M., {Romero}, A.~D.,
  {C{\'o}rsico}, A.~H., {Rohrmann}, R.~D., \& {Garc{\'{\i}}a-Berro}, E. 2010,
  \apj, 717, 183

\bibitem[{{Romero} {et~al.}(2015){Romero}, {Campos}, \& {Kepler}}]{romero15}
{Romero}, A.~D., {Campos}, F., \& {Kepler}, S.~O. 2015, \mnras, 450, 3708

\bibitem[{{Salaris} {et~al.}(2009){Salaris}, {Serenelli}, {Weiss}, \& {Miller
  Bertolami}}]{salaris09}
{Salaris}, M., {Serenelli}, A., {Weiss}, A., \& {Miller Bertolami}, M. 2009,
  \apj, 692, 1013

\bibitem[{{Schneider} {et~al.}(2010){Schneider}, {Richards}, {Hall}, {Strauss},
  {Anderson}, {Boroson}, {Ross}, {Shen}, {Brandt}, {Fan}, {Inada}, {Jester},
  {Knapp}, {Krawczyk}, {Thakar}, {Vanden Berk}, {Voges}, {Yanny}, {York},
  {Bahcall}, {Bizyaev}, {Blanton}, {Brewington}, {Brinkmann}, {Eisenstein},
  {Frieman}, {Fukugita}, {Gray}, {Gunn}, {Hibon}, {Ivezi{\'c}}, {Kent}, {Kron},
  {Lee}, {Lupton}, {Malanushenko}, {Malanushenko}, {Oravetz}, {Pan}, {Pier},
  {Price}, {Saxe}, {Schlegel}, {Simmons}, {Snedden}, {SubbaRao}, {Szalay}, \&
  {Weinberg}}]{schneider10}
{Schneider}, D.~P. {et~al.} 2010, \aj, 139, 2360

\bibitem[{{Schwarzschild} \& {H{\"a}rm}(1965)}]{schwarzschild65}
{Schwarzschild}, M., \& {H{\"a}rm}, R. 1965, \apj, 142, 855

\bibitem[{{Shu} {et~al.}(1987){Shu}, {Adams}, \& {Lizano}}]{shu87}
{Shu}, F.~H., {Adams}, F.~C., \& {Lizano}, S. 1987, \araa, 25, 23

\bibitem[{{Silvestri} {et~al.}(2006){Silvestri}, {Hawley}, {West}, {Szkody},
  {Bochanski}, {Eisenstein}, {McGehee}, {Schmidt}, {Smith}, {Wolfe}, {Harris},
  {Kleinman}, {Liebert}, {Nitta}, {Barentine}, {Brewington}, {Brinkmann},
  {Harvanek}, {Krzesi{\'n}ski}, {Long}, {Neilsen}, {Schneider}, \&
  {Snedden}}]{silvestri06}
{Silvestri}, N.~M. {et~al.} 2006, \aj, 131, 1674

\bibitem[{{Smith} {et~al.}(2009){Smith}, {Evans}, {Belokurov}, {Hewett},
  {Bramich}, {Gilmore}, {Irwin}, {Vidrih}, \& {Zucker}}]{smith09}
{Smith}, M.~C. {et~al.} 2009, \mnras, 399, 1223

\bibitem[{{Sokoloski} \& {Bildsten}(2010)}]{sokoloski10}
{Sokoloski}, J.~L., \& {Bildsten}, L. 2010, \apj, 723, 1188

\bibitem[{{Sweeney}(1976)}]{sweeney76}
{Sweeney}, M.~A. 1976, \aap, 49, 375

\bibitem[{{Sweigart} \& {Gross}(1978)}]{sweigart78}
{Sweigart}, A.~V., \& {Gross}, P.~G. 1978, \apjs, 36, 405

\bibitem[{{Tremblay} {et~al.}(2010){Tremblay}, {Bergeron}, {Kalirai}, \&
  {Gianninas}}]{tremblay10}
{Tremblay}, P.-E., {Bergeron}, P., {Kalirai}, J.~S., \& {Gianninas}, A. 2010,
  \apj, 712, 1345

\bibitem[{{Tremblay} {et~al.}(2013){Tremblay}, {Ludwig}, {Steffen}, \&
  {Freytag}}]{tremblay13}
{Tremblay}, P.-E., {Ludwig}, H.-G., {Steffen}, M., \& {Freytag}, B. 2013, \aap,
  559, A104

\bibitem[{{van Loon} {et~al.}(2005){van Loon}, {Cioni}, {Zijlstra}, \&
  {Loup}}]{vLoon05}
{van Loon}, J.~T., {Cioni}, M.-R.~L., {Zijlstra}, A.~A., \& {Loup}, C. 2005,
  \aap, 438, 273

\bibitem[{{Vassiliadis} \& {Wood}(1993)}]{vassiliadis93}
{Vassiliadis}, E., \& {Wood}, P.~R. 1993, \apj, 413, 641

\bibitem[{{Wachter} {et~al.}(2002){Wachter}, {Schr{\"o}der}, {Winters},
  {Arndt}, \& {Sedlmayr}}]{wachter02}
{Wachter}, A., {Schr{\"o}der}, K.-P., {Winters}, J.~M., {Arndt}, T.~U., \&
  {Sedlmayr}, E. 2002, \aap, 384, 452

\bibitem[{{Weidemann}(2000)}]{weidemann00}
{Weidemann}, V. 2000, \aap, 363, 647

\bibitem[{{Weidemann} \& {Koester}(1983)}]{weidemann83}
{Weidemann}, V., \& {Koester}, D. 1983, \aap, 121, 77

\bibitem[{{Weiss} \& {Ferguson}(2009)}]{weiss09}
{Weiss}, A., \& {Ferguson}, J.~W. 2009, \aap, 508, 1343

\bibitem[{{Williams} {et~al.}(2009){Williams}, {Bolte}, \&
  {Koester}}]{williams09}
{Williams}, K.~A., {Bolte}, M., \& {Koester}, D. 2009, \apj, 693, 355

\bibitem[{{Wood}(1995)}]{wood95}
{Wood}, M.~A. 1995, in Lecture Notes in Physics, Berlin Springer Verlag, Vol.
  443, White Dwarfs, ed. D.~{Koester} \& K.~{Werner}, 41

\bibitem[{{York} {et~al.}(2000){York}, {Adelman}, {Anderson}, {Anderson},
  {Annis}, {Bahcall}, {Bakken}, {Barkhouser}, {Bastian}, {Berman}, {Boroski},
  {Bracker}, {Briegel}, {Briggs}, {Brinkmann}, {Brunner}, {Burles}, {Carey},
  {Carr}, {Castander}, {Chen}, {Colestock}, {Connolly}, {Crocker}, {Csabai},
  {Czarapata}, {Davis}, {Doi}, {Dombeck}, {Eisenstein}, {Ellman}, {Elms},
  {Evans}, {Fan}, {Federwitz}, {Fiscelli}, {Friedman}, {Frieman}, {Fukugita},
  {Gillespie}, {Gunn}, {Gurbani}, {de Haas}, {Haldeman}, {Harris}, {Hayes},
  {Heckman}, {Hennessy}, {Hindsley}, {Holm}, {Holmgren}, {Huang}, {Hull},
  {Husby}, {Ichikawa}, {Ichikawa}, {Ivezi{\'c}}, {Kent}, {Kim}, {Kinney},
  {Klaene}, {Kleinman}, {Kleinman}, {Knapp}, {Korienek}, {Kron}, {Kunszt},
  {Lamb}, {Lee}, {Leger}, {Limmongkol}, {Lindenmeyer}, {Long}, {Loomis},
  {Loveday}, {Lucinio}, {Lupton}, {MacKinnon}, {Mannery}, {Mantsch}, {Margon},
  {McGehee}, {McKay}, {Meiksin}, {Merelli}, {Monet}, {Munn}, {Narayanan},
  {Nash}, {Neilsen}, {Neswold}, {Newberg}, {Nichol}, {Nicinski}, {Nonino},
  {Okada}, {Okamura}, {Ostriker}, {Owen}, {Pauls}, {Peoples}, {Peterson},
  {Petravick}, {Pier}, {Pope}, {Pordes}, {Prosapio}, {Rechenmacher}, {Quinn},
  {Richards}, {Richmond}, {Rivetta}, {Rockosi}, {Ruthmansdorfer}, {Sandford},
  {Schlegel}, {Schneider}, {Sekiguchi}, {Sergey}, {Shimasaku}, {Siegmund},
  {Smee}, {Smith}, {Snedden}, {Stone}, {Stoughton}, {Strauss}, {Stubbs},
  {SubbaRao}, {Szalay}, {Szapudi}, {Szokoly}, {Thakar}, {Tremonti}, {Tucker},
  {Uomoto}, {Vanden Berk}, {Vogeley}, {Waddell}, {Wang}, {Watanabe},
  {Weinberg}, {Yanny}, {Yasuda}, \& {SDSS Collaboration}}]{york00}
{York}, D.~G. {et~al.} 2000, \aj, 120, 1579

\bibitem[{{Zacharias} {et~al.}(2004){Zacharias}, {Urban}, {Zacharias},
  {Wycoff}, {Hall}, {Monet}, \& {Rafferty}}]{zacharias04}
{Zacharias}, N., {Urban}, S.~E., {Zacharias}, M.~I., {Wycoff}, G.~L., {Hall},
  D.~M., {Monet}, D.~G., \& {Rafferty}, T.~J. 2004, \aj, 127, 3043

\end{thebibliography}

\setlength{\baselineskip}{1.667\baselineskip}

\clearpage
\begin{deluxetable*}{lccccr@{$\pm$}lr@{$\pm$}lr@{$\pm$}lr@{$\pm$}lr@{$\pm$}l}
\tablecaption{Fit Results for the non-DA/DA Spectroscopic Sample}
\tablehead{
\colhead{Name} & 
\colhead{Telescope} & 
\colhead{\# of Fitted} &
\colhead{WD} &
\colhead{S/N\tablenotemark{a}} &
\multicolumn{2}{c}{\Teff} & 
\multicolumn{2}{c}{\logg} & 
\multicolumn{2}{c}{Distance} & 
\multicolumn{2}{c}{$M_{\rm WD}$} & 
\multicolumn{2}{c}{\tauc} \\
\colhead{} & 
\colhead{} & 
\colhead{Balmer Lines} &
\colhead{Type} &
\colhead{} &
\multicolumn{2}{c}{(K)} & 
\multicolumn{2}{c}{} & 
\multicolumn{2}{c}{(pc)} & 
\multicolumn{2}{c}{(\Msun)} & 
\multicolumn{2}{c}{(Myr)} 
}
\startdata
\cutinhead{DA/DB Systems\tablenotemark{b}}
J0849$+$4712A  &  APO  & 6 & DA & 28 & 11720 & 210  &  8.05 & 0.06  &  152 & 7  &  0.635 & 0.036  &  425 & 41  \\ 
J0849$+$4712B  &  APO  &\nodata& DB & \nodata & 17480 & 310 & 8.08 & 0.06 & \multicolumn{2}{c}{\nodata} & \multicolumn{2}{c}{\nodata} & \multicolumn{2}{c}{\nodata} \\ 
J2355$+$1708A  &  APO  & 5 & DA & 23 & 10160 & 160  &  8.31 & 0.07  &  131 & 7  &  0.797 & 0.045  &  927 & 109  \\ 
J2355$+$1708B  &  APO  &\nodata& DB & \nodata & 21470 & 440 & 8.12 & 0.03 & \multicolumn{2}{c}{\nodata} & \multicolumn{2}{c}{\nodata} & \multicolumn{2}{c}{\nodata} \\ 
\cutinhead{Candidate DA/DAH Systems\tablenotemark{c}}
PG 1258$+$593A\tablenotemark{d}  & \nodata & \nodata & DA & \nodata & 15160 & 240 & 8.00 & 0.05 & \multicolumn{2}{c}{65} & 0.61 & 0.03 & \multicolumn{2}{c}{\nodata} \\ 
PG 1258$+$593B  &  SDSS  &\nodata& DAH & \nodata & \multicolumn{2}{c}{\nodata} & \multicolumn{2}{c}{\nodata} & \multicolumn{2}{c}{\nodata} & \multicolumn{2}{c}{\nodata} & \multicolumn{2}{c}{\nodata} \\ 
J0002$+$0733A  &  APO  & \nodata & DA & \nodata & \multicolumn{2}{c}{\nodata} & \multicolumn{2}{c}{\nodata} & \multicolumn{2}{c}{\nodata} & \multicolumn{2}{c}{\nodata} & \multicolumn{2}{c}{\nodata} \\ 
J0002$+$0733B  &  APO  & \nodata & DAH & \nodata & \multicolumn{2}{c}{\nodata} & \multicolumn{2}{c}{\nodata} & \multicolumn{2}{c}{\nodata} & \multicolumn{2}{c}{\nodata} & \multicolumn{2}{c}{\nodata} \\ 
J0748$+$3025\tablenotemark{e}  &  SDSS  &\nodata& DAH & \nodata & \multicolumn{2}{c}{\nodata} & \multicolumn{2}{c}{\nodata} & \multicolumn{2}{c}{\nodata} & \multicolumn{2}{c}{\nodata} & \multicolumn{2}{c}{\nodata} \\ 
J1314$+$1732A  &  SDSS  &\nodata& DAH & \nodata & \multicolumn{2}{c}{\nodata} & \multicolumn{2}{c}{\nodata} & \multicolumn{2}{c}{\nodata} & \multicolumn{2}{c}{\nodata} & \multicolumn{2}{c}{\nodata} \\ 
J1314$+$1732B  &  APO  & 6 & DA & 50 & 12530 & 210  &  8.04 & 0.05  &  84 & 3  &  0.631 & 0.029  &  352 & 29  \\ 
J1412$+$4216A  &  APO  & 6 & DA & 104 & 15420 & 240  &  8.12 & 0.05  &  79 & 3  &  0.686 & 0.027  &  220 & 19  \\ 
J1412$+$4216B  &  APO, SDSS  &\nodata& DAH & \nodata & \multicolumn{2}{c}{\nodata} & \multicolumn{2}{c}{\nodata} & \multicolumn{2}{c}{\nodata} & \multicolumn{2}{c}{\nodata} & \multicolumn{2}{c}{\nodata} \\ 
J2044$+$4030A  &  APO  &\nodata& DAH & \nodata & \multicolumn{2}{c}{\nodata} & \multicolumn{2}{c}{\nodata} & \multicolumn{2}{c}{\nodata} & \multicolumn{2}{c}{\nodata} & \multicolumn{2}{c}{\nodata} \\ 
J2044$+$4030B  &  APO  & 6 & DA & 107 & 13590 & 240  &  8.00 & 0.05  &  61 & 2  &  0.607 & 0.026  &  264 & 22  \\ 
J2259$+$1404A  &  APO  & 5 & DA & 48 & 25200 & 390  &  8.68 & 0.05  &  103 & 5  &  1.042 & 0.028  &  137 & 15  \\ 
J2259$+$1404B\tablenotemark{e}  &  APO  &\nodata& DAH & \nodata & \multicolumn{2}{c}{\nodata} & \multicolumn{2}{c}{\nodata} & \multicolumn{2}{c}{\nodata} & \multicolumn{2}{c}{\nodata} & \multicolumn{2}{c}{\nodata} \\ 
\cutinhead{Candidate DA/DC Systems\tablenotemark{c}}
LP 549-32  &  APO  & \nodata & DC & \nodata & \multicolumn{2}{c}{\nodata} & \multicolumn{2}{c}{\nodata} & \multicolumn{2}{c}{\nodata} & \multicolumn{2}{c}{\nodata} & \multicolumn{2}{c}{\nodata} \\ 
LP 549-33  &  SDSS  & 5 & DA & 54 & 6660 & 110  &  7.84 & 0.10  &  31 & 2  &  0.499 & 0.052  &  1370 & 179  \\ 
LP 549-33  &  APO  & 5 & DA & 50 & 7090 & 110  &  8.41 & 0.09  &  24 & 2  &  0.855 & 0.057  &  3342 & 424  \\ 
GD 559A  &  APO  & 7 & DA & 51 & 18720 & 290  &  8.14 & 0.05  &  69 & 3  &  0.701 & 0.028  &  121 & 13  \\ 
GD 559B  &  APO  &\nodata& DC & \nodata & \multicolumn{2}{c}{\nodata} & \multicolumn{2}{c}{\nodata} & \multicolumn{2}{c}{\nodata} & \multicolumn{2}{c}{\nodata} & \multicolumn{2}{c}{\nodata} \\ 
J0029$+$0015A  &  SDSS  &\nodata& DC & \nodata & \multicolumn{2}{c}{\nodata} & \multicolumn{2}{c}{\nodata} & \multicolumn{2}{c}{\nodata} & \multicolumn{2}{c}{\nodata} & \multicolumn{2}{c}{\nodata} \\ 
J0029$+$0015B  &  SDSS  & 6 & DA & 33,17 & 9960 & 150  &  8.08 & 0.06  &  167 & 8  &  0.648 & 0.036  &  682 & 65  \\ 
J0344$+$1510A  &  APO  &\nodata& DC & \nodata & \multicolumn{2}{c}{\nodata} & \multicolumn{2}{c}{\nodata} & \multicolumn{2}{c}{\nodata} & \multicolumn{2}{c}{\nodata} & \multicolumn{2}{c}{\nodata} \\ 
J0344$+$1510B  &  APO  & 6 & DA & 36 & 8300 & 130  &  7.92 & 0.09  &  64 & 4  &  0.547 & 0.048  &  871 & 104  \\ 
J1544$+$2344A  &  SDSS  & 5 & DA & 15 & 9470 & 180  &  7.97 & 0.13  &  162 & 14  &  0.581 & 0.075  &  668 & 121  \\ 
J1544$+$2344B  &  SDSS  &\nodata& DC & \nodata & \multicolumn{2}{c}{\nodata} & \multicolumn{2}{c}{\nodata} & \multicolumn{2}{c}{\nodata} & \multicolumn{2}{c}{\nodata} & \multicolumn{2}{c}{\nodata} \\ 
\cutinhead{Candidate Triple Systems}
G 21-15\tablenotemark{f}  &  VLT  & 7 & DA+DC & 10,24,10,22 & 14460 & 210  &  7.66 & 0.04  & 41 & 2 &  0.448 & 0.018  &  244 & 10  \\ 
Gr 577  &  APO  & 7 & DA+DA & 85 & 9390 & 130  &  7.74 & 0.05  &  29 & 1  &  0.461 & 0.023  &  986 & 33  \\ 
Gr 576  &  APO  & 7 & DA & 73 & 14240 & 270  &  8.09 & 0.05  &  39 & 1  &  0.665 & 0.027  &  265 & 23  \\ 
PG 0901$+$140A  &  APO  &  6 & DA+DA? & 67 & 9100 & 140  &  7.78 & 0.08  &  59 & 3  &  0.474 & 0.041  &  585 & 57  \\ 
PG 0901$+$140B  &  APO  & 6 & DA & 39 & 8120 & 120  &  7.89 & 0.07  &  58 & 3  &  0.531 & 0.039  &  886 & 87  \\ 
J2047$+$0021A  &  SDSS  & 5 & DA & 51 & 14330 & 300  &  8.03 & 0.05  &  140 & 5  &  0.627 & 0.028  &  236 & 23  \\ 
J2047$+$0021B\tablenotemark{g}  &  SDSS  &\nodata& DQ+K7 & \nodata & \multicolumn{2}{c}{\nodata} & \multicolumn{2}{c}{\nodata} & \multicolumn{2}{c}{\nodata} & \multicolumn{2}{c}{\nodata} & \multicolumn{2}{c}{\nodata} \\
\cutinhead{Systems With Only One Spectrum}
J0000$-$1051A  &  SDSS  & 5 & DA & 13 & 8590 & 180  &  8.03 & 0.19  &  161 & 21  &  0.614 & 0.111  &  933 & 256  \\ 
J0117$+$2440B  &  SDSS  & 6 & DA & 12 & 17730 & 670  &  8.26 & 0.11  &  472 & 41  &  0.777 & 0.070  &  183 & 42  \\ 
J0139$+$1447A  &  SDSS  & 4 & DA & 7 & 8740 & 250  &  8.29 & 0.31  &  186 & 49  &  0.779 & 0.200  &  1316 & 880  \\ 
J0754$+$4950A  &  SDSS  & 4 & DA & 6 & 6710 & 340  &  7.39 & 0.82  &  133 & 55  &  0.328 & 0.326  &  1603 & 1090  \\ 
J1254$-$0218A  &  SDSS  & 7 & DA & 92 & 17260 & 260  &  8.11 & 0.05  &  120 & 5  &  0.681 & 0.027  &  151 & 14  \\ 
J1254$-$0218A  &  VLT  & 5 & DA & 8,8 & 15510 & 250  &  7.97 & 0.05  &  119 & 4  &  0.598 & 0.027  &  170 & 16  \\ 
J2115$-$0741A  &  SDSS  & 5 & DA & 27 & 8090 & 130  &  8.13 & 0.09  &  68 & 5  &  0.675 & 0.057  &  1264 & 198  \\ 
J2332$+$4917A  &  SDSS  & 6 & DA & 20 & 18490 & 450  &  7.99 & 0.07  &  400 & 21  &  0.614 & 0.041  &  93 & 17  \\
\cutinhead{Contaminants\tablenotemark{h}}
J2124$-$1620A  &  APO  & \nodata & non-WD & \nodata & \multicolumn{2}{c}{\nodata} & \multicolumn{2}{c}{\nodata} & \multicolumn{2}{c}{\nodata} & \multicolumn{2}{c}{\nodata} & \multicolumn{2}{c}{\nodata} \\ 
J2124$-$1620B  &  APO  & 5 & DA & 19 & 10290 & 180  &  8.06 & 0.09  &  127 & 8  &  0.640 & 0.055  &  611 & 85   
\enddata 
\tablenotetext{a}{Objects with more than one listed S/N indicate multiple spectra were used to fit for the WD parameters.}
\tablenotetext{b}{The DB WDs in our spectroscopic sample were fit by P.~Bergeron (pers.~communication).}
\tablenotetext{c}{Confirming the nature of the non-DA WDs in these pairs requires higher S/N spectra.}
\tablenotetext{d}{Spectroscopic data for PG 1258$+$593A were taken from \citet{gianninas11}. For a detailed discussion on this system see \citet{girven10}.}
\tablenotetext{e}{J0748$+$3025 and J2259$+$1404 have already been identified as DA+DAH pairs \citep[see][and references therein]{baxter14}. J0748$+$3025 is composed of a pair of WDs separated by $\approx$1\farcs5. However, the SDSS spectrum shows three cores in H$\beta$; see Figure \ref{fig:non_DA_spec}.}
\tablenotetext{f}{
\citet{farihi05} identified the companion to G 21-15 as a DC WD.}
\tablenotetext{g}{\citet{silvestri06} identified J2047$+$0021B as an unresolved binary composed of a carbon atmosphere WD (DQ) with a K7 companion.}
\tablenotetext{g}{J2124$-$1620 is composed of a DA WD and an A star.}
\label{tab:non_DA_spec}
\end{deluxetable*}

\end{document}